\documentclass[a4paper,11pt]{article}
\pdfoutput=1 % if your are submitting a pdflatex (i.e. if you have
\usepackage{packages/jcappub} % for details on the use of the package, please
\usepackage[T1]{fontenc} % if needed
\usepackage{amssymb,amsfonts,slashed,amsthm,amsmath,graphicx}
\usepackage[all]{xy}
\usepackage{dcolumn}
\usepackage{bm}
\usepackage{mathrsfs}
\usepackage{ucs}
\usepackage[utf8x]{inputenc}
\usepackage{hyperref}
\usepackage{hepunits}
\usepackage{packages/slashed}
\usepackage{floatrow}
\usepackage{subfigure}
\usepackage{indentfirst}
\usepackage{dsfont}
\renewcommand\({\left(}
\renewcommand\){\right)}
\renewcommand\[{\left[}
\renewcommand\]{\right]}

\renewcommand{\Re}{\operatorname{Re}}

\newcommand{\la}{\langle}
\newcommand{\ra}{\rangle}

\newcommand{\eref}[1]{\eqref{#1}}
\newcommand{\scN}{N}
\newcommand{\scM}{M}
\newcommand{\scS}{\sigma_{\textsc{d}}}

\newcommand{\be}{\begin{equation}}
\newcommand{\ee}{\end{equation}}
\newcommand{\bea}{\begin{eqnarray}} 
\newcommand{\eea}{\end{eqnarray}}
\newcommand{\ft}[2]{{\textstyle\frac{#1}{#2}}}

\def\pd{\ensuremath{\partial}}

\def\Re{\mathop{\rm Re}\nolimits}

\def\rme{{\mathrm e}}
\def\rmi{{\mathrm i}}

\newsavebox{\uuunit}
\sbox{\uuunit}
    {\setlength{\unitlength}{0.825em}
     \begin{picture}(0.6,0.7)
        \thinlines
        \put(0,0){\line(1,0){0.5}}
        \put(0.15,0){\line(0,1){0.7}}
        \put(0.35,0){\line(0,1){0.8}}
       \multiput(0.3,0.8)(-0.04,-0.02){12}{\rule{0.5pt}{0.5pt}}
     \end {picture}}
\newcommand {\unity}{\mathord{\!\usebox{\uuunit}}}

\newcommand{\U}{\mathop{\rm {}U}}

   \def\cG{{\cal G}} \def\cH{{\cal H}}
 \def\cI{{\cal I}}  \def\cK{{\cal K}} 
 \def\cM{{\cal M}} \def\cN{{\cal N}} \def\cO{{\cal O}} \def\cP{{\cal P}}
  \def\cR{{\cal R}}

    \def\cV{{\cal V}}

\def\atH0{|_{H_{0}}}
\def\AtH0{\bigg|_{H_{0}}}

\csname @addtoreset\endcsname{equation}{section}

\title{\LARGE {{Perturbative Stability along the Supersymmetric Directions of the Landscape\vspace{-1. cm}}}}

\author[a]{{\large Kepa Sousa,}}
\author[b,c]{{\large Pablo Ortiz,}}

\affiliation[a]{Department of Theoretical Physics and History of Science, University of the Basque Country UPV/EHU,
48080 Bilbao, Spain}
\affiliation[b]{Instituut-Lorentz for Theoretical Physics, Universiteit Leiden, 2333 CA Leiden, The Netherlands}
\affiliation[c]{Nikhef, Science Park 105, 1098 XG Amsterdam, The Netherlands}

\emailAdd{kepa.sousa@ehu.es}
\emailAdd{ortiz@lorentz.leidenuniv.nl}

\abstract{We consider the perturbative stability of non-supersymmetric configurations in $\cN=1$ supergravity models with a spectator sector not involved  in  supersymmetry breaking.  
Motivated by the supergravity description of complex structure moduli in Large Volume Compactifications of type IIB-superstrings, we concentrate on models where the interactions are consistent with  the supersymmetric truncation of the spectator fields, and we describe their couplings by a  random ensemble of generic supergravity theories. We characterise the mass spectrum of the spectator fields in terms of the statistical parameters of the ensemble and the  geometry of the scalar manifold. Our results show that   the non-generic couplings between the spectator and the supersymmetry breaking sectors  can stabilise all the tachyons which  typically appear in the spectator sector before including the supersymmetry breaking effects, and we find  large regions of the parameter space where the supersymmetric sector remains stable with probability close to one.
We discuss these results about the stability of the supersymmetric sector in two physically relevant situations: non-supersymmetric Minkowski vacua, and slow-roll inflation driven by the supersymmetry breaking sector.
For the class of models we consider, we have reproduced the regimes in which the KKLT and Large Volume Scenarios stabilise all supersymmetric moduli. We have also identified a new regime in which the supersymmetric sector is stabilised at a very robust type of dS minimum without invoking a large mass hierarchy.
}


\begin{document}

\maketitle
\flushbottom

%%%%%%%%%%%%%%%%%%%%%%%%%%%%%%%%%%
%%%%%%%%%%%%%%%%%%%%%%%%%%%%%%%%%%%%

\section{Introduction}
\label{sec:intro}

%%%%%%%%%%%%%%%%%%%%%%%%%%%%%%%%%%%%%
%%%%%%%%%%%%%%%%%%%%%%%%%%%%%%%%%%%%%%

In the last decade, many promising cosmological models have been derived in the framework of  String Theory   in order to explain the mechanisms responsible for the present acceleration of the universe (Dark Energy),  and inflation. The construction of those models is far from trivial, in particular,  due to the fact that the low energy supergravity description of string compactifications  typically involves hundreds of scalar fields, \emph{the moduli},  which characterise the size and shape of the extra dimensions.  Any cosmological model requires a good understanding  of the effective scalar potential  along all directions in field space, as a single tachyonic instability can easily spoil its predictions. However,  the large number of fields and complexity of the scalar potential, which, for example,  might contain more than $10^{500}$ vacua,  makes  impractical an exhaustive  stability analysis. Thus,  an alternative approach has been proposed:   to characterise the properties of the  effective potential  following a statistical treatment \cite{Bousso:2000xa,Ashok:2003gk,Douglas:2004kp,Denef:2004cf,Denef:2004ze,Marsh:2011aa,Bachlechner:2012at,Sumitomo:2012wa,Sumitomo:2012vx,Sumitomo:2012cf,BlancoPillado:2012cb,JJ,Brown:2013fba,Brown:2014sba}.  This intricate energy landscape associated to the effective scalar potential of string compactifications  is the so-called   \emph{string theory landscape}.

One particular method to characterise  the string landscape is to study the statistical properties of generic four dimensional  $\cN=1$ supergravity theories with a large number of fields, $\scN \gg1$, where  the couplings are treated  as random variables \cite{Denef:2004ze,Marsh:2011aa,Long:2014fba,Aazami:2005jf,Bachlechner:2012at,Marsh:2013qca,Pedro:2013nda}. This   approach is particularly useful to study the stability of field configurations, since, given that  the  Hessian of the scalar potential is a very large matrix, its statistical properties can be determined  using techniques from   the well developed  random matrix theory (see \cite{mehta1991random}). 
 In this framework it has been shown that the construction of viable cosmological models in the string landscape is very constrained.   In particular, 
 the probability of occurrence of stable de Sitter (dS) minima of the scalar potential, which are relevant for constructing  models of late-time cosmology,  is exponentially suppressed by the number of fields,
$\mathbb{P}_\text{min} \sim \rme^{- \scN^p}$, 
with $p$ being a number of order one  \cite{Marsh:2011aa}. Similarly, the possibility of constructing viable models of inflation in the string landscape has also been considered in several works following this approach \cite{Marsh:2013qca,Pedro:2013nda}.  In those papers it was argued  that prolonged periods of inflation  are very rare due to the large probability of encountering instabilities along the inflationary trajectory,  implying that the string landscape seems to favour  small field inflationary models. \\  

Compactifications of type-IIB string theory are a particularly interesting  framework for the construction of  cosmological models, since the physical mechanisms which induce the scalar potential  of the moduli are well understood. However, the couplings in the low ener\-gy effective supergravity description of these theories  are highly non-generic, and therefore the standard random supergravity approach cannot be  applied directly.  For instance, in the  scenario  proposed by Kachru, Kallosh, Linde and Trivedi (KKLT)  \cite{Kachru:2003aw},  a sector of the mo\-du\-li fields, the dilaton and complex structure moduli, are stabilised and integrated out at a large energy scale due to the action of background fluxes, and then  the K\"ahler moduli are stabilised at a lower energy scale (where supersymmetry is spontaneously broken) using non-perturbative effects. As a consequence of this hierarchical structure, the heavy moduli sector effectively decouples from the lighter K\"ahler moduli and the supersymmetry breaking effects, and therefore the stability of the two sectors can be studied independently. \\

On the one hand, as argued in \cite{Marsh:2011aa},  the stability of the K\"ahler moduli sector can be characterised using the standard random supergravity approach when the number of light moduli is large, $\scN_l\gg1$, and the results show that the probability of this sector to remain tachyon-free at a dS vacuum is still suppressed, $\mathbb{P}_l\sim\rme^{-\scN_l^p}$. On the other hand, the large mass hierarchy between the two sectors allows to study the stability of the dilaton and complex structure moduli neglecting the presence of the lighter fields and  supersymmetry breaking, and therefore  it is natural to search for supersymmetric AdS minima of the scalar potential induced by the background fluxes alone. Bachlechner \emph{et al.} \cite{Bachlechner:2012at} proved that the fraction of supersymmetric AdS minima in the heavy sector is also exponentially suppressed by the number of heavy fields, $\mathbb{P}_h \sim \text{exp}(-8\scN_h^2/m_h^2)$, where $m_h$ gives the typical ratio of the mass scale of the heavy sector  to the gravitino mass. This implies that, in general, the decoupling of heavy moduli provides only a little improvement over the fully generic  case.  The pro\-ba\-bi\-li\-ty that the full field configuration is tachyon-free, $\mathbb{P}_\text{min} = \mathbb{P}_{l} \cdot \mathbb{P}_{h}$, can still be made of order one when the number of K\"ahler moduli is small, $\scN_l \sim \cO(1-10)$,  and adjusting the statistical parameters so that the typical value for the gravitino mass is much smaller than the mass scale of the heavy sector, $m_h\gg1$. This is precisely the fine-tuning needed for consistency in KKLT scenarios, in other words,  this is equivalent to requiring the typical va\-cu\-um expectation value of the flux superpotential  to be suppressed with respect to the energy scale set by the background fluxes. Recently, it has been argued that this fine-tuning could arise in a natural way in the string theory landscape \cite{Sumitomo:2012wa,Sumitomo:2012vx,Sumitomo:2012cf}, but there is controversy  on how representative the models considered in these works are of Calabi-Yau compactifications\footnote{The authors of \cite{Cicoli:2013swa} also review indications against this fine-tuning  based on statistical arguments of the string theory landscape and on the non-perturbative stability properties of type-IIB vacua.} \cite{Cicoli:2013swa}.\\   

These results contrast with the situation in the large volume limit of type IIB flux compactifications, where it is possible to find a very robust and model-independent non-supersymmetric set of minima capable of stabilising all the moduli \cite{Balasubramanian:2005zx,Conlon:2005ki}. This regime, where the volume of the internal space is exponentially large, is the so-called  \emph{Large Volume Scenario} (LVS). In particular, this scenario does not require to fine-tune the expectation value of the flux superpotential, and hence in general there is no large hierarchy of masses between the dilaton and complex structure fields, and the K\"ahler moduli or the gravitino \cite{Conlon:2005ki}. Still, it is consistent to truncate the dilaton and complex structure moduli while preserving supersymmetry, since the couplings of this sector to the lighter moduli fields and the supersymmetry breaking effects are suppressed by powers of the volume of the internal space. As in the case of KKLT constructions, this decoupling allows  to study the stability of the two sectors independently. However,  when considering the stability of the truncated moduli,  due to the absence of a mass hierarchy, it is no longer justified to neglect completely the presence of the K\"ahler moduli or the supersymmetry breaking effects. As a consequence, stabilising  the truncated moduli at  supersymmetric AdS minima of the scalar potential induced by the fluxes is no longer  required to construct stable dS vacua. As we shall discuss in detail in the present paper, the necessary conditions to construct supersymmetric AdS minima are more restrictive than those needed to stabilise the supersymmetric sector  of a theory where  supersymmetry is spontaneously broken. Therefore, by considering a general configuration to stabilise the supersymmetric sector (not necessarily a supersymmetric minimum), it is possible to reconcile the results obtained from the random supergravity approach with the known properties of the scalar potential of LVS. In other words, in this setting it is possible to find generic configurations where the truncated sector is free of tachyons with order one probability, $\mathbb{P}_h \sim \cO(1)$. If, in addition, we consider models with just a few K\"ahler moduli, as in the effective theory of type-IIB compactified in the orientifold of $\mathbb{P}_{[1,1,1,6,9]}^4$, the probability  of the full field configuration being stable can also be made order one, $\mathbb{P}_\text{min} \sim \cO(1)$.\\

 The main objective of  the present work is to study the  perturbative stability of a decoupled supersymmetric sector with a large number of fields  in $\cN=1$ supergravity theories. In particular, we will concentrate on supergravity models  with couplings compatible with the \emph{consistent supersymmetric truncation} of the decoupled sector \cite{Binetruy:2004hh,Achucarro:2007qa,Achucarro:2008sy,Achucarro:2008fk,thesis}.  For simplicity, we will consider only models without gauge interactions, and where supersymmetry is spontaneously broken as a result of the interactions of the chiral fields in the sector surviving the truncation, i.e. $F-$term breaking.
In a consistent supersymmetric truncation, the solutions to the equations of motion obtained  after truncating the supersymmetric sector are also exact solutions to the equations of the full model, and moreover, supersymmetry is exactly preserved in the reduced theory.  The type of models satisfying these conditions can be characterised in a compact way using the K\"ahler function $G$, which in terms of the K\"ahler potential $K$ and the superpotential $W$ reads  $G= K + \log|W|^2$. Denoting by $H$ the fields in the supersymmetric (to be truncated) sector and by $L$ the set of surviving fields, the type of theories consistent with the exact truncation of the $H-$sector at a configuration $H=H_0$ have a K\"ahler function satisfying \cite{Binetruy:2004hh,Achucarro:2007qa,Achucarro:2008sy,Achucarro:2008fk,thesis}
\be
\pd_H G(H , \bar H, L,\bar L)|_{H_0} =0 \qquad \text{for all} \qquad L. \nonumber
\ee
Although this type of theories are more restrictive than those where the truncation leads only to approximate solutions \cite{Gallego:2008qi,Gallego:2009px,Brizi:2009nn}, they constitute the simplest class of models which describe a  decoupled supersymmetric sector.   More importantly, when the action has this type of structure,  it is consistent to study  the stability of the truncated sector alone, since the Hessian of the scalar potential  is block-diagonal in the truncated and surviving sectors. As we shall review here, the  effective Lagrangian of type-IIB string compactifications has \emph{exactly} this type of structure at zero-order in $\alpha'$ and non-perturbative corrections, with the $H-$sector identified with the complex structure and dilaton fields, and the $L$ fields with the K\"ahler moduli \cite{Gallego:2011jm}. Moreover, it was shown in \cite{Gallego:2011jm}  that the couplings in the LVS can also be seen as a small deformation of this class of K\"ahler functions. Our results are of interest for scenarios similar to the ones discussed in \cite{Balasubramanian:2004uy,Westphal:2006tn,Rummel:2011cd}, which consider the branch of dS vacua obtained after the K\"ahler uplifting (induced by $\alpha'$ corrections) of a LVS-type of non-supersymmetric AdS vacuum.\footnote{For a detailed discussion about the validity of the supergravity approximation in these scenarios see \cite{Baumann:2014nda} and references therein.} \\

 We will characterise  the stability of the truncated sector following the random supergravity approach, that is, treating the couplings in the supersymmetric sector as random variables, and modelling the corresponding block of the Hessian using Random Matrix Theory. All the parameters in the Hessian affected by the physics of the supersymmetry breaking sector, such as the supersymmetry breaking scale and the gravitino mass, will be treated as fixed constants and studied in a case by case basis. Moreover, motivated by the situation in type-IIB compactifications, where the choice of the Calabi-Yau determines the geometry of the moduli space, we will also assume that the K\"ahler manifold is fixed (not random), as in \cite{Ashok:2003gk,Douglas:2004kp,Denef:2004cf,Denef:2004ze}. 
In this setting, we will discuss the conditions for the decoupled sector to remain tachyon-free with order one probability\footnote{We will restrict ourselves to the study of the perturbative stability of these configurations, and thus we will not consider the possibility of tunnelling instabilities.}, $\mathbb{P}_h \sim \cO(1)$, both when the full field configuration is a critical point of the scalar potential with a small cosmological constant, i.e. Minkowski, or a point of an inflationary trajectory during a phase of slow-roll inflation. In the latter case, we will assume that the decoupled fields act as spectators, and that it is  the dynamics in the sector surviving the truncation the one driving inflation. We will show that the necessary conditions for the stability of the truncated sector translate into constraints on the geometry of the moduli space, (or equi\-va\-lent\-ly, on the K\"ahler potential, $K$), and on the statistical properties of the fermion mass spectrum, which are determined by the superpotential, $W$. In particular, we will argue that  the stability depends crucially on the ratio between the mass scale of the truncated sector and the gravitino mass, which we denote by $m_h$.\\
  
Regarding non-supersymmetric Minkowski vacua, we will show that  there is a broad range of parameter space where the configuration of the truncated sector is typically stable, $\mathbb{P}_h\sim \cO(1)$. Thus, there is no need of  fine-tuning, as opposed to KKLT scenarios, where the hierarchical structure requires $m_h\gg 1$. We have also found that, when the mass scale of the truncated sector is smaller that the gravitino mass, $m_h\lesssim1$, the configuration of the truncated fields typically corresponds to a class of very robust  non-supersymmetric minima. We shall argue that in this regime, for supergravity models with  a similar structure to the supergravity description of Large Volume Scenarios of type-IIB flux compactifications,  the lightest scalar field of the supersymmetric sector has a finite positive mass. In other words, the spectrum has a mass gap and the probability of finding massless modes or tachyons is exponentially suppressed, and decreases exponentially with the number of truncated fields.  
Note that, as shown in \cite{Rummel:2011cd}, this situation is not guaranteed in the case of  Large Volume Scenarios. For instance, in the model considered in \cite{Rummel:2011cd}, part of the complex structure moduli axions remain massless after including the effects from background fluxes, which could lead to the appearance of tachyons when additional contributions, such as non-perturbative corrections, are taken into account. Our results indicate that the gap in the spectrum can be as large as the gravitino mass, and therefore this new branch of vacua would survive to small deformations of the model, in particular to those associated to $\alpha'$ and string loop ($g_s$) corrections, provided they do not dominate over the supersymmetry breaking effects induced by the no-scale structure of the K\"ahler sector.
In the case of inflation, we shall show that the same region of the parameter space, $m_h\lesssim1$, is also specially favoured by our stability analysis in models where the Hubble scale is much larger than the gravitino mass, $H \gg m_{3/2}$.\\

In order to perform the  stability analyses in the present work we have derived a set of necessary   conditions \eqref{constraints} for the meta-stability of non-supersymmetric configurations in  \emph{arbitrary} $\cN=1$ supergravity theories involving only chiral multiplets.  In particular, these constraints   are required for the directions orthogonal to the sGoldstino to be tachyon-free, and therefore complementary to those studied in  \cite{GomezReino:2006dk,GomezReino:2006wv,GomezReino:2007qi,Covi:2008ea,Covi:2008cn,Covi:2008zu}. The conditions are expressed  in terms of  the ratio of the Hubble parameter to the gravitino mass, the distribution of fermion masses, and certain geometrical objects associated to the K\"ahler manifold:  the bisectional curvatures along the planes  defined by the  sGoldstino  and each of the   mass eigenstates of the chiral fermions. The set of  conditions  \eqref{constraints}, together with their graphical representation in Fig. \ref{perturbedDiagram}, are one of the main results of this work, as they can be used to test the viability of a broad range of  cosmological models  which extends beyond the class of models considered here.   \\

The structure of the paper is the following. In section \ref{SUGRA} we will review the properties of the scalar potential in $\cN=1$ supergravity models. In section \ref{sec:conditions} we will discuss the  perturbative stability of non-supersymmetric field configurations in generic supergravity theories.  In section \ref{sec:themodel} we define  the class of supergravity theories under study, that is, we summarise the properties of consistent supersymmetric truncations and the implementation of the random supergravity approach.  In section \ref{sec:beforeuplifting} we review the predictions of the random supergravity approach about the stability of the supersymmetric sector when  is considered in isolation, i.e.  not  taking into account the couplings to the supersymmetry breaking sector. In section \ref{sec:afteruplifting} we analyse how these results change when the couplings to the supersymmetry-breaking fields are included. In particular, we discuss the viability of constructing Minkowski vacua and slow-roll inflationary models in this class of theories. Finally we give a summary of our results in section \ref{sec:summary}.

%%%%%%%%%%%%%%%%%%%%%%%%%%%%%%
%%%%%%%%%%%%%%%%%%%%%%%%%%%%%%

\section{Aspects of $\cN=1$ supergravity}
\label{SUGRA}

%%%%%%%%%%%%%%%%%%%%%%%%%%
%%%%%%%%%%%%%%%%%%%%%%%%%%

We follow the conventions in \cite{Binetruy:2004hh}. In particular we use the Minkowski metric with signature $(-,+,+,+)$, and we work in units of $c=\hbar=1$,  so that the reduced Planck mass reads $M_p^{-2}= 8 \pi G$, which is also set to unity $M_p = 1$.

%%%%%%%%%%%%%%%%%%%%%%%%%%%%%%%%%
 \subsection{Critical points of the scalar potential}
 %%%%%%%%%%%%%%%%%%%%%%%%%%%%%%%%%
\label{criticalPoints}

In order to set our notation we start reviewing the properties of critical points of the scalar potential in $\cN=1$ supergravity theories. The  class of supergravity actions we study in the present work  only involve complex scalar fields $\xi^I$ and their superpartners, the Weyl fermions $\chi^I$ (chiral multiplets with no gauge interactions). The fields are labeled with the index $I$ running in $I=1,\ldots,\scN$ for $\scN$ chiral multiplets. Since we are interested in characterising the perturbative stability of purely  bosonic configurations, we only need to consider the bosonic part of the action, which is constructed with  the Ricci scalar $R$, the kinetic terms of the scalar fields $T$, and the scalar potential $V$:
\begin{equation}
  S = \int d^{4}x\sqrt{-g} \(\ft{1}{2}R + T  - V \) \label{Sdef}.
\end{equation}
In the absence of gauge fields, the couplings can be expressed entirely in terms of two functions of the scalars: the K\"ahler potential  $K(\xi ,\bar{\xi})$ and the holomorphic superpotential  $W(\xi)$. These two functions are only defined up to \emph{K\"ahler transformations}:  
\be
 K \to K + h(\xi) + \bar{h}(\bar{\xi}) \qquad \text{and} \qquad  W \to W \rme^{-h(\xi)},
\ee
with $h(\xi)$ being an arbitrary holomorphic function.  For convenience we will use the K\"ahler invariant formulation of supergravity, where the full action and the supersymmetry transformations are written in terms of a single function\footnote{This formulation can be related to the one used in \cite{Marsh:2011aa,Bachlechner:2012at} and in \cite{Ashok:2003gk}, by choosing  a K\"ahler gauge where the modulus of the superpotential is set to a constant $|W|\to 1$, and then fixing the remaining  $\U(1)$ R-symmetry by requiring the  term  $\rme^{\frac{K}{2} } W$  to be real.} $G= K + \log |W|^2$, which is well defined for non-vanishing superpotential $W\neq0$. In particular, the kinetic terms of the scalar fields are characterised by a non-linear  sigma model  with target space on a K\"ahler-Hodge manifold, and the corresponding metric can be expressed in terms of the derivatives of the K\"ahler function $G_{I\bar J} \equiv \partial_{I} \partial_{\bar{J}} G$.  In a theory with $\scN$ chiral multiplets the kinetic terms read
\be 
  T = G_{I\bar{J}}\; \pd_{\mu}\xi^{I}\pd^{\mu} \xi^{\bar J},  \qquad \text{where} \qquad I,J = 1,\ldots, \scN.
  \label{Skinetic} 
  \ee 
In general we will denote the partial derivatives with respect to  $\xi^I$ and $\xi^{\bar I}= (\xi^I)^*$ with subindices, and we will rise and lower the indices using the metric $G_{I \bar J}$ and its inverse $G^{I \bar J}$.  As we are not considering  gauge interactions, the scalar potential has the simple form
\begin{equation} 
 V = \rme^{G} ( G^IG_{I} - 3).
  \label{Gaction}   \end{equation} 
A given configuration $\xi_0$ is an  extremum of the scalar potential if it satisfies the  set of stationarity conditions  
\be
V_K|_{\xi_0}=0 \qquad \Longleftrightarrow \qquad \[(\nabla_K G_I)\,  G^I -  (2-G^IG_I) \, G_K\]_{\xi_0}=0,
\label{stationarity}
\ee  
for all $K= 1,\ldots, \scN$. In the previous expression the covariant derivative $\nabla_I$ is  the Levi-Civita connection associated to the metric $G_{I\bar J}$.   \\

Supersymmetry is spontaneously broken at a critical point of the scalar potential whenever the expectation value of the supersymmetry transformations is non-zero. In a bosonic configuration, only the supersymmetry transformations of the fermions can be non-zero. In particular, those of the chiral fermions for homogeneous configurations read
\be
\delta \chi^I= - \ft12 \, \rme^{G/2} G^{I \bar J} 
G_{\bar J} \; \epsilon, 
\label{SUSYtrans} 
\ee
 where $\epsilon$ is the parameter of supersymmetry transformations. At critical points where supersymmetry is spontaneously broken, the gradient of the K\"ahler function $G_I|_{\xi_0}$ defines a direction in field space known as the \emph{sGoldstino direction}. The sGoldstino corresponds to the supersymmetric partner of the would-be Goldstone fermion associated to broken supersymmetry. We will also describe this direction in terms of the unit vector $z_X$ with coordinates
 \be
z_{X,I} = \frac{G_I}{\sqrt{G_K G^K}}.
 \ee
From the supersymmetry transformations \eqref{SUSYtrans}, it follows that a homogeneous bosonic field configuration $\xi_0$ where supersymmetry is unbroken must satisfy the set of necessary conditions 
\be
G_I|_{\xi_0} =0, \qquad \text{for all} \qquad I=1,\ldots, \scN. \label{SUSYcond0}
\ee  
Actually, it is easy to check that supersymmetric configurations are also critical points of the scalar potential, since they satisfy (\ref{stationarity}), and thus they are called \emph{supersymmetric critical points}. Due to the form of the scalar potential (\ref{Gaction}), supersymmetric critical points are always anti-de Sitter (AdS): 
\be
V|_{\xi_0} = -3 \rme^{G}<0, %\qquad \Longrightarrow \qquad \gamma = -1,
\ee
except in those cases where the superpotential vanishes, for which they are Minkowski vacua, $V|_{\xi_0} =0$.

\subsection{The structure of the Hessian}
%%%%%%%%%%%%%%%%%%%%%%%%%%%%%%%%%
\label{sec:structureH}

In order to determine the stability properties of an extremum of the scalar potential, we need to study the eigenvalue spectrum of the corresponding Hessian,
\be\cH=  \begin{pmatrix}
    \nabla_{I} V_{\bar J} & \nabla_{I} V_J \\
     \nabla_{\bar I} V_{\bar J} &  \nabla_{\bar I} V_J
  \end{pmatrix}, 
   \ee
 which determines the squared-masses of the scalar fields at Minkowski and de Sitter critical points. In this subsection we will describe the different contributions of the Hessian and will relate them to the masses of the fermions and to the geometry of the K\"ahler manifold. \\
 
 After using the stationarity conditions \eqref{stationarity}, the second covariant derivatives of the scalar potential at the extremum $\xi_0$ of $V$ read\footnote{To make contact with the notation of \cite{Achucarro:2007qa,Achucarro:2008fk},  note that at any supersymmetric critical point $\xi_0$ the covariant  and regular derivatives of  $G_I(\xi,\bar \xi)$ coincide, e.g. $\nabla_J G_I|_{\xi_0} = \pd_J G_I|_{\xi_0}$, and similarly for $G_{\bar I}(\xi,\bar \xi)$.} 
\bea
\nabla_{I}V_{\bar J}&=& \(G_{I\bar J}-G_IG_{\bar J}\)V+ \rme^G\[ G^{K\bar L}(\nabla_KG_I)(\nabla_{\bar L}G_{\bar J})+G_{I\bar J}-R_{I\bar JK\bar L}G^KG^{\bar L}\]\ ,\nonumber\\
\nabla_IV_{J}&=&\(\nabla_{I}G_{J}-G_{I}G_{J}\) V+\rme^G\[2\nabla_{I}G_{J}+G^K\nabla_{K}\nabla_{I}G_J\]\ .\label{hessiangeneral}
\eea
In these expressions it is straightforward to  identify the mass of the gravitino $m_{3/2}$ and the  mass matrix of the chiral fermions $M_{IJ}$ 
 \be
m_{3/2} \equiv \rme^{G/2}, \qquad  M_{IJ} \equiv \rme^{G/2} \nabla_I  G_J.
\label{fermionMasses}
\ee
To simplify the notation we will perform the rescalings
\bea
 \cH \to m_{3/2}^2\,  \cH, \qquad \text{and} \qquad M_{IJ} \to  m_{3/2}\,  M_{IJ}\ ,
\label{reescalings}
\eea
so that in what follows all the masses of the scalar fields and fermions will be expressed in units of the gravitino mass\footnote{Note that this implies a slight abuse of notation, since everywhere else we use natural units $M_p=1$. However, it will be clear from the context which units we are using in each case.}.
Similarly, we will parametrise the expectation value of the scalar potential at an extremum $\xi_0$ by the quantity
 \be
\gamma \equiv  \frac{V}{ 3 \, m_{3/2}^2}\simeq \frac{H^2}{m_{3/2}^2},
\ee
which is essentially the square of the Hubble parameter  $H$  in units of the gravitino mass. The structure of the Hessian becomes particularly clear when we choose the fields $\xi^I$ so that they  have canonical kinetic terms at the critical point, i.e. $G_{I\bar J}|_{\xi_0} = \delta_{I \bar J}$. Moreover, we will require that one of the axis of the local frame points along the sGoldstino direction, i.e. $G_I \equiv  G_X\,  \delta_{I X}$,  this is the so-called \emph{sGoldstino basis}. In these coordinates it is straightforward to show that the Hessian reads\footnote{We also choose the component $G_X$ of the sGoldstino   to be real, which results into $G_X|_{\xi_0}=\sqrt{3(\gamma+1)}$.}  
\vspace{.1cm}
\be
\cH =( \cM + \unity) \big(\cM + (3 \gamma + 1 )\unity\big)
+\sqrt{3(\gamma+1)} \,D _X \cM  - 3(\gamma+1) \cR- 9 \gamma(\gamma+1) \cP_X.
\vspace{.1cm}
\label{HessianDecomp}
\ee
Here  $\cM$ and $D_X \cM$ are the (rescaled) fermion mass matrix and  its derivative along the sGoldstino direction written in the $2\scN-$vector notation, $\cR$ is a matrix built from the components of the Riemann tensor, and $\cP_X$ is the projector along the sGoldstino direction:
\bea
\cM &= \begin{pmatrix}
    0 & \nabla_{I} G_J \\
     \nabla_{\bar I} G_{\bar J} &  0
  \end{pmatrix}, \qquad
D_X  \cM &= \begin{pmatrix}
    0 & \nabla_X \nabla_{I} G_J \\
     \nabla_{\bar X}  \nabla_{\bar I} G_{\bar J} &  0
  \end{pmatrix}, \nonumber \\
&&  \nonumber \\
   \cR &= \begin{pmatrix}
    R_{X\bar X I  \bar J} & 0 \\
     0&   R_{\bar X X \bar I   J}
       \end{pmatrix}, \qquad \cP_X &=  \begin{pmatrix}
    \delta_{IX} \delta_{\bar J \bar X} &\delta_{I X} \delta_{J X}\\
     \delta_{\bar I\bar X} \delta_{\bar J\bar  X}&    \delta_{\bar I \bar X} \delta_{J X}
  \end{pmatrix}.
  \label{matrixnotation}
\eea
For convenience, we will define the following shorthand to refer to the first term of the Hessian in \eqref{HessianDecomp}: %we define $\cH_\gamma$ as
\be
\cH_\gamma \equiv  ( \cM + \unity) \big(\cM + (3 \gamma + 1 )\unity\big).
\ee

%%%%%%%%%%%%%%%%%%%%%%%%%%%%
%%%%%%%%%%%%%%%%%%%%%%%%%%%%%%%%
\section{Necessary conditions for metastability}
%%%%%%%%%%%%%%%%%%%%%%%%%%%%
%%%%%%%%%%%%%%%%%%%%%%%%%%%%%%%%%%%%%%%%%%%

\label{sec:conditions}

In this section we will present our approach to characterise the perturbative stability of a consistently truncated supersymmetric sector. We will discuss the stability of non-supersymmetric configurations on generic supergravity theories including only chiral multiplets.
In particular, we will derive a set of constraints on the geometry of the moduli space and the spectrum of fermion masses necessary  for  the perturbative  stability along the directions of field space preserving supersymmetry, i.e. orthogonal to the sGoldstino, and therefore  complementary to those studied in \cite{GomezReino:2006dk,GomezReino:2006wv,GomezReino:2007qi,Covi:2008ea,Covi:2008cn,Covi:2008zu}.
 This analysis is both applicable to the study of the perturbative stability of critical points of the scalar potential, or points of an inflationary trajectory.   
When studying the perturbative stability of a consistently truncated supersymmetric sector it might seem reasonable to ignore completely the supersymmetry breaking effects, as it is done in the case of the complex structure/dilaton sector in KKLT scenarios. The analysis presented in this section provides a systematic way to test the consistency of this procedure. 
In the case of KKLT scenarios it is well known that this approach is justified due to the presence of a large hierarchy between the mass scale of the complex structure/dilaton (supersymmetric) sector and the supersymmetry breaking scale \cite{Gallego:2008qi,Gallego:2009px}, and therefore it is safe to identify the metastable configurations of the supersymmetric sector with supersymmetric minima of the scalar potential. However, this hierarchical structure is absent in Large Volume Scenarios and therefore a more detailed analysis is required in these type of models. 
Actually, we shall see that stable non-supersymmetric configurations in LVS will correspond in general to saddle points and AdS ma\-xi\-ma in the supersymmetric limit, that is, before the spontaneous breaking of supersymmetry is included. This discussion is important to understand the results in the following sections, where we will show that the fraction of field configurations where the supersymmetric sector remains tachyon-free in certain LVS can be made of order one without fine-tuning the parameters, while the probability of occurrence of supersymmetric AdS minima (required in KKLT constructions) is exponentially suppressed in general.\\ 
 
For a Minkowski or de Sitter field configuration to be metastable, all the eigenvalues of the  Hessian matrix \eqref{HessianDecomp} have to be positive. Ideally, one would like to express all the eigenvalues of the Hessian in terms of the K\"ahler function and its derivatives to characterise the type of couplings leading to stable dS vacua. However, a generic expression is too involved to extract any useful information, and thus a different strategy has to be followed. In the series of papers \cite{GomezReino:2006dk,GomezReino:2006wv,GomezReino:2007qi,Covi:2008ea,Covi:2008cn,Covi:2008zu}, they made use of the following observation: if the Hessian is positive definite, so it is its projection along any vector  $Z= (z,\bar z)^T$:
\be
\la Z, \cH\,  Z \ra \ge 0.
\label{necessCondition}
\ee
In particular, the authors of  \cite{GomezReino:2006dk,GomezReino:2006wv,GomezReino:2007qi,Covi:2008ea,Covi:2008cn,Covi:2008zu} studied the condition obtained from imposing this requirement along the (complex) sGoldstino directions 
\be
Z_{+X}=\frac{1}{\sqrt{2}}\begin{pmatrix}
z_X\\
z_{\bar X}
\end{pmatrix}
 \qquad  \text{and} \qquad Z_{-X}=\frac{\rmi }{\sqrt{2}}\begin{pmatrix}
  z_X\\
 -  \bar z_{ X}
 \end{pmatrix}.
\ee
As was discussed in detail in \cite{Covi:2008ea,Covi:2008cn}, the corresponding constraint \eqref{necessCondition} is particularly restrictive due to the stationarity conditions \eqref{stationarity}, which imply that the vectors $Z_{\pm X}$ are  eigenvectors of the fermion mass matrix $\cM$:
\be
\cM \, Z_{\pm X} = \mp (3 \gamma +1) \, Z_{\pm X}.
\label{eigenvalueSG}
\ee
Indeed, fixing the geometry of  the K\"ahler manifold, while most of the eigenvalues of the Hessian  could be made arbitrarily positive by tuning the mass matrix $\cM$ (i.e. the superpotential $W$),  the projection of the Hessian along the sGoldstino direction cannot be adjusted so easily due to the above constraint on $\cM$. Combining the necessary conditions associated to the vectors $Z_{\pm X}$, it is possible to find a restriction on the geometry of the K\"ahler manifold which, when expressed in terms of the \emph{sectional curvature} $S[X]\equiv - R_{X \bar X X \bar X}$, reads     
\be
S[X]\ge -\frac{2}{3} \frac{1}{1+\gamma}.
\label{sGoldstinoStab}
\ee
We will now derive a set of complementary conditions obtained when considering the other $2 \scN-2$ real directions orthogonal to the sGoldstino, that is, those preserving supersymmetry. Thus, in the rest of our analysis the term $\cP_X$ in \eqref{HessianDecomp}  will always be absent.

%%%%%%%%%%%%%%%%%%%%%%%
\subsection{Metastability conditions}
%%%%%%%%%%%%%%%%%%%%%%%

\label{sec:Constraints}

To characterise the eigenvalue spectrum of the Hessian, it is convenient to work in a local frame where the fermion mass matrix  $\cM$ is diagonal, since in this basis the term $\cH_\gamma$ of the Hessian \eqref{HessianDecomp} is also diagonal. Due to the special structure of $\cM$, it is possible to show that it  has $2\scN$  real eigenvalues arranged in  pairs of the form\footnote{Note the change of notation with respect to \cite{Achucarro:2007qa,Achucarro:2008fk,Achucarro:2012hg}, where the masses of the fermions were denoted by $|x_\lambda|\equiv m_\lambda$.} 
\be
 \cM \, Z_{\pm \lambda} = \pm m_\lambda \,   Z_{\pm \lambda},\qquad \text{with} \qquad \lambda=1,\ldots,\scN.
\label{Meigenvectors}
 \ee
 The corresponding normalised eigenvectors are given by $Z_{+ \lambda} =\tfrac{1}{\sqrt{2}} (z_\lambda,\bar z_\lambda)^T$ and $Z_{- \lambda}=\tfrac{1}{\sqrt{2}}(\rmi z_\lambda,-\rmi \bar z_\lambda)^T$, where $z_\lambda$ solves
 \be
M \,\bar   z_ \lambda = m_\lambda z_ \lambda,
 \ee
and we choose  $m_\lambda\ge0$. Since $M$ is symmetric, we can always find  a set of orthonormal vectors $z_\lambda$ which satisfy the previous equation. Indeed, after requiring the fields to have canonical kinetic terms, it is still possible to redefine them using a unitary transformation of the form $\tilde \xi =\xi\, U$. Performing these transformations we can bring the matrix $M$ to a diagonal form  $M = U D U^T$, where $U$ is unitary and $D=\mathrm{diag}(m_\lambda)$, with $m_\lambda \in \mathbb{R}^+$.  This result,  known as Takagi's factorisation, applies to any complex symmetric matrix, and the eigenvectors $z_\lambda$ can be read from the columns of the unitary matrix $U_{I \lambda} =z_{\lambda,I}$. Note that this diagonalisation is also consistent with the choice of the sGoldstino basis, since the  vectors $Z_{\pm X}$  associated to the sGoldstino direction are also  eigenvectors of the matrix $\cM$, \eqref{eigenvalueSG}. 
 The corresponding eigenvalue $m_X$ is related to the  
unphysical Goldstone fermion of broken supersymmetry, and thus it does not have the interpretation of a mass.  The rest of the parameters $m_\lambda$, with $\lambda=1,\ldots,\scN-1$,  determine the mass spectrum of the chiral fermions $\chi^I$.\\

In general, the contributions to the Hessian proportional to $D_X \cM$ and $\cR$ will not be diagonal in the basis formed by $Z_{\pm \lambda}$, but their diagonal elements in this frame\footnote{The details can be found in Appendix \ref{appendix1}.} 
\be
\la Z_{\pm \lambda}, D_X \cM\,  Z_{\pm \lambda} \ra\equiv\pm \frac{\pd m_{\lambda}}{\pd X},\qquad\qquad    \la Z_{\pm \lambda} \,  \cR  \,  Z_{\pm \lambda} \ra = - B[X,\lambda],
\ee
have a simple physical interpretation. First, it can be shown that the real parameters $\pd m_\lambda / \pd X$ are closely related to the derivatives of the fermion masses along the sGoldstino direction  (see Appendix \ref{appendix1}), and for simplicity we will refer to them in this way. Second, the set of $\scN-1$ quantities $B[X,\lambda] \equiv -  R_{X \bar X \lambda \bar \lambda}$ are the so-called \emph{bisectional curvatures}  along the planes formed by the sGoldstino direction $z_X$  and each of the eigenvectors $z_{\lambda}$.  The bisectional curvature, first introduced in  \cite{goldberg},  has  been proven to be an important phenomenological quantity, since it determines the  size of supersymmetry breaking-induced soft masses in the visible sector \cite{Kaplunovsky:1993rd,Marsh:2011ud,Dutta:2010sg,Dutta:2012mw}. In the framework of inflation, it has also been used to characterise the stability of the inflationary trajectory \cite{Kallosh:2010xz}.  In these works, the viability of the studied models translates into constraints on the geometry of the K\"ahler manifold through the bisectional curvature.\\

In order to derive a set of simple necessary constraints, we will use the projection of the Hessian along all the $2 \scN -2$ supersymmetric $Z_{\pm \lambda}$ directions, $\mu_{\pm \lambda}^2 \equiv \la Z_{\pm \lambda }, \cH\,  Z_{\pm \lambda } \ra$. Collecting the results above we find the following conditions for Minkowski and dS vacua:
\be
\mu_{\pm \lambda}^2 = ( m_\lambda \pm 1) \big(m_\lambda \pm (3 \gamma + 1 )\big) 
  \pm \,\sqrt{3 (\gamma+1)} \, \frac{\pd  m_\lambda}{\pd X}+3(\gamma+1) \,   B[X,\lambda]  \ge 0,
\label{constraints}
\ee
for all $\lambda = 1, \ldots, \scN -1$.  Similarly to  \cite{GomezReino:2006dk,GomezReino:2006wv,GomezReino:2007qi,Covi:2008ea,Covi:2008cn,Covi:2008zu}, we can find a necessary condition which does not depend on the derivatives of the fermion masses by adding together the quantities $\mu_{+\lambda}^2$ and   $\mu_{-\lambda}^2$, which for $\gamma\ge0$ reads
\be
\mu_{+\lambda}^2 + \mu_{-\lambda}^2 \ge0\qquad  \Longrightarrow \qquad B[X,\lambda] \ge - \frac{ m_\lambda^2 + 3 \gamma +1}{3 (\gamma +1 )}.
\label{mildConstraint}
\ee
A particular case of the above condition \eqref{mildConstraint} was also derived in \cite{Covi:2008zu} for su\-per\-gra\-vi\-ty models involving only two chiral superfields, i.e. $\scN=2$.\\

In the case of AdS critical points, the requirement of stability implies that all the squared-masses of the scalar fields have to satisfy  the Breitenlohner-Freedman bound  \cite{Breitenlohner:1982bm}, and therefore all the previous conditions have to be modified accordingly. For instance, taking into account that we work in units of the gravitino mass, the set of conditions \eqref{constraints} become
\be
\mu_{\pm \lambda}^2 \ge \frac{3}{4 }\;  \frac{V(\xi_0)}{m_{3/2}^2 }  = \frac{9}{4}\gamma.
 \label{BF2}
\ee 
To understand  the implications of the set of constraints \eqref{constraints} and their dependence on the different parameters of the theory (the spectrum of fermion masses and their derivatives, the geometry of the K\"ahler manifold, and the supersymmetry breaking scale), we will now discuss them in two  different contexts. First, we will analyse the metastability of supersymmetric AdS critical points, and we will check that we recover known results. In that situation, the parameters $\mu_{\pm \lambda}^2$ are the exact eigenvalues of the Hessian, and therefore the corresponding  constraints \eqref{BF2} are both necessary and sufficient to guarantee the perturbative stability of the configuration. Then we will  discuss  non-supersymmetric  configurations, focusing the study on  the perturbative stability of the fields preserving supersymmetry. This analysis is both applicable to the case when the field configuration represents a non-supersymmetric vacuum, and when it corresponds to a point in the inflationary trajectory. In the latter case, the perturbative stability of all the fields not related to the inflaton or the sGoldstino is a requirement for the viability of the model, since the presence of any large tachyonic instability would spoil the slow-roll conditions.\\

 Let us emphasise that, in general, the conditions (\ref{constraints}-\ref{BF2}) are necessary but cannot guarantee the perturbative stability along the supersymmetric directions of a non-su\-per\-sym\-me\-tric configuration. However, as we shall discuss in later sections, there are interesting situations where these conditions become both necessary and sufficient. For instance, whenever the term $\cH_\gamma$ dominates over the rest of contributions to the Hessian, since then
  the quantities $\mu_{\pm \lambda}^2$  can be identified as the  eigenvalues of the Hessian to first order in perturbation theory. 
  As we shall discuss in section  \ref{sec:themodel}, this is the case for the low energy  supergravity description of type-IIB flux compactifications, and in particular of Large Volume Scenarios.\\

For simplicity, in the following analyses we will assume that all the parameters involved in the constraints \eqref{constraints} are independent from each other and can be varied freely. More complicated situations are possible, for example when two or more of the parameters have a functional dependence on each other, but we will not consider them here.

%%%%%%%%%%%%%%%%%%%%%%%%%%%%%%%%
\subsection{Supersymmetric vacua and uplifting to dS}
\label{subsec:uplift}
%%%%%%%%%%%%%%%%%%%%%%%%%%%%%%%%%%%%%%%%%%%

As we discussed in section \ref{criticalPoints}, supersymmetric critical points are AdS as long as the superpotential is non-zero, $W\neq0$, and in particular they satisfy  $\gamma=-1$. Therefore, the Hessian \eqref{HessianDecomp} is simply given by:
\be
\cH = (\cM +\unity) (\cM -2 \, \unity) \quad \Longrightarrow \quad \mu _{\pm \lambda}^2 = (m_\lambda \pm 1) (m_\lambda\mp 2 )= \( m_{\lambda}\mp\ft12 \)^2-\ft94 \ge -\ft{9}{4}\ .
\label{spectrum2}
\ee
This implies that  the Hessian is also diagonal in the basis that diagonalises $\cM$, and therefore  the parameters $\mu_{\pm \lambda}^2$ can be identified with the complete set of  eigenvalues of $\cH$. This means in particular that the set of conditions \eqref{BF2} are \emph{necessary and sufficient}, and also that they can be applied to all directions $Z_{\pm \lambda}$, with $\lambda = 1, \ldots, \scN$, since there is no sGoldstino at supersymmetric critical points. From the eigenvalues \eqref{spectrum2} one can see what type of extremum the supersymmetric critical point $\xi_0$ is, namely:  
 \bea
  m_\lambda>2 \quad \text{for all $\lambda$} &\Rightarrow& \text{local AdS minimum,} \nonumber\\
  m_\lambda<1 \quad \text{for all $\lambda$} &\Rightarrow& \text{local AdS maximum,} 
  \label{Vstability}
\eea 
and any other combination corresponds to AdS saddle points ($m_{\lambda}=1,2$ give flat directions). However, supersymmetric critical points are always perturbatively stable regardless of the possible negative curvature of the potential, since they are AdS and the Breitenlohner-Freedman bound \eqref{BF2} is always satisfied, as can be seen from eq. \eqref{spectrum2}.\\

As discussed in the introduction, supersymmetric vacua play an important r\"ole in the construction of de Sitter vacua in cosmological models derived from superstrings. As was originally proposed in the  KKLT scenario \cite{Kachru:2003aw}, it is possible to engineer a dS vacuum by the  \emph{uplifting}  of a supersymmetric AdS vacuum $\xi_0$ to dS, which consists in introducing a physical mechanism to break supersymmetry. Such uplifting mechanisms can be implemented by adding new fields and interactions to the model which lead to the spontaneous breaking of supersymmetry, or by introducing explicit breaking terms. Ideally, these mechanisms add a positive definite correction to the scalar potential $\delta \, V_\text{uplift}$, possibly field-dependent, so that the vacuum expectation value of $V$ becomes positive at $\xi_0^I$, 
\be
V = V_\text{susy}|_{\xi_0}+ \delta V_\text{uplift}|_{\xi_0}\ge 0,
\ee
while the supersymmetric configuration is still a metastable critical point of the potential.
In general, the supersymmetric field configuration $\xi_0$ is not a critical point of the uplifting term $\delta V_\text{uplift}$, and thus the critical points of the final  potential typically shift away from $\xi_0$, or disappear completely for sufficiently large values of the final cosmological constant. That is why in general one should ensure that the original supersymmetric critical point $\xi_0$ is a minimum, demanding all chiral fermions to have masses larger than twice the gravitino mass, cf. eq. \eqref{Vstability}. In the case of the dilaton and complex structure moduli in  KKLT constructions this condition is granted since, for consistency of the effective field theory, the expectation value of the flux superpotential has to be tuned to be small compared to the typical mass scale set by the fluxes. This implies, expressed in a K\"ahler independent way,  that the gravitino mass is  small with respect to the masses of the fermion superpartners of the dilaton and  complex structure moduli, $m_\lambda \gg1$.  \\
\vspace{-.3cm}

In the present work, we intend to explore the necessary conditions for metastability of a supersymmetric sector in situations where such mass hierarchy is not present, as in Large Vo\-lume Scenarios. As we shall see  in the following subsection, in general, it is no longer necessary to require the supersymmetric sector to be stabilised at a supersymmetric AdS minimum. In particular, we will prove that a stable configuration of the  embedded supersymmetric sector may correspond to any type of AdS critical point (even a saddle point or maximum) when it is considered in isolation, that is, neglecting the presence of other fields and of supersymmetry breaking, i.e.  $\gamma\to -1$.  In other words, the interactions between the supersymmetric and the supersymmetry-breaking sector may turn a would-be AdS maximum or a saddle point in the supersymmetric limit into a metastable dS configuration.
Alternatively, one can interpret the coupling between the  supersymmetric sector and the fields breaking supersymmetry as  a non-generic type of $F-$term uplifting mechanism, like those studied in \cite{Achucarro:2007qa,Achucarro:2008fk}.

%%%%%%%%%%%%%%%%%%%%%%%%%%%
\subsection{Non-supersymmetric configurations}
\label{nonsusyconf}
%%%%%%%%%%%%%%%%%%%%%%%%%%

In this subsection we will analyse the set of constraints \eqref{constraints} in the case when the field configuration  is  non-supersymmetric. As we mention above, these constraints are both applicable to the case where the  configuration is  an extremum of the scalar potential, and where the vacuum energy of the fields is driving a phase of slow-roll inflation\footnote{To derive the Hessian \eqref{HessianDecomp} we assume that the field configuration $\xi_0$ is a critical point of the potential. If $\xi_0$ is a point of the inflationary trajectory, there is a small gradient along the inflaton, but provided the first slow-roll parameter is sufficiently small, the corrections to  \eqref{HessianDecomp}  become a subleading effect \cite{Covi:2008cn}.}. In order to proceed, we analyse the different contributions to the Hessian separately:  first we will discuss the term depending on the fermion mass matrix  $\cH_\gamma$, and then we will characterise the effect of including the  contributions  associated to the derivatives of the fermions masses  $D_X \cM$, and the curvature of the K\"ahler manifold,  $\cR$.

\addtocontents{toc}{\protect\setcounter{tocdepth}{1}}
\subsubsection{Dependence on the fermion masses, $\cH=\cH_\gamma$}
\addtocontents{toc}{\protect\setcounter{tocdepth}{2}} 
\label{subsec:nonsusy1}
%%%%%%%%%%%%%%%%%%%%%%%%%%%%%%%%%%%%%%%%%%%

\begin{figure}[t]
 \vspace{-1cm}
 \centering \includegraphics[width=0.4\textwidth]{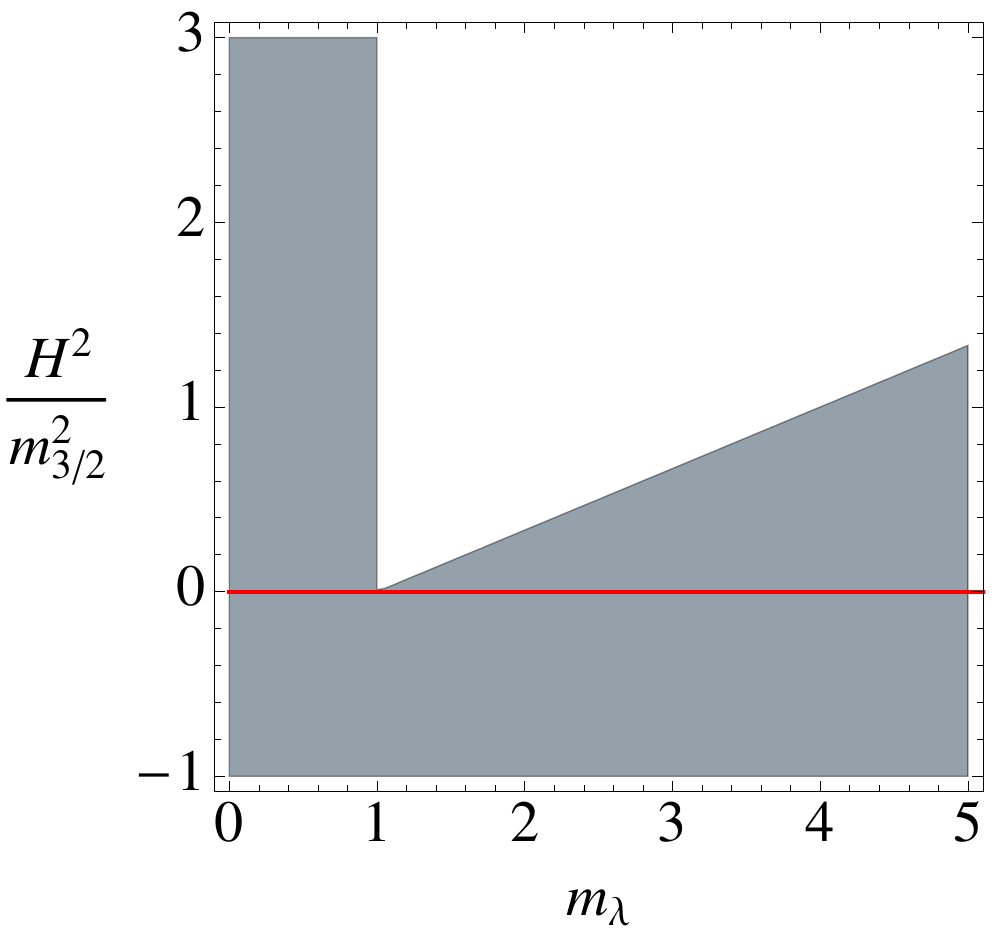}
  \caption{Stability diagram of a non-supersymmetric field configuration $\xi_0$ for a  Hessian of the form \eqref{separableHessian}, with $\pd_X \cM= \cR=0$.   On the vertical axis  $\gamma\equiv H^2/m_{3/2}^2$ is the squared ratio of the Hubble parameter to the gravitino mass. The horizontal axis represents the mass of one of the  chiral  fermions in units of the gravitino mass, $m_\lambda$.  
In this case  the conditions \eqref{constraints}  are both necessary and sufficient,  
  and the shaded area represents the region of parameter space where the field configuration $\xi_0$ has no tachyons along the direction defined by $z_\lambda$, the  eigenstate  of the fermion mass matrix  $M \bar z_\lambda = m_\lambda z_\lambda$.
Configurations where $m_\lambda<1$ remain stable for arbitrary large values of the  ratio $H^2/m_{3/2}^2\gg1$.}
 \label{separableDiagram}
\end{figure}

We begin studying the simplest case where $\cH_\gamma$ is the only non-zero contribution to the Hessian along the  field space directions preserving supersymmetry:
\be
D_X \cM = \cR= 0\qquad \Longrightarrow \qquad  \cH  =\cH_\gamma= ( \cM + \unity) \big(\cM + (3 \gamma + 1 )\unity\big).
\label{separableHessian}
\ee
 Then, as in the case of  supersymmetric critical points, the Hessian is diagonal in the basis of eigenvectors $Z_{\pm \lambda}$ of the fermion mass matrix and the quantities $\mu_{\pm \lambda}^2$ can be identified with the eigenvalues of $\cH$, which read: 
\be 
 \mu_{\pm \lambda}^2 =( m_\lambda \pm 1) \big(m_\lambda \pm (3 \gamma + 1 )\big) = \( m_{\lambda}\pm\ft12 (3 \gamma+2)\)^2-\ft94\gamma^2.
  \label{upliftedmasses}
\ee 
Therefore, the conditions (\ref{constraints}-\ref{BF2}) are both necessary and sufficient to ensure the  stability of a configuration along the supersymmetric directions, which is entirely determined by the fermion mass spectrum $m_\lambda$ and the parameter $\gamma$.
 Since all the eigenvalues of the Hessian are bounded below by $-\ft94\gamma^2$, it follows that, when the configuration is an AdS critical point, $ \gamma\in [-1,0)$, these eigenvalues always satisfy the Breitenlohner-Freedman bound \eqref{BF2}, and thus it is always stable.  However, if the configuration  is either Minkowski or de Sitter ($\gamma\ge0$), stability demands that  the scalar potential has a minimum along the supersymmetric directions, which corresponds to situations where all fermionic masses satisfy:
\bea
\mu_{\pm \lambda}^2 \ge0 \qquad\Longrightarrow \qquad   m_\lambda<1 \quad \text{or} \quad  m_\lambda>3 \gamma + 1 \quad  \text{for all $\lambda$}.
  \label{VupStability}
\eea  
Thus, in Minkowski vacua ($\gamma=0$) the supersymmetric sector is always metastable, possibly with flat directions   if one or more  of the fermion masses equals the gravitino mass, $m_\lambda=1$. An interesting consequence for de Sitter configurations (either a vacuum or at a point of the inflationary trajectory) is that, if all fermion masses are smaller than the gravitino mass, i.e. $m_\lambda < 1$, the supersymmetric sector remains tachyon-free for arbitrary large values of the cosmological constant.  Conversely,  if the fermion spectrum contains any mass larger than $m_{3/2}$, the critical point will always become unstable for sufficiently large values of  the Hubble   parameter. These results are illustrated in Fig. \ref{separableDiagram}, which shows the  stability diagram of a non-supersymmetric configuration along a direction orthogonal to the sGoldstino. The horizontal axis  is related to the mass of the corresponding  fermionic partner $m_\lambda$, and the quantity on the vertical axis is the parameter $\gamma$.  In the diagram, the perturbatively stable configurations  are represented by the grey shaded area. 
\\
  %%%%%%%%%%%%%%%%%%%%%%%%%%%%%%%%%%%%
%%%%%%%%%%%%%%%%%%%%%%%%%%%%%%%%%%%%%%

 This simple example already illustrates the claim made in the previous subsection: in general, the condition $m_\lambda > 2$ necessary for a supersymmetric critical point to be a minimum,  is neither necessary or sufficient  when the supersymmetric sector is embedded in a larger model.  
 As we will show in section \ref{sec:afteruplifting}, the Hessian of the complex structure and dilaton sector of the tree-level scalar potential of type-IIB flux compactifications has the structure given by \eqref{separableHessian}, with $\gamma=0$.
More generally, this type of couplings arises naturally in models of $F-$term uplifting  where some heavy moduli are truncated while preserving supersymmetry \cite{Achucarro:2007qa,Achucarro:2008fk}, and in the context of inflation it has also  been considered in \cite{Achucarro:2012hg}. We shall discuss more about this class of models in section  \ref{sec:afteruplifting}.

%%%%%%%%%%%%%%%%%%%%%%%%%%%%
\subsubsection[Fermion mass derivatives and curvature]{Dependence on the fermion mass derivatives and the curvature}

 \label{subsec:nonsusy2}

 %%%%%%%%%%%%%%%%%%%%%%%%%%%%%%%%%%%%
 \begin{figure}[t]
 \vspace{-1cm}
 \centering \includegraphics[width=0.4\textwidth]{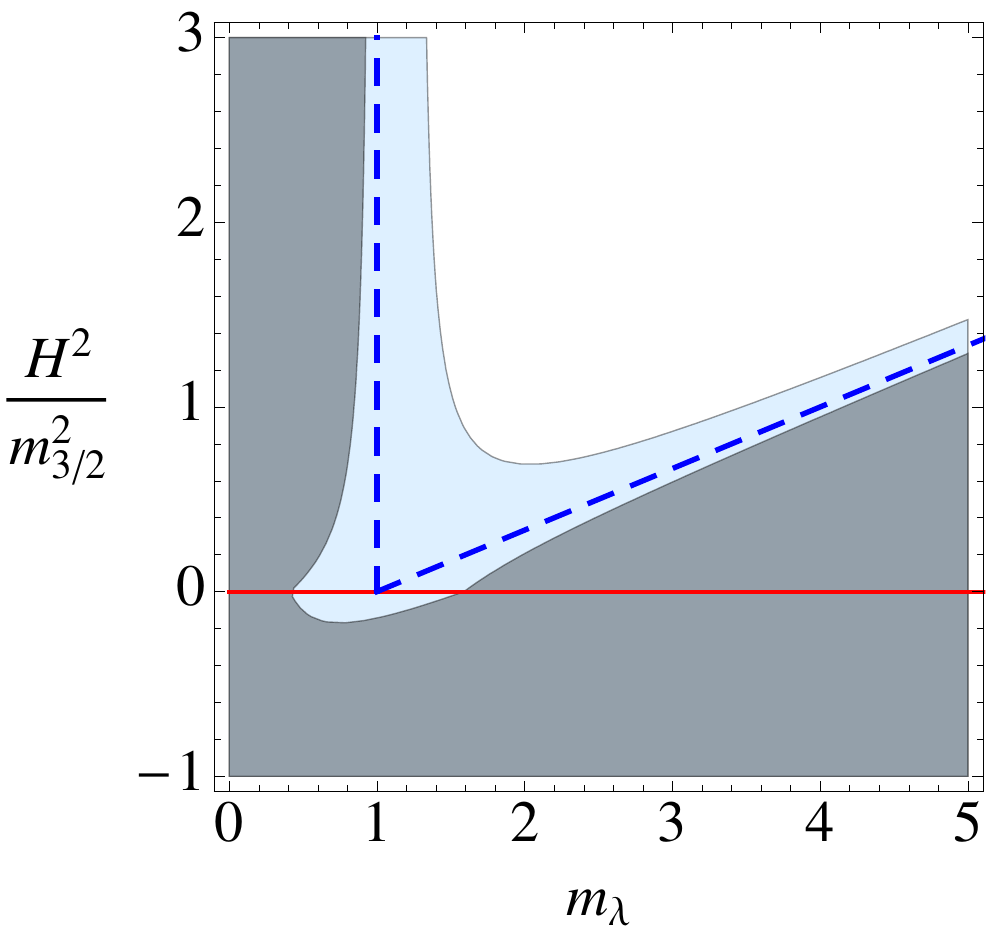}
\hspace{2cm}
\centering \includegraphics[width=0.4\textwidth]{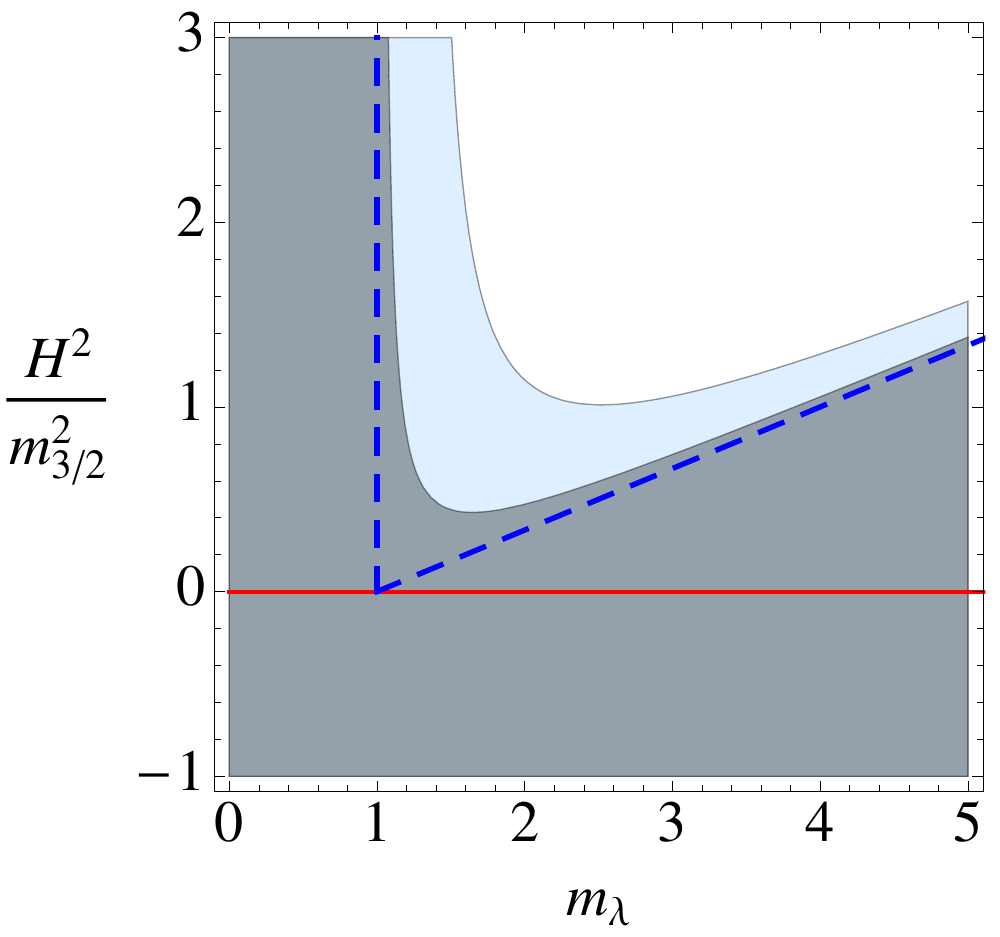}
  \caption{Stability diagram of a non-supersymmetric  field configuration $\xi_0$ associated  to the set of necessary conditions  (\ref{constraints}). On the vertical axis  $\gamma\equiv H^2/m_{3/2}^2$ is the squared ratio of the Hubble parameter to the gravitino mass. The horizontal axis represents the mass of one of the  chiral  fermions in units of the gravitino mass, $m_\lambda$.   Coloured areas show, for different choices of parameters, the regions where the stability conditions are satisfied, while in the white area $\xi_0$ has tachyonic instabilities. LEFT: $\pd m_\lambda /\pd X=0.2$, with $B[X,\lambda]=0$ (grey area) and $B[X,\lambda]=0.3$ (light blue area). RIGHT: $\pd m_\lambda /\pd X=-0.2$, with $B[X,\lambda]=0$ (grey area), and $B[X,\lambda]=0.3$ (light blue area). The dashed line represents the constraints in the case $B[X,\lambda]=\pd m_\lambda /\pd X=0$, showed in figure \ref{separableDiagram}.}
 \label{perturbedDiagram}
\end{figure}
%%%%%%%%%%%%%%%%%%%%%%%%%%%%%%%%%%%%%%

In general, when the terms $D_X \cM$  and $\cR$ are taken into account, the Hessian will not be diagonal in the basis formed by the vectors $Z_{\pm \lambda}$, and the set of necessary  conditions \eqref{constraints} will not be sufficient to ensure the stability of the field configuration. \\

Let us first  focus on the contribution coming from the term in the Hessian proportional to the derivative of the fermions mass matrix $D_X \cM$, while keeping the curvature term set to zero $\cR=0$. In  Fig. \ref{perturbedDiagram}, we have represented  the region of parameter space satisfying the stability conditions \eqref{constraints} for a particular  direction in field space $z_\lambda$ (shaded grey area), setting two  different constant values for the fermion mass derivatives, $\pd m_\lambda /\pd X=0.2$ (left plot) and  $\pd m_\lambda /\pd X = -0.2$ (right plot), and with a zero bisectional curvature $B[X,\lambda]=0$. Since these conditions are in general necessary but not sufficient, a field configuration located at the grey shaded region of the diagram cannot be  guaranteed to be stable, but those out of the shaded area will definitely contain one or more tachyonic directions in the spectrum. According to these diagrams, the stability conditions relax (the grey shaded area grows) with respect to the case studied above, $D_X\cM=0$,  when the derivatives of the fermion masses $\pd m_\lambda /\pd X$ take  negative values, and they become more restrictive otherwise. To understand this point, note that, when the derivatives of the fermion masses satisfy 
\be
\frac{\pd m_\lambda}{\pd X}\Big|_\text{opt} = - \frac{3 \gamma +2}{\sqrt{3 (\gamma +1)}} m_\lambda,
\label{optimumDm}
\ee   
the two parameters $\mu_{\pm \lambda}^2$ become equal, and both  constraints \eqref{constraints}  reduce to the less restrictive condition \eqref{mildConstraint}. Therefore, in the case of dS and Minkowski configurations ($\gamma\ge0$),  as the derivatives of the fermion masses approach  this optimum (negative) value, the stability constraints on
the fermion masses $m_\lambda$ and the parameter $\gamma$  become milder, as observed in the diagrams. It would be interesting to find a deeper physical interpretation of why the precise negative value $\pd m_\lambda /\pd X|_\text{opt}$ displayed in \eqref{optimumDm} maximises the stability along the supersymmetric directions. Indeed, we intend to further investigate the underlying physical reason for this in future work.\\

The effect of having a non-zero term $\cR$ is simpler to analyse. The set of necessary conditions \eqref{constraints}   clearly favour positive values of the    bisectional curvature  $B[X,\lambda]>0$. We can check that this is indeed the case in the plots of  Fig. \ref{perturbedDiagram}, where we have  displayed the region of parameter space satisfying the stability conditions \eqref{constraints} for two different constant values of the bisectional curvature,  $B[X,\lambda] = 0$ (grey area),  and $B[X,\lambda]=0.3$ (light blue area).  \\

When   the Hubble scale  is large compared to any of the fermion masses, $H \gg m_{3/2}$ and $H \gg m_\lambda\,  m_{3/2}$\footnote{Recall that we measure the chiral fermion masses $m_\lambda$ in units of the gravitino mass, $m_{3/2}$.},   the bisectional curvatures also play a fundamental r\"ole determining the stability of the inflationary trajectory. In that limit  (keeping $\pd m_\lambda /\pd X$ fixed)  the range of fermionic masses where the field configuration is  free of tachyons  is:
  \be
  0 \le m_\lambda \lesssim 1 + B[X,\lambda].
\label{largeGbound}
 \ee 
Then,  when the bisectional curvature is zero, we recover the limit discussed  above and only configurations where the largest mass of the chiral fermions satisfies $m_\lambda|_\text{max} <1$,  remain stable for sufficiently large values of $\gamma$.  However, for non-vanishing $B[X,\lambda]$ the situation changes.  On the one hand, from eq. \eqref{largeGbound} it is easy to  see that positive values improve the stability,
 as shown by the light blue regions of Fig. \ref{perturbedDiagram}. On the other hand, negative values of the bisectional curvature shrink the range of fermion masses compatible with  stable dS configurations.  Actually, when $B[X,\lambda]< -1$, the field associated to the direction  $z_{\lambda}$ always becomes tachyonic  for sufficiently large values of the Hubble parameter, $\gamma\gg1$, and therefore the corresponding  field configuration is necessarily unstable.   These constraints are of in\-te\-rest both for the construction of de Sitter vacua with small cosmological constant $\gamma\approx 0$ (as in the present vacuum), and for models of slow-roll inflation, to  study the stability of the inflationary trajectory. The study of the viability of inflationary models using the presented constraints  deserves further consideration, and we hope to report on this issue in a subsequent publication \cite{Geodesics}. \\
 
As we shall discuss in section \ref{sec:afteruplifting}, in supergravity models where the K\"ahler function has a similar structure as in the Large Volume Scenario of type-IIB flux compactifications,  the parameters $\mu_{\pm \lambda}^2$ give a good approximation to the masses in the  supersymmetric sector, in that case the dilaton and complex structure moduli. When this is the case, the conditions \eqref{constraints} are both necessary and sufficient to guarantee the stability of the supersymmetric sector. Then, from the previous discussion it follows that, in this framework, \emph{the conditions for a given configuration to correspond to a supersymmetric AdS minimum in the supersymmetric limit are more restrictive than requiring the  supersymmetric sector to be stable at a dS va\-cu\-um}. Indeed, on the one hand, supersymmetric minima require all the chiral fermion masses in the supersymmetric sector to be larger than twice the gravitino mass.
On the other hand, it is possible to satisfy all the  conditions \eqref{constraints} for the supersymmetric sector to be metastable  at the dS vacuum  regardless of the fermion mass spectrum. For this, we can make use of the extra freedom given by the possibility to tune the derivatives of the fermion masses, that is, the superpotential $W$, and the bisectional curvatures, which are determined by the K\"ahler potential $K$. \\

As a final comment, note the  close relationship between  the present discussion and the recent developments in the construction of dS vacua using a single-step approach which does not rely on the uplifting of supersymmetric AdS minima (see, for instance, \cite{Westphal:2006tn,Covi:2008zu,Danielsson:2012by,Blaback:2013qza,Kallosh:2014oja}). So far, this method has been proven more fruitful than previous attempts involving uplifting mechanisms, and in particular it has led to the construction of a large number of stable dS solutions, using both numerical and analytic methods. A particularly interesting example is the work by Kallosh \emph{et al.} \cite{Kallosh:2014oja}, where they construct a large number of analytic dS vacua in nearly no-scale models. For this purpose, the authors make use of the freedom to choose the sGoldstino direction, which in general affects the magnitude of $\pd m_\lambda / \pd X$ and $B[X,\lambda]$, and the cosmological constant, which in our case is set by  $\gamma$. Thus, these works also illustrate the main point of this section: by relaxing the condition that a particular sector of the theory is stabilised at a supersymmetric AdS minimum before including the supersymmetry breaking effects, we can improve significantly the chances of finding stable dS vacua. This will be reflected in the results of the random supergravity analysis that we will perform, which show a large increase on the fraction of stable dS configurations with respect to the KKLT type of constructions.   

\subsubsection{Summary}

As a summary of the main results of this section, we have found a set of  necessary conditions for the perturbative stability of non-supersymmetric field configurations in a generic supergravity theory, which are displayed in \eqref{constraints}. In particular, these conditions are required for the directions orthogonal to the sGoldstino to be tachyon-free. They involve the distribution of fermion masses and  their derivatives along the sGoldstino direction, the bisectional curvatures along the planes formed by the sGoldstino and the mass eigenstates of the chiralini, and the amount of supersymmetry breaking, which is expressed in terms of the ratio of the Hubble parameter to the gravitino mass. 
 We have characterised the effect of the bisectional curvatures and the derivatives of the fermion masses in the stability conditions \eqref{constraints}, finding the values that maximise the stability along the directions in field space which preserve supersymmetry. These effects are illustrated in figure \ref{perturbedDiagram}. In later sections we will discuss supergravity theories where a fraction of the field content can be consistently truncated while preserving supersymmetry. In order to analyse the stability of the truncated sector in those models,  we will characterise the fermion mass distributions using random matrix theory techniques, and we will derive more precise constraints on the couplings making use of the conditions \eqref{constraints}.\\

In the next section, we will introduce the class of random supergravity models which we shall use later on to  characterise the perturbative stability of a consistently truncated supersymmetric sector.  Those readers not interested in the technical details may continue reading in section \ref{sec:beforeuplifting}, and return to section \ref{sec:themodel} to find the relevant formulae.

%%%%%%%%%%%%%%%%%%%%%%%%%%%%%%
%%%%%%%%%%%%%%%%%%%%%%%%%%%%%%%%
\section{Modeling the supersymmetric sector}
%%%%%%%%%%%%%%%%%%%%%%%%%%%%%%%
%%%%%%%%%%%%%%%%%%%%%%%%%%%%%%%%%

\label{sec:themodel}

Having presented the main tool for our analysis, the set of constraints \eqref{constraints}, we now turn to the problem at hand: the study of the perturbative stability of a decoupled supersymmetric sector, such as the complex structure moduli in the KKLT constructions, and Large Volume Scenarios of Type IIB flux compactifications  \cite{Giddings:2001yu,Kachru:2003aw,Balasubramanian:2005zx}.  Motivated by these scenarios, where stabilisation of the complex structure moduli can be studied using a statistical treatment \cite{Denef:2004cf,Denef:2004ze,Marsh:2011aa,Bachlechner:2012at}, we will focus on theories where the supersymmetric sector contains a large number of fields, and  we will  characterise the couplings using a statistical description.  In the following two subsections  we will define the simplest class of supergravity models which implements these two features.  On the one hand, we will require the interactions  of the model to be consistent with the \emph{exact supersymmetric truncation} of the decoupled sector (section \ref{consistentTruncations}), and  on the other hand we will assume that the couplings of the supersymmetric sector are generic and can be treated as random variables.  In particular, we will characterise the spectrum of fermion masses of the supersymmetric sector  $m_\lambda$, and their derivatives, $\pd m_\lambda/\pd X$, using standard techniques from random matrix theory (section \ref{randomSugra}). 

 %%%%%%%%%%%%%%%%%%%%%%%%%%%%
\subsection{Supersymmetric  decoupling}
%%%%%%%%%%%%%%%%%%%%%
\label{consistentTruncations}

As mentioned above, the problem of supersymmetric decoupling of a sector of the field content  is specially relevant when considering moduli stabilisation in Type IIB flux com\-pac\-ti\-fi\-ca\-tions. For instance, in KKLT constructions  it is assumed that the dilaton and complex structure moduli are stabilised at some large energy scale and integrated out, leaving behind an effective field theory described by $\cN=1$ supergravity. In general, the solutions to these effective field theories are only approximate solutions of the full theory, and are valid at energies much lower than the mass scale of the fields that have been integrated out. The consistency of this approach has been discussed extensively in the literature \cite{Choi:2004sx,deAlwis:2005tf,deAlwis:2005tg,Brizi:2009nn,Rummel:2011cd}, and was finally settled by Gallego \emph{et al.} in \cite{Gallego:2008qi,Gallego:2009px}. As was discussed in these works, the fine-tuning of the vacuum expectation value of the flux superpotential is essential for the decoupling of the heavy moduli sector. However,  moduli stabilisation in Large Volume Scenarios does not require fine-tuning the   vacuum expectation value of the flux superpotential, and in general there is no large mass hierarchy between the  complex structure and dilaton fields  and the K\"ahler moduli. Despite of this,  it is still possible to truncate the complex structure and the dilaton while preserving supersymmetry approximately, since the couplings between these fields and the K\"ahler moduli  and the supersymmetry breaking effects are suppressed by inverse powers of the volume of the internal manifold. In this case, to leading order in inverse powers of the volume, the  couplings are  consistent with the \emph{supersymmetric truncation} of the dilaton and complex structure  sector \cite{Gallego:2011jm,Gallego:2013qe}.\\

   In a consistent truncation of a given theory (not necessarily supersymmetric), the heavy fields are frozen at a critical point of the scalar potential, in such a way that the solutions of the corresponding  reduced theory are  \emph{exact} solutions of the full theory. If the original theory is a  supergravity model, a supersymmetric  truncation satisfies the additional requirement that the reduced theory is also described by $\cN=1$ supergravity, which in particular implies that the vacuum expectation values of the truncated fields must preserve supersymmetry exactly \cite{Binetruy:2004hh,Achucarro:2007qa,Achucarro:2008sy,Achucarro:2008fk,thesis}\footnote{The  problem  of truncating  a locally supersymmetric theory while preserving a fraction of the supersymmetries has also  been thoroughly studied  in the context of reductions of $\cN$ extended supergravity theories, to lower $\cN'$  supergravities \cite{Andrianopoli:2001zh,Andrianopoli:2001gm}. Here we are interested in the case $\cN=\cN'$.}. This type of constructions is interesting for our purposes because they also ensure the consistency of studying the perturbative stability of the truncated sector on its own. On the one hand, as we will review below,  the perturbative stability analysis decouples in the truncated and surviving sectors. And on the other hand, the truncated fields will not be excited due to the possible space-time dependence of the fields in the reduced theory, which is specially relevant to in the context of inflationary models. In this sense, we can say that the truncated sector is \emph{supersymmetrically decoupled} from the fields in the reduced theory. \\

We will now review the features of a $\cN=1$ supergravity model which allows for the supersymmetric truncation of a sector. Consider a supergravity theory where the fields can be split in two different sets, the ``heavy'' (supersymmetric) sector $H^\alpha$, and the ``light'' (supersymmetry breaking) sector  $L^i$: 
 \be
\xi^I\quad \longrightarrow \quad \(H^\alpha, L^i\) \qquad \text{where} \qquad \alpha=1\ldots \scN_h, \quad \text{and} \quad
i=1 \ldots \scN_l.
 \ee 
In the absence of gauge couplings, the truncation is defined by fixing the heavy scalar fields at a supersymmetric critical point of the scalar potential $H^\alpha = H^\alpha_0$, and the corresponding effective action is obtained substituting the vacuum expectation values of the heavy fields in the original action
 \be
S_{eff}(L,\bar L) = S(H_0, \bar H_0, L,\bar L).
 \ee
 In order for the truncated fields to preserve supersymmetry regardless of the configuration of the surviving sector $L^i$, the K\"ahler function has to satisfy the following set of constraints: 
\be
G_\alpha(H_0,\bar H_0,L,\bar L)=0 \qquad \text{for all} \qquad L^i \qquad \text{and} \qquad \alpha=1\ldots \scN_h.
\label{conditionTruncation}
\ee
Note  that,  in general,  supersymmetry will be spontaneously broken in the reduced theory, i.e. $\pd_i G|_{H_0}\neq0$, and the sGoldstino must always belong to the light sector. The previous equations are sufficient to guarantee that the heavy fields are at a critical point of the scalar potential. In fact, it can be shown  that if the heavy fields leave supersymmetry unbroken for any configuration of the light fields, the equations of motion derived from the effective theory coincide with those obtained from the full action. Therefore, the truncation $H^\alpha = H^\alpha_0$ can be seen as an ansatz to solve the full equations of motion, and thus, in particular the heavy fields can never be sourced by the space-time dependence of the light fields.\\

 Since, by assumption, the equations \eqref{conditionTruncation} must hold for any value of the light fields $L^i$,  the couplings between the two sectors are very constrained. Taking  derivatives with respect to the  coordinates $L^i$ and $L^{\bar i}$ we find the following implications:
\begin{itemize}
\item   The K\"ahler manifold of the 
reduced theory is a totally geodesic submanifold of the K\"ahler manifold of the full theory, i.e. the Levi-Civita connection satisfies $\Gamma_{ij}^\alpha=0$. Moreover, the sigma model metric and the Riemann tensor have the following block-diagonal structure in the two sectors:
\be
G_{i \alpha} (H_0,\bar H_0,L,\bar L) = 0, \qquad R_{i\bar j k \bar \alpha }= 0.
\ee  
In addition, the conditions \eqref{conditionTruncation} imply that the sGoldstino is parallel to the reduced manifold at every point. 
\item  The fermion mass matrix $M$ and its derivative along the sGoldstino direction  $\nabla_X M$ are both block-diagonal in the heavy and light  sectors:
\be
M_{ i \alpha}|_{H_0} =0, \qquad \nabla_X M_{i \alpha}|_{H_0} =0.
\ee
\end{itemize}
Totally geodesic submanifolds of complex dimension larger than one, i.e. other than geodesics themselves, are rare in generic K\"ahler manifolds. However they appear naturally in the scalar manifold of supergravity models where the chiral sector is invariant under a global or local symmetry. Indeed, local and global symmetries of the chiral sector are always associated to an isometry of the K\"ahler manifold (see \cite{Freedman:2012zz}), and it is well known that the fixed point of an isometry always defines a totally geodesic submanifold \cite{opac-b1099770}. \\

Note that, since supersymmetry is broken in general by the light fields $L^i$, we cannot take for granted the stability of the supersymmetric sector, and a detailed perturbative stability analysis is needed. The previous conditions imply that all the $2 \scN\times 2 \scN-$matrices involved in the Hessian of the scalar potential have a block-diagonal structure in the heavy and light sectors:
\be
\cM = \cM_{h} \oplus  \cM_{l}, \qquad D_X \cM= D_X \cM_{h}  \oplus  D_X \cM_{l}, 
\qquad \cR = \cR_{h} \oplus \cR_{l}.
\ee
 and then the Hessian itself is block-diagonal in the two sectors  $\cH = \cH_{h} \oplus \cH_{l}$. Therefore, as we anticipated at the beginning of the section, if the couplings are compatible with the consistent  truncation of the supersymmetric sector, it is consistent to study the perturbative stability of the supersymmetric sector independently of  the sector surviving the truncation, i.e.  it is sufficient to consider the block of the Hessian $\cH_h$: 
 \be
\cH_h =( \cM_h + \unity) \big(\cM_h + (3 \gamma + 1 )\unity\big)
+\sqrt{3(\gamma+1)} \, D_X \cM_h  - 3(\gamma+1) \cR_h.
\vspace{.1cm}
\label{HessianDecompSusy}
\ee
In the following sections we consider models with a supersymmetrically decoupled sector, in the sense described above, focusing on the perturbative stability of the configuration defining the truncation, $H_0^\alpha$, along the supersymmetric directions $H^\alpha$.

%%%%%%%%%%%%%%%%%%%%%%%%%%%%%%%%%%%%%%%%
\subsection{Statistical description}
\label{randomSugra}
%%%%%%%%%%%%%%%%%%%%%%%%

In order to study the perturbative stability of the supersymmetrically decoupled sector, we need to be able to characterise the eigenvalue spectrum of the fermion mass matrix $\cM_h$, and the properties of its derivative $D_X \cM_h$.  At a generic point of the reduced theory, where $H^\alpha= H_0^\alpha$,  these matrices read: 
\be
\cM_h = \begin{pmatrix}
0 & M_{\alpha \beta} \\
\bar  M_{\bar \alpha \bar \beta} & 0,
\end{pmatrix}, \qquad \qquad D_X \cM_h = \begin{pmatrix}
0 & \nabla_X M_{\alpha \beta} \\
 \nabla_{\bar X}  \bar M_{\bar \alpha \bar \beta} & 0
\end{pmatrix},
\label{MdM}
\ee
where the tensors $M_{\alpha \beta} =  \nabla_\alpha G_\beta$ and $\nabla_X M_{\alpha \beta } = \nabla_X (\nabla_\alpha G_\beta)$,  will depend in general on the configuration $H_0^\alpha$ and on the light fields $L^i$.   As discussed in the introduction, the present analysis is motivated by the type of supergravity theories which arise in type-IIB flux compactifications. In that framework, given the large number of moduli fields and the complexity of the couplings, it is impractical to obtain a detailed  knowledge of the matrices involved in the Hessian. However, as was originally proposed by Denef and Douglas \cite{Denef:2004cf,Denef:2004ze} and later developed in \cite{Marsh:2011aa,Bachlechner:2012at}, given the large size of these matrices, it is still possible to obtain ge\-ne\-ral features of the spectrum of eigenvalues of the Hessian   following a statistical approach. Indeed, treating the couplings associated to the supersymmetric sector as random variables, more precisely the components of the tensors $M_{\alpha \beta}$ and $\nabla_X M_{\alpha \beta}$, it is possible to characterise the properties of the matrices \eqref{MdM} using standard techniques from random matrix theory (see \cite{mehta1991random}).  Since the two matrices in \eqref{MdM} have the same structure and symmetries, it will be sufficient to present the relevant results focusing on the fermion mass matrix $\cM_h$, leaving the discussion of the properties of $D_X \cM$ for the end of this section.

\subsubsection{Probability distribution of the couplings}

Following \cite{Marsh:2011aa,Bachlechner:2012at}, we will assume that these random variables $M_{\alpha \beta}=M_{\beta \alpha}$ are characterised by a unique probability distribution $\Omega$, with zero mean and standard deviation $\sigma$ (possibly field-dependent) for $\alpha\neq \beta$ and $\sqrt{2} \sigma$ for $\alpha=\beta$:
  \be
M_{\alpha \beta} \in \Omega(0,\sigma) \quad \text{if} \quad \alpha < \beta, \quad \text{and} \qquad M_{\alpha \alpha} \in \Omega(0,\sqrt{2} \sigma).
\label{distribution}
\ee
It is important to check that the corresponding joint probability distribution is invariant under the symmetries of the supergravity theory: supersymmetry and K\"ahler transformations, and the diffeomorphisms on the K\"ahler manifold. The consistency with supersymmetry transformations is a trivial check since we are only considering the supersymmetric sector\footnote{The components of the fermion mass matrix along the Goldstino direction are not invariant under supersymmetry, and actually in the supersymmetric unitary gauge, the fermion mass matrix has a zero mode associated to the Goldstino, (see \cite{Freedman:2012zz}). This ambiguity is avoided in \cite{Denef:2004ze,Marsh:2011aa} by imposing the constraint \eqref{sGoldstinoStab} on the matrix $\cM$.}. The invariance under K\"ahler transformations is also granted, as we work in a manifestly K\"ahler invariant formalism. Finally, although it is not evident, this definition is also consistent with the invariance under diffeomorphisms. Indeed, the associated covariance matrix can be written as    
\begin{equation}
\mathbb{E}\[M_{\alpha \beta} \,M_{\bar \gamma \bar \delta}\] =\sigma^2 \, (\delta_{\alpha \bar  \gamma}\, \delta_{\beta \bar  \delta} + \delta_{\alpha \bar  \delta}\, \delta_{\beta \bar  \gamma} )  
\ .
\label{GcovarianceMatrix}
\end{equation}
Then, noting that we are working in the local frame where the eigenvalues of $\cH$ can be identified with the masses of the scalar fields, we can obtain an expression manifestly covariant under diffeomorphisms  replacing the delta functions by the metric tensor, $\delta_{\alpha \bar \beta}= G_{\alpha \bar \beta}$ (see  \cite{Ashok:2003gk}).  Moreover, it is easy to check that the previous expression is the only choice of the covariance matrix consistent with the symmetry of $M_{\alpha \beta}$ and is also isotropic, that is,  invariant under the $\U(\scN_h)$ transformations of the local frame. In the limit where the size of the matrices is very large, i.e. the number of $H^\alpha$ fields satisfies $\scN_h \gg 1$, due to the universality of random matrix theory, the results we will now present do not depend on any higher moments of the distribution $\Omega$, provided they are appropriately bounded. \\

The  quantities appearing in the Hessian \eqref{HessianDecompSusy}  which are associated to the supersymmetry breaking sector, i.e. $\gamma$ and the gravitino mass\footnote{As explained in section \ref{sec:structureH}, the gravitino mass has been hidden for convenience in the units. Its value depends in general on all the fields, and in particular on the supersymmetry breaking sector.} $m_{3/2}$, will be regarded as parameters and studied in a case by case basis. Moreover, following the works in \cite{Ashok:2003gk,Douglas:2004kp,Denef:2004cf,Denef:2004ze} we also assume that the geometry of the K\"ahler manifold is determined, e.g. in type-IIB compactifications by the choice of the Calabi-Yau, 
and therefore the bisectional curvatures $B[X,\lambda]$ will also be treated as constant parameters. Note that, in general, all these quantities will be dependent on the configuration $H_0^\alpha$ and on the light fields $L^i$. As a result, we will be able to give predictions about the stability of the supersymmetric sector depending on the parameters defining the underlying distribution of the couplings, the geometry of the moduli space, and the supersymmetry breaking scale. \\
 
In the following subsection, we will use a few basic results from random matrix theory to characterise the spectrum of masses of the fermions $m_\lambda$ and their derivatives along the sGoldstino direction $\pd m_\lambda / \pd X$. We will use these spectra to characterise the perturbative stability of the supersymmetric sector in this class of models via the analysis of the constraints \eqref{constraints}.

%%%%%%%%%%%%%%%%%%%%%%%%%%%%%%
\subsubsection{The Altland-Zimbauer CI-ensemble}
%%%%%%%%%%%%%%%%%%%%%%%%%%%%%%% 

The set of hermitian matrices with the same structure as \eqref{MdM}, and random complex entries drawn from a probability distribution with the properties given in \eqref{distribution}, form the so-called \emph{Altland-Zimbauer} or \emph{CI-ensemble}. As we already discussed in section \ref{sec:Constraints}, the eigenvalues of the fermion mass matrix $\cM_h$ come in pairs $\pm m_\lambda$, with $0 \le m_1 \le m_2\le \ldots \le m_{\scN_h}$, and taking the distribution $\Omega$ to be gaussian, the joint probability density for the fermion masses $m_\lambda$ will be given by 
\be
f(m_1, \ldots, m_{\scN_h}) = {\cal C}\ \exp\left(- \frac{1}{2 \sigma^2} \sum_{\lambda=1}^{\scN_h} m_\lambda^2 \ +\sum_{\lambda < \nu}^{\scN_h}{\rm{ln}}|m_{\nu}^2 - m_\lambda^2| +\,\sum_{\lambda=1}^{\scN_h} {\rm{ln}}\,m_\lambda \right) \, . \label{eq:CI}
\ee
In random matrix theory, the spectrum of eigenvalues is characterised by the spectral density $\rho(m) \, dm $, which in this case gives the average number of fermion masses $m_\lambda$ in the interval $[m, m + dm)$. In the limit when $\scN_h\to\infty$, the spectral density is closely related to the Wigner's semicircle law (SC) \cite{Wigner:1955}, and it reads\footnote{Actually, the spectral density of the CI-ensemble presents a characteristic cleft of width $1/\scN_h$ near $m=0$, where it behaves as $\rho_{SC}(m)\sim m$, but we will neglect it, as it is a subleading effect in the large $\scN_h$ limit.}:
\be
\rho_\text{SC}(m)  =  \frac{4 \, \scN_h}{\pi  m_h^2}\sqrt{m_h^2 -  m^2} ,\qquad\qquad \rho_\text{MP}(m^2)  =  \frac{2 \, \scN_h}{\pi  m_h^2 \, m}\sqrt{m_h^2 -  m^2} , 
\label{SCMP}
\ee
%%%%%%%%%%%%%%%%%%%%%%%%%%%%%%%
\begin{figure}[t]
\vspace{-1cm}
\centering \includegraphics[width=0.45\textwidth]{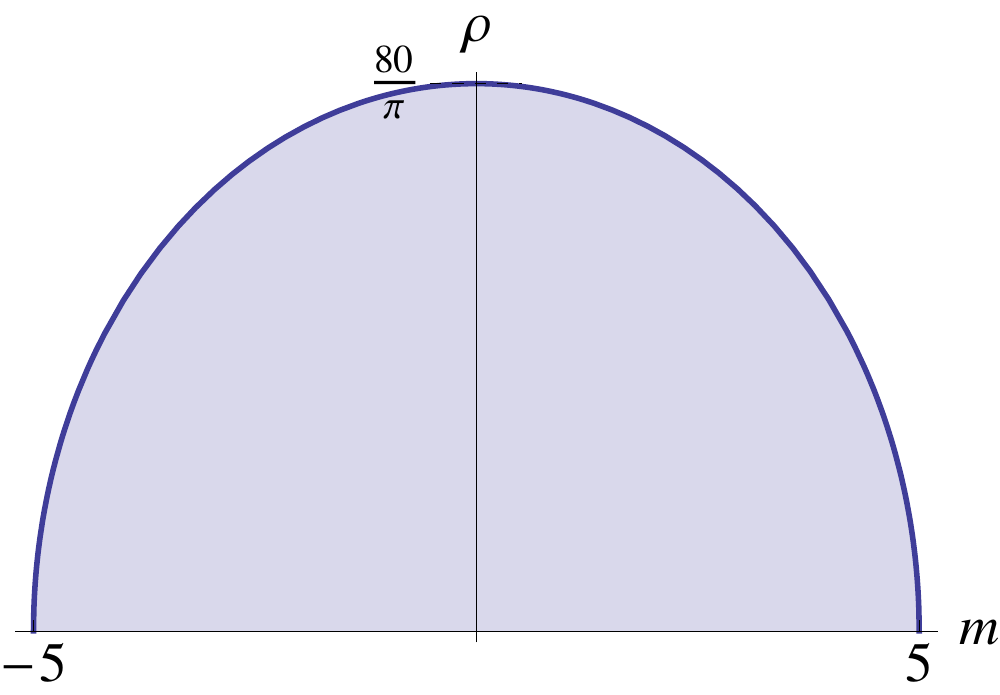}
\hspace{1cm}
\centering \includegraphics[width=0.45\textwidth]{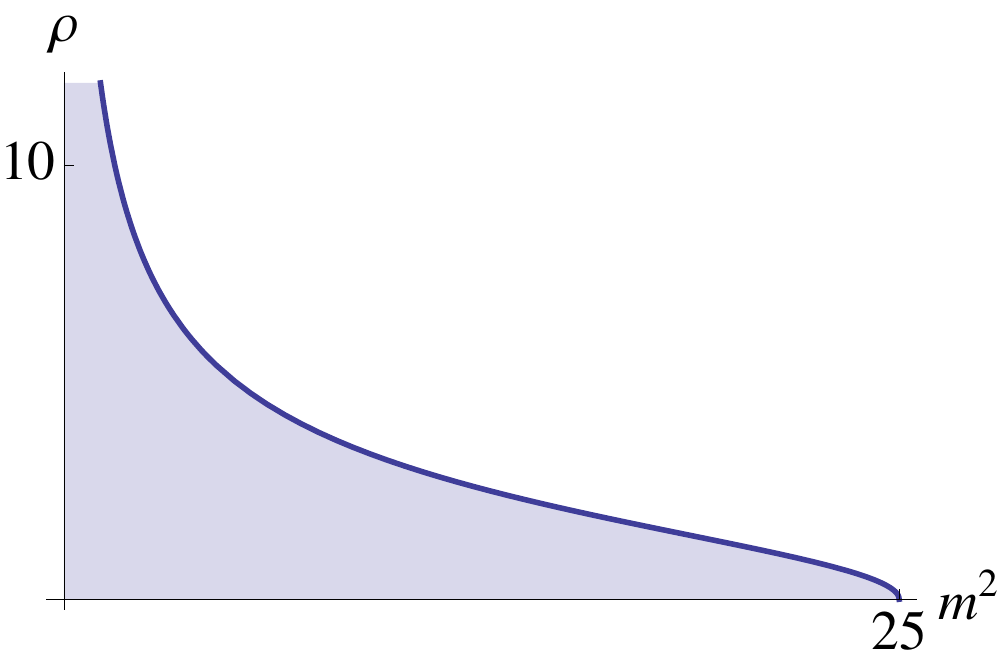} 
 \caption{Typical spectrum of fermion masses  for $m_h=5$ and $\scN_h=100$. LEFT: The spectral density of the fermion mass matrix $\cM$  resembles Wigner's semicircle law to leading order in $1/\scN$. RIGHT:  The Mar\v{c}enko-Pastur law gives the typical distribution for  the square of the fermion masses $m^2$.}
 \label{MPSCdist}
\end{figure} 
%%%%%%%%%%%%%%%%%%%%%%%%%%%%%%%%%
for $m \le m_h$ and zero otherwise. Here we have defined $m_h^2 \equiv 4 \scN_h \, \sigma^2 $, which sets the  mass scale of the truncated sector in units of the gravitino mass. For later convenience, we have also written the distribution of the square of the fermion masses $\rho_\text{MP}(m^2)\,  dm^2$, which is a particular case of the so-called  Mar\v{c}enko-Pastur law (MP) \cite{MPLaw}.  An example of both distributions is displayed in Fig. \ref{MPSCdist}. The fact that the spectral density \eqref{SCMP} has a compact support in the large $\scN_h$ limit does not imply that the probability of finding eigenvalues out of the specified range is zero. The previous expression only gives the typical spectrum of a large matrix from the CI-ensemble, but other atypical spectra are also possible at the cost of having a suppressed probability (see appendix \ref{RMT}). Consider for example the  position of the limiting eigenvalues $m_1$ and $m_{\scN_h}$, which will be of interest for our discussion, since they give the mass of the lightest and the heaviest fermions, respectively. To leading order in $1/\scN_h$, their expectation values are given by:
\be
\mathbb{E}[m_1] =0,  \qquad \qquad \mathbb{E}[m_{\scN_h}]=m_h.
\label{limitingEig}
\ee
However, for large but finite values of $\scN_h$, there is a non-zero probability of finding $m_1$ and $m_{\scN_h}$ away from these values, which is determined by the so-called Tracy-Widom distribution \cite{Tracy:1992rf} in a region of size $\cO(\sigma^2 \, \scN_h^{-1/3})$ around the expected values. To leading order in $1/\scN_h$,  the corresponding cumulative probability distributions have the form  
\be
 \mathbb{P}(m_{1}\ge t) \sim  \rme^{- \frac{2 \, t^2}{m_h^2}\, \scN^2_h  }, \qquad \qquad\mathbb{P}(m_{\scN_h}\le t \le m_h) \sim \, \rme^{- \ft{1}{6}x^3 \scN^2_h}\ ,\label{fluctuations}
\ee
where $x\equiv \big(t^2-m_h^2  \big)/m_h^2$. The first expression gives the probability that the lightest fermionic mass is larger than a given value $t\ge0$. The second expression  represents the probability that all the fermionic masses are bounded above by a value $t$ smaller than $m_h$, i.e. the typical size of the largest fermion mass. As we can see, these atypical spectra are also possible, but at the cost of an exponentially suppressed probability.\\

The matrix $D_X \cM$ has similar symmetries and structure as $\cM$, and assuming its entries to be distributed as in \eqref{distribution} (but with a different standard deviation), it can also be identified as an element from the CI-ensemble. Choosing the covariance matrix to be given by 
\begin{equation}
 \mathbb{E}\[\nabla_X M_{\alpha \beta} \, \nabla_{\bar X} M_{\bar \gamma \bar \delta}\]= \frac{\scS^2}{ 4 \scN_h} (\delta_{\alpha \bar  \gamma}\delta_{\beta \bar  \delta} + \delta_{\alpha \bar  \delta} \delta_{\beta \bar  \gamma} )  
\ ,
\label{dxGcovarianceMatrix}
\end{equation}
the eigenvalues of a typical spectrum of $D_X \cM_h$ will be contained in the interval $\[-\scS,\scS\]$ with probability close to one. Therefore, using standard linear algebra it can be proven that the diagonal elements of $D_X \cM_h$ in any orthonormal basis are also bounded by the same limits. In particular, in the basis of eigenvectors of $\cM_h$, we have:
\be
\la Z_{\pm \lambda}, D_X \cM_h\,  Z_{\pm \lambda}\ra = \pm \frac{\pd m_\lambda }{\pd X}\in \[ - \scS, \scS\].
\label{boundDerivatives}
\ee
Note that, in general, it cannot be taken for granted that the matrices $\cM_h$ and $D_X \cM_h$ are statistically independent from each other, since this depends on the details of the couplings between the $H^\alpha$ fields and the sGoldstino, but this does not affect any of the previous results.\\

In the following sections, we will analyse in detail the perturbative stability of the supersymmetrically truncated sector combining the constraints \eqref{constraints} with the statistical cha\-rac\-te\-ri\-sa\-tion of the fermion mass spectrum just discussed. 

%%%%%%%%%%%%%%%%%%%%%%%%%%%%%%%%%%%%%%%
%%%%%%%%%%%%%%%%%%%%%%%%%%%%%%%%%%%

\section{Statistics of supersymmetric vacua}
\label{sec:beforeuplifting}

Our starting point in this section are the results of section \ref{subsec:uplift} regarding the character of AdS supersymmetric critical points. We now incorporate the statistical properties of the fermionic mass spectrum given by random matrix theory as discussed in the previous section. We will recall the results obtained by Bachlechner \emph{et al.} in \cite{Bachlechner:2012at} which show that, in a generic supergravity theory, the fraction of supersymmetric AdS  critical points which are minima of the scalar potential is exponentially suppressed by the number of fields, $\mathbb{P}_\text{min} \sim\text{exp}(-8\scN^2/m_h^2)$, where, in this case,  the constant $m_h$ sets the ratio between the typical scale of the supersymmetric masses  and the gravitino mass. This result is particularly relevant for the construction of dS vacua in KKLT scenarios, where the hierarchy between the masses of the dilaton and complex structure fields and the supersymmetry breaking scale allows to study the stability of these moduli neglecting the supersymmetry breaking effects. In this framework, it is assumed that the heavy moduli are stabilised at a supersymmetric  minimum of the flux superpotential with large supersymmetric masses, so that they remain stabilised after the spontaneous breaking of supersymmetry.    Then, the results in \cite{Bachlechner:2012at} imply that in a generic supergravity theory, unless the parameter $m_h$ is fine-tuned so that the typical mass scale of the heavy moduli is much larger than the gravitino mass, i.e. $m_h\gg \scN$, the fraction of critical points of the scalar potential   consistent with  a KKLT construction is exponentially suppressed.
 Afterwards, in section \ref{sec:afteruplifting}, we will explore more general settings where the dS vacuum is constructed without requiring   one sector   of the fields to be stabilised at an AdS minimum, which may occur in moduli stabilisation mechanisms where there is no  mass hierarchy between the moduli fields and the supersymmetry breaking scale.  We show that the probability of the decoupled supersymmetric sector  being tachyon-free can still be made of order one for certain values of the parameters which determine the distribution of the couplings and the geometry of the moduli  space.  

%%%%%%%%%%%%%%%%%%%%%%%%%%%%%%%%%%%%

\subsection{Eigenvalue spectrum of the Hessian}

%%%%%%%%%%%%%%%%%%%%%%%%%%%%%%%%%
   
Supersymmetric AdS critical points are extrema of the K\"ahler function and, as we shall review here, at   AdS supersymmetric  critical points the Hessian of the scalar potential is closely related to the Hessian of the K\"ahler function $G$, which we shall denote by $\cG$.  Indeed, taking into account that the fields are canonically normalised, it can be shown that
\be
\cG = \unity + \cM \qquad \Longrightarrow \qquad \cH =\cG \big(\cG - 3\,  \unity\big).
\label{HessianDecomp2}
\ee 
As it was pointed out in \cite{Achucarro:2007qa,Achucarro:2008fk}, this relation implies a  one-to-one correspondence between the supersymmetric AdS maxima  and the minima of the gravitino mass, which holds in full generality when gauge interactions are included \cite{Achucarro:2008fk}. To see this point, note that   $\cG$ is also diagonal in the basis of $Z_{\pm \lambda}$ and the eigenvalues are given by  $g_{\pm \lambda} = 1 \pm m_\lambda$. Thus,  the K\"ahler function $G$ is minimised for field configurations where all the fermion masses satisfy $m_\lambda <1$, which corresponds precisely to supersymmetric AdS maxima \eqref{Vstability}.\\ 

\begin{figure}[t]
\vspace{-1cm}
\centering \includegraphics[width=0.48\textwidth]{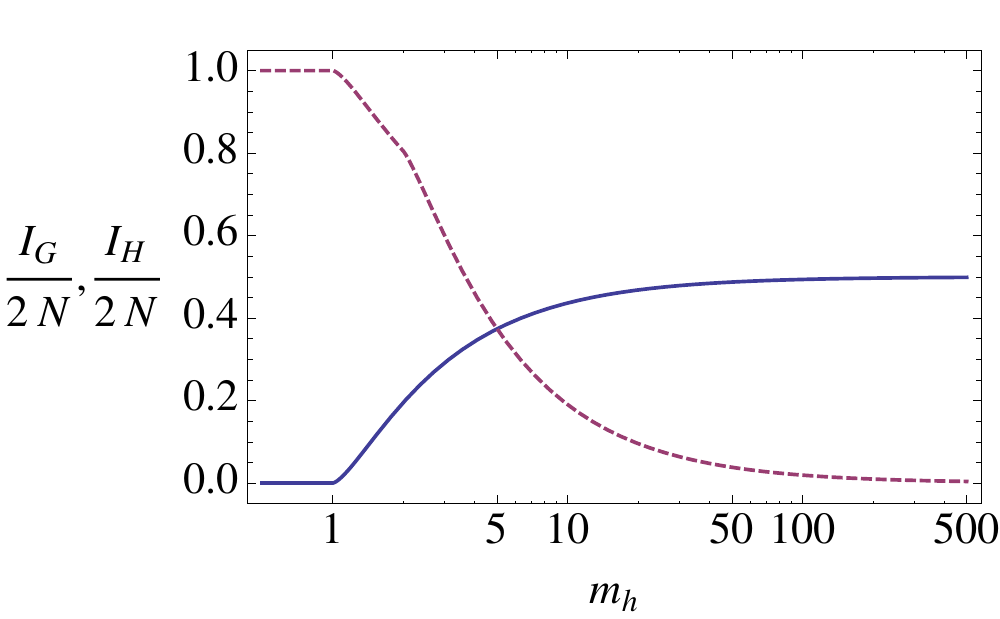}
 \caption{Expectation value of the Morse index (divided by $2 \scN$) of the Hessian of the K\"ahler function $\cI_G$ (solid), and of the scalar potential $\cI_H$ (dashed) as a function of the standard deviation of the fermion masses $m_h$. The index measures the number of negative eigenvalues of the corres\-pon\-ding Hessian. Notice the one-to-one correspondence between minima of $G$ and supersymmetric AdS  maxima of $V$, for $m_h<1$.}
 \label{Fig:index}
\end{figure}

Given the relation between $\cH$ and the Hessian of the K\"ahler function,  let us start  cha\-rac\-te\-ri\-sing the dependence of the spectral density of $\cG$ on the parameter $m_h$ which   determines the  scale of the supersymmetric masses, and in particular it gives the expected mass of the heaviest fermion (\ref{limitingEig}). For that purpose, it is convenient to study the index $\cI_G$, which represents the number of negative eigenvalues in its spectrum, i.e. the Morse index. In Fig.\@\ref{Fig:index} we have plotted the index $\cI_G$ (solid line) as a function of the parameter $m_h$, which can be easily calculated from the spectral density of the fermion masses (\ref{SCMP}): 
\begin{equation}
\cI_G(m_h)= \left\{ \begin{array}{ll}
0 &\qquad  \textrm{if $m_h \le 1$},\\
\scN - \ft{2 \scN}{ \pi m_h} \sqrt{1-m_h^{-2}} -\ft{2 \scN}{ \pi} \arctan\left(\ft{1}{\sqrt{ m_h^2-1}}\right) &\qquad \textrm{otherwise.}
\end{array} \right. 
\label{indexG}
\end{equation}
When the supersymmetric mass scale is smaller than the gravitino mass, $m_h\le1$, at a typical  supersymmetric critical point, all  fermion masses  are also bounded above by the gravitino mass, $m_\lambda \le 1$, and thus  the  critical point corresponds to a minimum of the K\"ahler function $G$, possibly with flat directions. In any other situation, the dominant type of supersymmetric critical points is a saddle point,  and in fact, in the limit when $m_h$ is very large,  $\cI_G \to \scN$, only half of the eigenvalues  of $\cG$ become positive\footnote{In the absence of supersymmetry breaking, the number of fields in the supersymmetric sector coincides with the total number of fields, thus in the formulas from section \ref{randomSugra} we set $\scN_h\to\scN$.}.\\

%%%%%%%%%%%%%%%%%%%%%%%%%%%%%%%%%%%%%%%
%%%%%%%%%%%%%%%%%%%%%%%%%%%%%%%%%%%%%%%

Similarly, we can define the index of the Hessian of the  scalar potential $\cI_H$ as the number of eigenvalues which are negative.  In figure \ref{Fig:index} we have plotted the value of $\cI_H$ (dashed line) as a function of the supersymmetric mass scale $m_h$, which we have calculated using the spectral density of the fermions  (\ref{SCMP}) and the relation (\ref{spectrum2}):
\begin{equation}
\cI_H(m_h)= \left\{ \begin{array}{ll}
2\scN &\qquad  \textrm{if $m_h\le 1$},\\
2\scN-\cI_G(m_h)&\qquad \textrm{if $1\le m_h\le 2$}, \\
2\scN- \cI_G(m_h) -  \cI_G(\ft{m_h}{2})&\qquad  \textrm{otherwise}.
\end{array} \right. \end{equation}
 Thus, for $m_h <1$ the typical supersymmetric critical point is an AdS maximum, and as $m_h$ becomes larger, $\cI_H$ decreases towards zero. It is interesting to note that  saddle points are the dominant type of critical point for most values of $m_h$, and supersymmetric AdS minima become the dominant type only when the  scale of the supersymmetric masses  is tuned to be much larger than the gravitino mass, $m_h\to \infty$.   \\

\begin{figure}[t]
\vspace{-1cm}
\centering \hspace{-.5cm}\raisebox{-0.07\height}{\includegraphics[width=4.9cm]{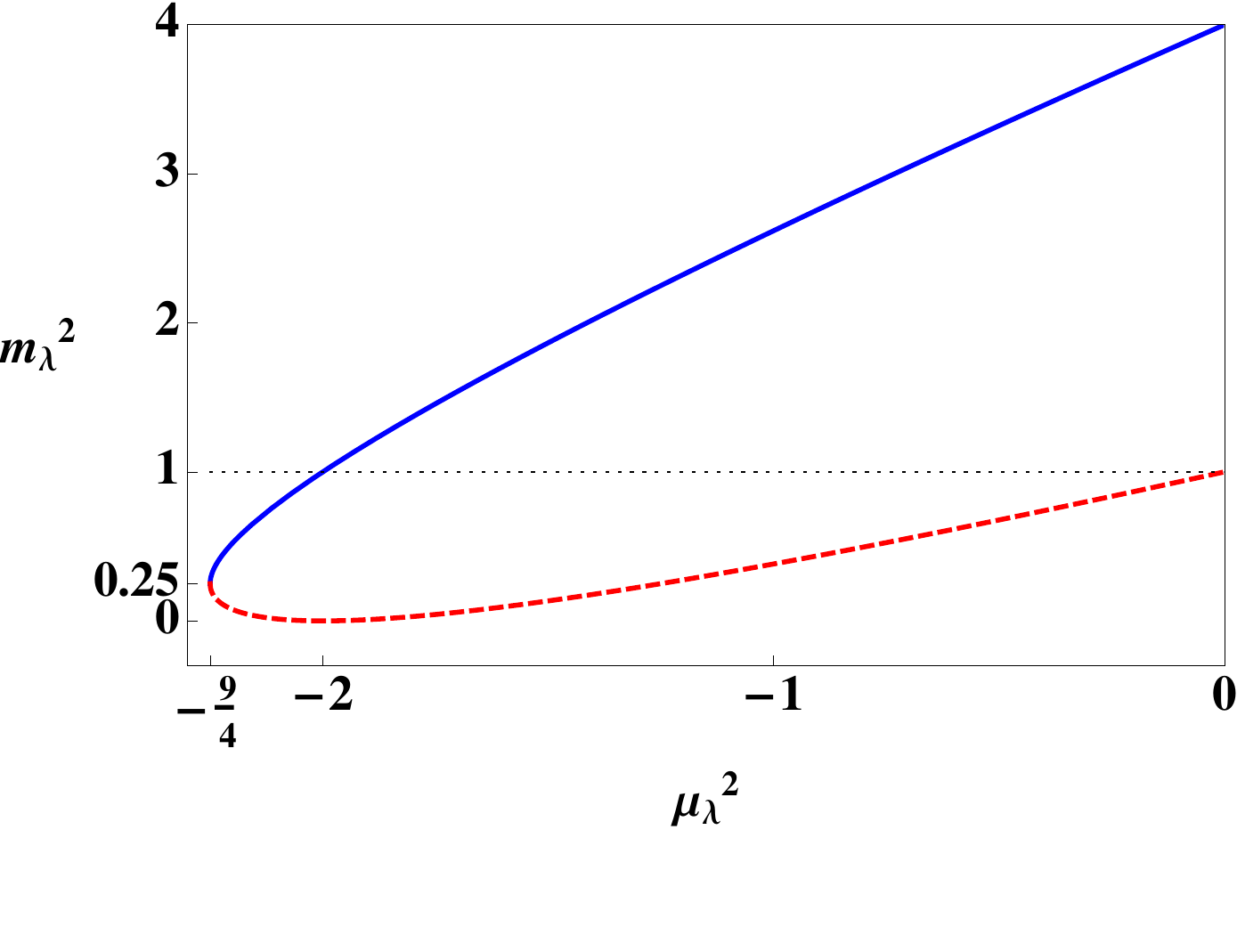}}
\hspace{.1cm}
\centering \raisebox{0.15\height}{\includegraphics[width=0.305\textwidth]{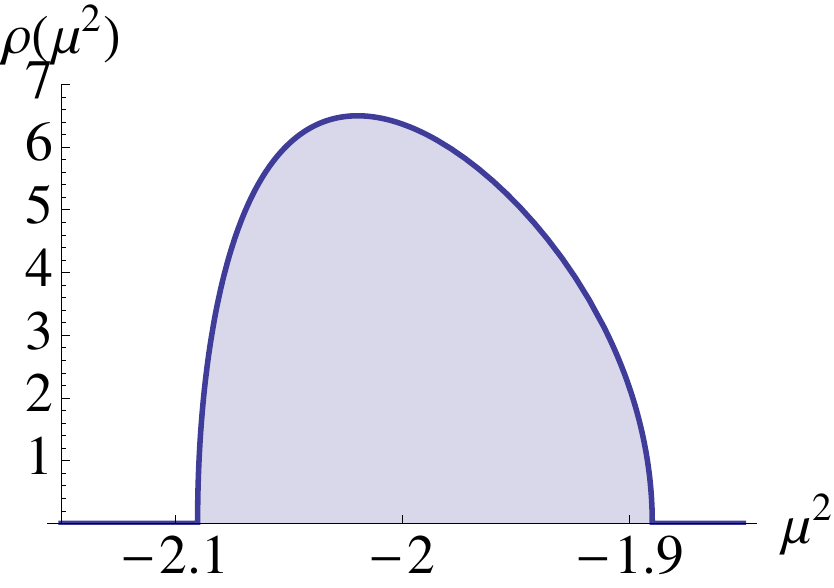}} 
\hspace{.1cm}
\centering \raisebox{0.14\height}{\includegraphics[width=0.305\textwidth]{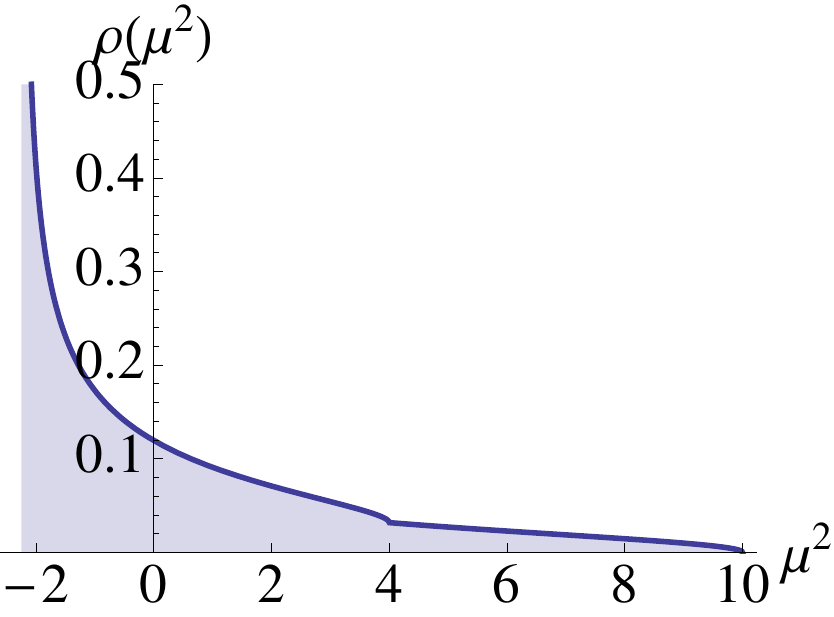}} 
 \caption{LEFT: The two branches of the multi-valued function obtained when inverting eq. \eref{spectrum2}, $m_+^2(\mu^2)$ (solid)  and $m_-^2(\mu^2)$ (dashed). When all fermion masses are smaller than the gravitino mass, $m_\lambda<1$, the configuration is a supersymmetric AdS maximum. CENTRE: Typical spectral density of the Hessian $\cH$, eq.  \eref{totaldensity},  for $m_h=0.1$ where all modes are BF-allowed tachyons. RIGHT: Spectral density of the Hessian $\cH$  for $m_h=3$. In the plots the spectral densities  are normalised to unity.}
 \label{fig1}
\end{figure}

Let us now calculate the scalar mass spectrum for a typical supersymmetric  critical point. The expression (\ref{spectrum2})  relating the eigenvalues of the Hessian $\cH$ to the fermion masses can be used in combination with the Mar\v{c}enko-Pastur law \eqref{SCMP} to determine the typical spectral density of $\cH$. First, expressing the square of the fermion masses  in terms of the eigenvalues of the Hessian $\mu^2$, cf. eq. \eqref{spectrum2}, we find a multiple-valued function with two branches, which we denote by $m_\pm^2(\mu^2)$, and are displayed in Fig. \ref{fig1}. Then, the contribution from each of the branches to the spectral density of $\cH$ reads simply:
\be
\rho_\text{MP}\(m_{\pm}^2\)\Bigg|\frac{d m_{\pm}^2}{d\mu^2}\Bigg| = \Theta (m_h^2-m_{\pm}^2) \,  \frac{2}{\pi m_h^2} \sqrt{\frac{m_h^2 -m_{\pm}^2}{\mu^2+\ft{9}{4}}}\ ,
\label{susycritdist}
\ee
where $m_{\pm}^2$ should be understood as functions of $\mu^2$, and the Heaviside theta functions $\Theta$ are a reflection of the support of the  Mar\v{c}enko-Pastur distribution for  each of the branches. The total eigenvalue density function, given the spectrum of the Hessian, can then be written as \cite{Bachlechner:2012at}:
\be
\rho(\mu^2) d\mu^2 =\[\rho_\text{MP}\(m_{+}^2\)\Bigg|\frac{d m_{+}^2}{d\mu^2}\Bigg|+\rho_\text{MP}\(m_{-}^2\)\Bigg|\frac{d m_{-}^2}{d\mu^2}\Bigg|\,\]d\mu^2 \ .\label{totaldensity}
\ee
%%%%%%%%%%%%%%%%%%%%%%%%%%%%%%%%%%%%%%%
%%%%%%%%%%%%%%%%%%%%%%%%%%%%%%%%%%%%%%%
 Note that  this distribution is normalised to $2 \scN$ in contrast with the spectral density of the fermions \eqref{SCMP} because, while the number of fermion masses is $\scN$, the number  eigenvalues of  the Hessian $\cH$ is $2\scN$. Illustrative examples of the distribution \eqref{totaldensity} are given in figure \ref{fig1}.  In those plots we see that this spectrum interpolates between a shifted Wigner semicircle law for $m_h\to 0$, corresponding to AdS maxima, and a shifted version of the Mar\v{c}enko-Pastur distribution for spectrum for $m_h\gg 1$, corresponding to supersymmetric critical points which are \emph{typically} AdS saddle points.

%%%%%%%%%%%%%%%%%%%%%%%%%%%%
\subsection{Uplifting a supersymmetric sector}
%%%%%%%%%%%%%%%%%%%%%%%%%%

In KKLT constructions, the dilaton and the complex structure moduli are assumed to be stabilised at a supersymmetric minimum of the flux superpotential. As long as the supersymmetric moduli are stabilised with sufficiently large  masses,   the supersymmetry breaking effects necessary to lift the vacuum to dS will not induce the appearance of instabilities. The fact that there is no region of parameter space where supersymmetric AdS minima are the dominant type of critical points does not imply that they do not exist. Supersymmetric mi\-ni\-ma correspond to field configurations where all the fermion masses are bounded below by two times the gravitino mass \eqref{Vstability}, which requires an atypical fluctuation of the smallest fermion mass,  $m_1 \ge 2$. Although in \cite{Denef:2004cf} it is argued that metastability is a relatively mild constraint, the probability of such fluctuation was calculated in \cite{Bachlechner:2012at}, and it was found to be exponentially suppressed. Indeed, using \eqref{fluctuations} we recover\footnote{To be precise, the authors of  \cite{Bachlechner:2012at} calculated this probability  using the results in \cite{Edelman}, which does not rely on the assumption of  the fluctuations being  small. A more detailed discussion about large fluctuations of the spectrum  can be found in Appendix \ref{RMT}.}:
 \be
 \mathbb{P}(m_1>2) \sim  \rme^{- \frac{8\scN^2}{m_h^2}}.
\label{SUSYsup}
 \ee 
This result implies that when the parameter $m_h$ determining the distribution of  fermion masses satisfies $m_h \lesssim \scN$, supersymmetric minima are very rare,  and then the vacua obtained using standard uplifting mechanisms will typically lead to tachyonic instabilities. For instance, if the number of fields of the supersymmetric sector is of the order of hundreds  $\scN \sim 100$, this regime corresponds to configurations where all the chiral fermions are lighter than about a  hundred times the mass of the gravitino, $m_\lambda \lesssim 100$ in our units.  Nevertheless, as argued in \cite{Bachlechner:2012at}, when the masses of the fermions are typically much larger that the gravitino, $m_h \gg \scN$, the AdS vacua are typically tachyon-free, and thus they are good candidates  to construct stable dS vacua using an uplifting mechanism.\\

This result might lead to confusion when naively applied to the  case of complex structure moduli in large volume compactifications of Type IIB superstrings. In \cite{Balasubramanian:2005zx} it was shown that, under very general circumstances, it is possible to find a non-supersymmetric minimum of the scalar potential capable of stabilising all the moduli of the compactification. In this scenario, as in  KKLT constructions, the dilaton and complex structure  sector can be regarded as an  approximately supersymmetric sector. This class of minima can  be found in compactifications with a large number of  complex structure fields $h^{2,1}$, such as in the $\mathbb{P}^4_{[1,1,1,6,9]}$  model with   $h^{2,1} = 272$, where the  statistical treatment would  be appropriate to study their stability. The analysis that  Conlon \emph{et al.} made of this model  \cite{Conlon:2005ki} shows that, at the minimum, the chiral fermions in the complex structure sector  have a mass of the same order as the mass of the gravitino, $m_\lambda \sim 1$. Therefore, the result (\ref{SUSYsup}) would seem to indicate  that such a minimum for the supersymmetric moduli sector should be extremely difficult to find, in clear contradiction with arguments given in \cite{Balasubramanian:2005zx,Conlon:2005ki}. This apparent discrepancy is due to the fact that in Large Volume Scenarios there is no large mass hierarchy between the supersymmetric moduli and the supersymmetry breaking scale, and in consequence, it is not possible to study the supersymmetric sector  neglecting the supersymmetry breaking effects.  In the next  section we will show that, in the absence of this hierarchical structure, the couplings  between the complex structure and K\"ahler sectors  can stabilise the BF-allowed tachyons which appear when the dilaton and complex structure are considered in isolation.  The possibility of such an effect was  discussed in detail in the context of $F-$term uplifting mechanisms consistent with the supersymmetric truncation of the supersymmetric sector,  \cite{Achucarro:2007qa,Achucarro:2008fk}.

%%%%%%%%%%%%%%%%%%%%%%%%%%%%%%%%%%%%%
%%%%%%%%%%%%%%%%%%%%%%%%%%%%%%%%%%%%

\section{Stability of non-supersymmetric configurations}
\label{sec:afteruplifting}

%%%%%%%%%%%%%%%%%%%%%%
%%%%%%%%%%%%%%%%%%%%%%%%%%%%

In the present section we study the stability of non-supersymmetric configurations des\-cri\-bing either vacua of the scalar potential with a realistic cosmological constant, or points of an inflationary trajectory. For this purpose we will make use of the random matrix theory techniques described in section \ref{randomSugra}.  We will consider  supergravity models   where the couplings are consistent with the supersymmetric truncation  of a sector of the theory with a large number of fields $H^\alpha$, where $\alpha=1, \ldots, \scN_h$, and we will concentrate on the stability analysis of the truncated sector (see section \ref{consistentTruncations}). We will discuss the conditions under which  the supersymmetric sector remains tachyon-free with order one probability, $\mathbb{P}_\text{stable}\sim \cO(1)$, with no exponential suppression. These conditions will appear as restrictions on the parameter fixing the  mass scale of this sector $m_h$, the statistical properties of the derivatives of the fermions masses, and on the geometry of the K\"ahler   manifold. In addition, we will derive the typical mass spectrum of the supersymmetric sector in those situations where the parameters $\mu_{\lambda}^2$ in eq. \eqref{constraints} can be identified with the squared-masses of the truncated scalar fields. This is the case for supergravity models where 
the K\"ahler function is separable in the truncated and surviving sectors, $G = G_{h}(H,\bar H) + G_{l}(L,\bar L)$, as it happens for instance in certain no-scale models, and in models which are small deformations of this class of structure, similar to the situation in Large Volume Scenarios  \cite{Gallego:2011jm}.

%
%

%%%%%%%%%%%%%%%%%%%%%%%%%%%%%%%

\subsection{Stability of non-supersymmetric Minkowski vacua}
\label{section:upliftedVacua}

%%%%%%%%%%%%%%%%%%%%%%%%%

We begin studying the stability of   field configurations $(H_0^\alpha,L_0^i)$ representing vacua of the scalar potential appropriate to describe the present value of the  cosmological constant. 
The expectation value of the scalar potential in these vacua needs to be  extremely small, $V\sim 10^{-120}$, and even if supersymmetry is unbroken at a relatively low energy scale, e.g. $m_{3/2}\sim 10^3$  TeV, the constant $\gamma$ which determines the ratio of the Hubble parameter  to the gravitino mass has a  value of order $\gamma  \lesssim 10^{-90}$. Therefore, for the present purpose it will be sufficient to concentrate on the stability of non-supersymmetric Minkowski vacua, $\gamma\approx0$. In this case  the Hessian of the scalar potential along the decoupled supersymmetric sector \eqref{HessianDecompSusy} reads
\be
\cH_h =\cG^2_h
+\sqrt{3} \, D_X \cM_h  - 3\cR_h,
\vspace{.1cm}
\label{HessianDecomp3}
\ee
where $\cG_h=\unity + \cM_h$ is the Hessian of the K\"ahler function along the truncated sector.\\

 In general, the parameters $\mu_{\pm\lambda}^2$  along the supersymmetric directions can not be identified with the squared-masses  of the truncated scalar fields,  and therefore the conditions \eqref{constraints} are not sufficient to  guarantee the  stability of the supersymmetric sector. However, since they are still necessary,  they can provide useful information about the couplings between the   supersymmetric and non-supersymmetric sectors. In the case of non-supersymmetric Minkowski vacua, the conditions \eqref{constraints} and \eqref{mildConstraint} reduce to
\be
 \mu_{\pm \lambda}^2 = (m_\lambda\pm 1)^2 \pm\sqrt{3}\frac{\pd m_\lambda}{\pd X} + 3 B[X,\lambda] \ge 0, \quad \qquad B[X,\lambda] \ge - \frac{ m_\lambda^2 +1}{3}\ .
\label{M4constraints}
\ee
Let us first consider the second set of constraints, which only involve  the  bisectional curvatures and the fermion masses. {\it A priori} $B[X,\lambda]$ can take any real value, but  if the K\"ahler manifold is \emph{regular} at $H_0$, it will always be bounded by two finite constants $B_\text{min}$ and $B_\text{max}$ such that $B[X,\lambda] \in\[B_\text{min}, B_\text{max} \]$  for all  $\lambda = 1, \ldots , \scN_h$ and a fixed choice of the sGoldstino direction.\footnote{The existence of the constants $B_\text{min}$ and $B_\text{max}$ is granted, since the bisectional curvatures are always bounded by the lowest and largest eigenvalues of a regular hermitian matrix $B=B^\dag$ with components $B_{\alpha \bar \alpha} \equiv -R_{X \bar X \alpha \bar \alpha }$.}
For the vacuum to be tachyon-free with order one probability, the bound \eqref{M4constraints} on the bisectional curvatures must be satisfied for all possible values of the fermion masses in a typical spectrum, and in particular for the smallest one, $m_1$,  which has zero expectation value to leading order in $1/\scN_h$, cf. eq. \eqref{limitingEig}.
In that case the range of bisectional curvatures is constrained by the following bound:
 \be
B_\text{max}\geq-\tfrac13. 
 \ee
As discussed in section \ref{randomSugra}, stable field configurations where these conditions are not satisfied are also possible, provided that the smallest fermion mass is non-zero, but they occur  with an exponentially suppressed probability, see eq. \eqref{fluctuations}. This suppression can be made mild if there is a large  hierarchy between the  mass scale of the supersymmetric sector and the gravitino mass, $m_h\gg\scN\gg1$, as in KKLT type of constructions. However, in the present work are mostly interested in  situations where this fine-tuning is not present. \\

 From the first set of constraints in \eqref{M4constraints} it is possible to derive stronger bounds on the bisectional curvatures, but they depend in general on the details of the fermionic spectrum of masses  and their derivatives. Nevertheless, in a typical critical point the range of derivatives of fermion masses is also bounded as in \eqref{boundDerivatives}. Noting that the strongest constraint is obtained when one of the fermions has the same mass as the gravitino, $m_\lambda=1$, it is easy to check that when all the bisectional curvatures have sufficiently large positive values
 \be
 B_\text{min} \ge \frac{1}{\sqrt{3}}\scS,
\label{sufficientCondition}
\ee 
all  the necessary  stability conditions  \eqref{constraints}  are satisfied by typical vacua. Any other situation  requires a case by case analysis. Let us now consider two classes of models consistent with the supersymmetric truncation of a heavy sector:  the case of a separable K\"ahler function, which includes no-scale models, and small deformations of this type of structure, similar to the case of supergravity Lagrangians in Large Volume Scenarios of type-IIB compactifications.

\subsubsection{Separable K\"ahler function}
%%%%%%%%%%%%%%%%%%%%%%%%%
\label{sec:M4Separable}

The simplest class of theories consistent with the exact truncation of a supersymmetric sector fixed at a configuration $H_0^\alpha$ is characterised by  separable K\"ahler functions of the form:
\be
G (H ,\bar H , L, \bar L ) = G_h (H ,\bar H )+G_l ( L, \bar L ), \qquad \text{with} \qquad \partial_\alpha G_h|_{H_0}=0.
\label{separableG1}
\ee 
The properties of supergravity models characterised by this class of K\"ahler functions, which were first discussed in \cite{Binetruy:2004hh},   have been analysed in detail in  \cite{Achucarro:2007qa,Achucarro:2008sy,Achucarro:2008fk}. In this section we consider the case where  the light field configuration $L_0^i$ satisfies 
\be
G_l^{i \bar j} G_{l|i} G_{l|\bar j} \, \big|_{L_0}=3,
\label{M4condition}
\ee
so that the point in field space $(H_0^\alpha, L_0^i)$ has a vanishing vacuum energy, $\gamma=0$.
 As was shown  in  \cite{Gallego:2011jm,Gallego:2013qe},  the  effective Lagrangian of  Type-IIB flux compactifications, to leading order in loop and non-perturbative corrections, is also described by the ansatz  above,  where  the $H^\alpha$ fields would  correspond to the dilaton and complex structure moduli, and the  $L^i$ fields would be identified with the K\"ahler moduli. More generally, the ansatz \eqref{separableG1} includes a large class of no-scale models, where the sector surviving the truncation satisfies  the condition  \eqref{M4condition}  regardless of the configuration of $L^i$ \cite{Freedman:2012zz}.  We now  begin recalling some of the properties of this type of models defined by \eqref{separableG1}.\\

  It is straightforward to check that the supergravity Lagrangians characterised by se\-pa\-ra\-ble K\"ahler functions satisfy the condition \eqref{conditionTruncation}, which allows the consistent truncation of the $H^{\alpha}$ fields. In particular,  the K\"ahler manifold has a cross-product structure in the truncated and surviving sectors, $\cK=\cK_h\otimes\cK_l$, and therefore the reduced manifold $\cK_l$ is clearly a totally geodesic submanifold. With this type of couplings the stability analysis of the truncated sector is particularly simple. Since the condition \eqref{separableG1} ensures that the sGoldstino direction lies along the reduced manifold $\cK_l$, the cross-product structure of the K\"ahler manifold implies that the contribution $\cR_h$ to the Hessian along the supersymmetric sector \eqref{HessianDecomp3} vanishes identically, $\cR_h=0$. Moreover, it can be checked that the fermion mass matrix of the truncated sector is independent of the  configuration of the surviving fields $L^i$, and in particular of the sGoldstino, implying that the contribution $D_X \cM_h$ in \eqref{HessianDecomp3} is also zero.  As a consequence, the block of the Hessian in the truncated directions has the simple structure discussed in section \ref{subsec:nonsusy1}, and the corresponding stability analysis applies here as well.  In the case of non-supersymmetric Minkowski vacua ($\gamma=0$) the block of the Hessian corresponding to the supersymmetric sector has a very simple expression:
\be
\cH_{h} =\cG^2_h \qquad \Longrightarrow \qquad \mu_{\pm \lambda}^2 = (m_\lambda \pm 1)^2.
\vspace{.1cm}
\label{HessianDecomp3b}
\ee
The first important consequence is that the block $\cH_h$ is diagonalised in the same basis as the fermions mass matrix $\cM_h$, implying that  the parameters $\mu_{\pm \lambda}^2$  are exactly the  squared-masses of the scalar fields of the supersymmetric sector, and the conditions \eqref{M4constraints} are both necessary and sufficient to ensure the stability of the truncated fields. Thus, from equation \eqref{HessianDecomp3b} it is clear that the supersymmetric sector is always metastable (see also fig. \ref{Fig:index2}), possibly with zero-modes whenever the Hessian of the K\"ahler function  $\cG_h$ has a zero eigenvalue, i.e. for each fermionic  mass satisfying, $m_\lambda=1$ \cite{Achucarro:2007qa,Achucarro:2008fk}.\\

Note that, in contrast to KKLT type of models, in general there is no large  hierarchy  between the mass scale of the supersymmetric sector and the supersymmetry breaking effects. Moreover, it is no longer necessary that the truncated sector is stabilised at a supersymmetric AdS minimum when is considered in isolation, that is, in the supersymmetric limit $\gamma\to -1$. Indeed, while supersymmetric AdS minima correspond to field configurations where all the fermion masses are greater than twice the gravitino mass $m_\lambda\ge 2$,  we see from \eqref{HessianDecomp3b} that the mass parameters are always positive regardless of the fermionic spectrum of the truncated sector $\mu_{\pm \lambda}^2\ge 0$. Therefore, this class of models constitute an explicit example where stabilising the supersymmetric sector at a supersymmetric AdS minimum is in general more restrictive than requiring the final non-supersymmetric vacuum to be stable.\\

\begin{figure}[t]
\vspace{-1cm}
 \centering \includegraphics[width=0.45\textwidth]{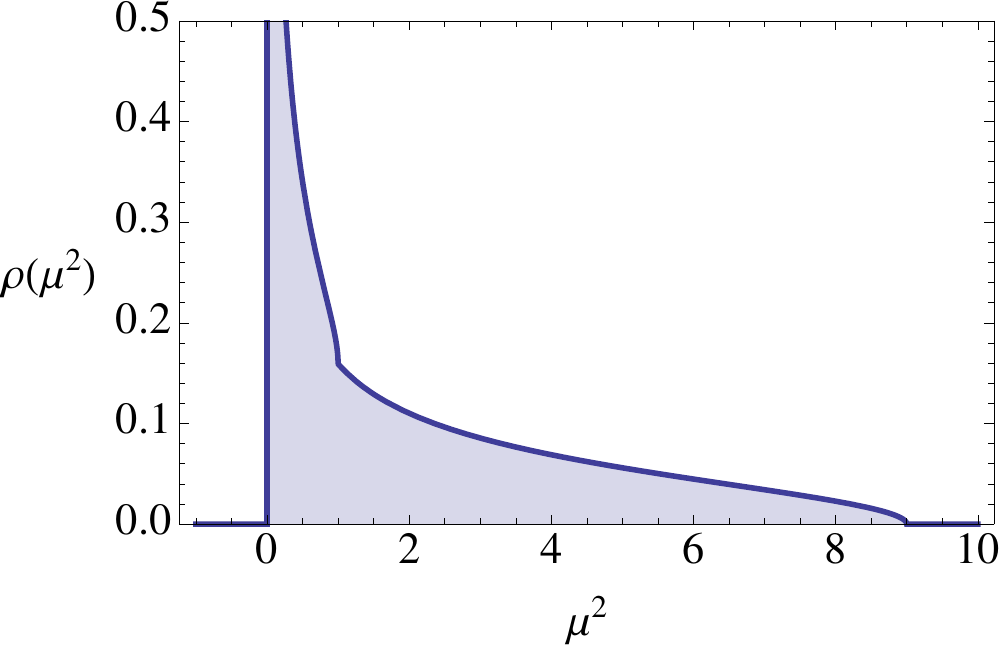}
\hspace{1cm}
\centering \includegraphics[width=0.45\textwidth]{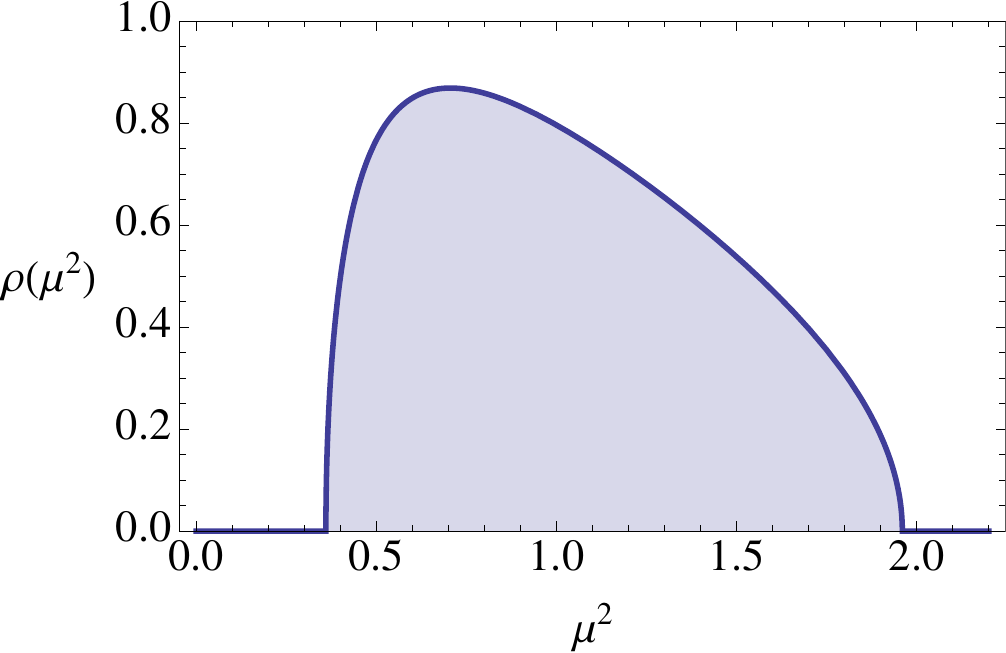}
    \caption{Spectral density of the Hessian of the supersymmetric sector $\cH_h$, eq. \eqref{totaldensityM4}, at a Minkowski configuration $(H_0^\alpha,L_0^i)$ when the K\"ahler function has the separable  form \eqref{separableG1}. In this case the  spectrum is  always tachyon-free. LEFT: When  the mass scale of the supersymmetric sector is $m_h>1$ the spectral density  diverges as $\rho(\mu^2) \sim 1/\mu$ near $\mu=0$. The plot shows the case  $m_h=2$. RIGHT: When the mass scale is  $m_h<1$ the stability of the configuration is protected by a gap in the mass spectrum of size $\mu_\text{min}^2 = (1- m_h)^2$. In the plot $m_h=0.4$. }
 \label{fig:sepM4}
\end{figure}

Proceeding as in the case of supersymmetric critical points, it is possible to calculate the typical spectrum of masses for the scalar fields in a Minkowski background. First, by inverting the expression in  \eref{HessianDecomp3b} for the squared-masses of the scalar fields, we can write the fermion masses $m^2$ in terms of the eigenvalues of the Hessian $\mu^2$.  As a result, we find a multivalued function with two branches which we denote by $m_{\pm}^2(\mu)$:
\be
m_{\pm}^2(\mu) = \(1\pm \mu\,\)^2\ , \qquad \qquad  \Bigg|\frac{d\,  m_{\pm}^2}{ d \, \mu^2}\Bigg| = \frac{ |1\pm \mu\,|}{ \mu}\ .
  \label{spectrum3}
\ee 
Using these expressions in combination with equation \eqref{totaldensity} we can derive the eigenvalue density function of the scalar masses from the Mar\v{c}enko-Pastur law \eref{SCMP},  which determines the spectral density of the fermion masses. The resulting expression, accurate  to leading order in $1/\scN_h$, depends on a single parameter $m_h$ associated to  the mass scale of the truncated sector:
\bea
\rho(\mu^2) d\mu^2 =\frac{2 \, \scN_h}{\pi  m_h^2 \, \mu} \Big[&&\Theta \left(m_h^2-(1+ \mu)^2\right) \sqrt{m_h^2 - (1+ \mu)^2}\nonumber\\
+&&\Theta \left(m_h^2-(1- \mu)^2\right) \sqrt{m_h^2 -  (1- \mu)^2} \;\Big]d\mu^2 \ .\label{totaldensityM4}
\eea
  In  Fig. \ref{fig:sepM4} we  display two  examples of the spectral density of the square-masses of the scalar fields for  different values of $m_h$. \\

 When the mass scale of the supersymmetric sector is larger than the gravitino mass, in our units $m_h > 1$, the spectral density of a typical critical point has a  shape which resembles the Mar\v{c}enko-Pastur law, with all the scalar fields having  masses lying  in the range $\mu\in [0 , 1 +m_h]$ (see left plot in Fig. \ref{fig:sepM4}). Note that in this regime the scalar mass density function diverges as $\rho(\mu^2) \sim 1/\mu$ near $\mu=0$, which indicates that a significant fraction of the scalar fields may  have  a mass  much lighter than the gravitino.  As we shall see in the next subsections, this set of light scalar fields in the truncated sector are susceptible to becoming tachyonic under small deformations of this class of models.   \\

%%%%%%%%%%%%%%%%%%%%%%%%%%%%%%%%%%%%
%%%%%%%%%%%%%%%%%%%%%%%%%%%%%%%%%%%

In the regime where $m_h<1$,  all the masses are  contained in the range $\mu \in [1-m_h, 1+m_h]$   centred on the gravitino mass,  (right plot of  Fig. \ref{fig:sepM4}), in other words: since the expectation value of the  lightest scalar field has a finite positive mass $\mu_{min}=1 - m_h>0$ the mass spectrum develops a  gap. Note that in this range of the parameter $m_h$, the spectral density of scalar masses resembles the Wigner's semicircle law, and when the typical mass scale of the heavy sector is small $m_h\to 0$, all the scalar fields have masses close to the gravitino mass. This is consistent with the mass sum rules in supergravity theories \cite{Freedman:2012zz}, which indicate that the gravitino mass sets the average mass splittings between fermions and bosons within a chiral multiplet. Therefore, the scalar fields can have  masses of the order of the gravitino mass, while the fermions masses remain low, i.e. $m_\lambda\le m_h \ll 1$. \\

 The fact that in the regime $m_h<1$ the expectation value of the mass of the lightest scalar field has a   finite positive value, does not prevent the occurrence of atypical minima where one or more scalars have an arbitrarily small mass, but the probability of finding such spectra is exponentially suppressed. From the expression \eqref{HessianDecomp3b}, which relates the fermion mass spectrum to the masses of the scalar fields, it is easy to see that in the regime $m_h<1$, the smallest value of the squared-mass parameters is associated to the fermion with the largest mass $m_{\scN_h}$:
 \be
\mu_\text{min}^2 = (1- m_{\scN_h})^2.
\label{massGap}
 \ee
Then, if the lightest scalar field has a mass smaller than a fraction $\alpha$ of the gravitino mass, i.e. $\mu_\text{min} < \alpha<1$, the heaviest fermion must have a mass satisfying $m_{\scN_h}>1-\alpha$.  From the Tracy-Widom distribution for small fluctuations of the largest eigenvalue of $\cM_h$ (see appendix \ref{RMT}), it is easy to check that the probability of a fermion mass spectrum with that property is exponentially  suppressed by the number of heavy fields for small values of the mass parameter  $m_h$, that is:  
\be
\text{If} \ \ m_h  < 1-\alpha: \qquad   \mathbb{P}(\mu_\text{min} < \alpha) \sim \rme^{-\ft{4}{3}\scN_h x^{\frac{3}{2}}  } \ \   \qquad \text{with}\qquad  x = \ft{(1-\alpha)^2}{m_h^2} -1 .
\label{PlightFields}
\ee
The fact that the probability of finding light scalar fields in the supersymmetric sector is exponentially suppressed already indicates that the stability of this branch of vacua is particularly robust against deformations of this type of models, and we will show that this is the case in the next subsections. It is also interesting to note that these vacua are also easy to identify since, as was discussed in section \ref{sec:beforeuplifting}, those configurations where all the fermions are lighter than the gravitino, $m_\lambda <1$, are in one-to-one correspondence with the minima of the K\"ahler function $G$ along the truncated sector, which is much simpler to analyse than the scalar potential.

%%%%%%%%%%%%%%%%%%%%%%%%%%%%%% 
\subsubsection{Quasi-separable K\"ahler function}
%%%%%%%%%%%%%%%%%%%%%%%%
\label{sec:Qseparable}

In the present subsection we will discuss the stability properties  of the supersymmetric sector  in  models where the K\"ahler function has an \emph{almost} separable structure of the form  
\be
G(H,\bar H, L, \bar L)=G_{h}(H,\bar H)+G_{l}( L, \bar L) +\epsilon\, G_{mix}(H,\bar H, L, \bar L)\ ,
\label{quasiSeparableG}
 \ee
where $\epsilon$ is a small parameter. This type of structure can be found in  the effective Lagrangians describing the  Large Volume Scenarios of type-IIB flux  compactifications, where the form of the K\"ahler function is a small deformation of the ansatz  \eqref{separableG1}, and the size of the correction is suppressed by the inverse volume of the internal space $\epsilon \sim 1/\cV$   \cite{Gallego:2011jm}. The ansatz above can also be of use to study the   class of nearly no-scale models which   have been recently discussed in \cite{Kallosh:2014oja}.  Here, in addition to $\pd_\alpha G_h|_{H_0} = 0$ we also  require that
\be
\pd_\alpha G_{mix}(H_0,\bar H_0,L,\bar L) = 0 \quad \text{for all $\ L^i$}\ ,
\label{Gmix}
\ee  
so that the full K\"ahler function $G$ is consistent with the \emph{exact truncation} of the su\-per\-sym\-me\-tric sector at a configuration $H_0^{\alpha}$ regardless of the value of $\epsilon$,  \eqref{conditionTruncation}.  To be more precise, we will consider models where the parameters satisfy the  following hierarchy: $m_h\sim\cO(1)$ and $|1 -m_h| \gg \epsilon$. Note that, in the context of string compactifications, the contribution $G_{mix}$ induced by $\alpha'$ and non-perturbative corrections will not satisfy in general eq. \eqref{Gmix}. However as discussed in section \ref{consistentTruncations}, by restricting ourselves to this simplified class of models, we have a significant payoff in terms of calculational control. Moreover, as more realistic mo\-dels can be seen as small deformations of those satisfying \eqref{Gmix}, we can still extract valuable information from their study.\\

In general, the block of the Hessian associated to the truncated sector, $\cH_h$, will contain all the terms present in \eqref{HessianDecomp3}, but due to this hierarchical structure the first term $\cG_h^2$, which is the only one appearing in the fully separable case \eqref{HessianDecomp3b}, will be the dominant one and the other contributions will act only as small corrections: 
\be
\cH_h = \cG_h^2 + \delta \cH_h, \qquad \quad  \delta \cH_h \equiv \sqrt{3}\, D_X \cM_h - 3 \cR_h.
\label{pertHessian}
\ee
The correction  $\delta \cH_h$ is the contribution given  by  the non-separable part of  the K\"ahler function $G_{mix}$, and therefore its entries have a size determined by the magnitude of $\epsilon$\footnote{
In large volume scenarios $\rme^{G/2} \sim 1/\cV \sim \epsilon$, and hence $\epsilon$ also sets the mass scale of the gravitino  \cite{Gallego:2011jm}. Here the factor $\epsilon$ for the gravitino mass is absent due to the rescalings   \eqref{reescalings}. Without this rescaling the entries of $\cH_h^{0}$ would be  of order $\epsilon^2$ and those of $\delta \cH_h$ of order $\epsilon^3$ or higher.}.  
In this setting it is possible to  derive an approximate expression for the   eigenvalues of the Hessian on the supersymmetric sector using perturbation theory (see appendix \ref{appendix1} for details).  Since the zero order term $\cG^2_h$ is diagonal in the basis of eigenvectors of $\cM_h$ \eqref{Meigenvectors},  $Z_{\pm \lambda}$ with $\lambda=1,\ldots,\scN_h$,  to first order in perturbation theory we find  $\mu_{\pm\lambda}^2 \approx (\mu_{\pm\lambda}^{2})^{0} +\delta \mu_{\pm\lambda}^2$, where $(\mu_{\pm\lambda}^{2})^{0}$ are the squared-masses  in the separable case \eqref{HessianDecomp3b} and 
\be
  \delta \mu_{\pm\lambda}^2\equiv \la Z_{\pm\lambda}, \delta \cH_h Z_{\pm\lambda}\ra = \pm \sqrt{3} \, \frac{\pd m_\lambda}{\pd X} + 3B[X,\lambda]
\label{massShiftsM4}
\ee
the first order corrections (see Appendix \ref{appendix1}). From this discussion it follows that the pa\-ra\-me\-ters $\mu_{\pm \lambda}^2$ defined in \eqref{constraints} coincide with the first order approximation of the squared-masses of the truncated fields $H^{\alpha}$, implying that the conditions
\eqref{constraints} are both  necessary and sufficient to guarantee the stability of the truncated sector  to  this  level of approximation.
In particular, the stability of the supersymmetric sector is entirely determined by the analysis of section \ref{subsec:nonsusy2} restricted to the case $\gamma=0$.\\

In general the shifts on the mass parameters \eqref{massShiftsM4} induced by the non-separable contribution to the K\"ahler function will have different values for  each eigenvector $Z_{\pm \lambda}$. However, at typical critical points, both the magnitude of  fermion mass derivatives (eq. \eqref{boundDerivatives}) and the bisectional curvatures are bounded (see  beginning of  section \ref{section:upliftedVacua}), and therefore  it follows that   the expectation value of $\delta \mu_{\pm\lambda}^2$ is in the range:
\be
\delta \mu_{\pm\lambda}^2 \in \[ - \sqrt{3} \, \scS + 3B_\text{min} \, , \, \sqrt{3} \, \scS + 3 B_\text{max} \],
\label{shiftsBounds}
\ee
where  the upper and lower bounds are determined by the size of $\epsilon$:   $ \scS \sim |B_\text{min}|\sim |B_\text{max}| \sim \cO(\epsilon)$. Since the zero order mass parameters are non-negative $(\mu_{\pm\lambda}^2)^{0} \ge 0$,  if the bisectional curvatures are sufficiently large to satisfy the  conditions  \eqref{sufficientCondition}, then the  supersymmetric sector is guaranteed to be stable with order one probability, $\mathbb{P} \sim \cO(1)$. Any other situation requires a more  detailed analysis of the stability conditions \eqref{constraints}. \\

%%%%%%%%%%%%%%%%%%%%%%%%%%%%%%%%%%%%
\begin{figure}[t]
\vspace{-1cm}
 \centering \includegraphics[width=0.45\textwidth]{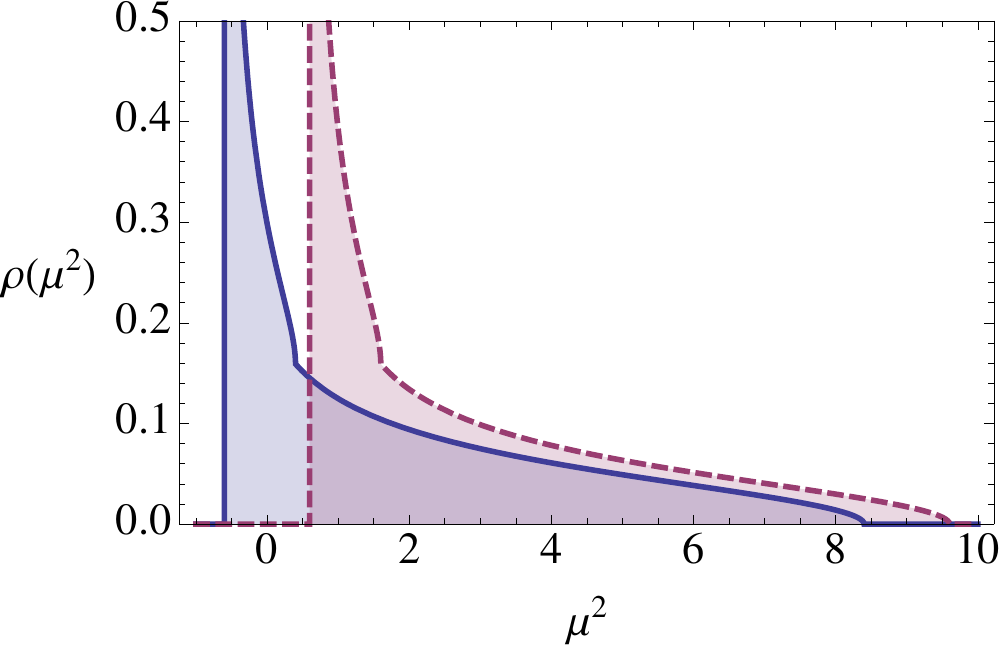}
\hspace{1cm}
\centering \includegraphics[width=0.45\textwidth]{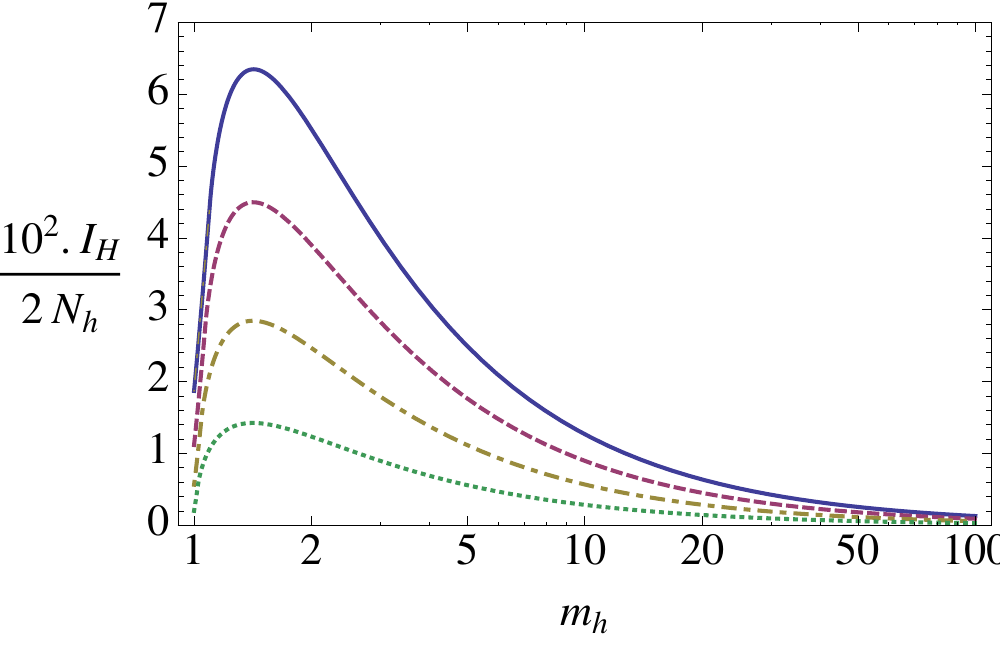}
    \caption{LEFT: Spectral density of the Hessian of the supersymmetric sector $\cH_h$ for an almost separable  K\"ahler function of the form \eqref{quasiSeparableG} at a Minkowski configuration $(H_0^\alpha,L_0^i)$. The parameters are set to  $m_h=2$ and $\pd m_\lambda/\pd X=0$, with  two different constant values of the bisectional curvatures: $B[X,\lambda]=-0.2$ (solid), $B[X,\lambda]=0.2$ (dashed). RIGHT: Lower bound for the index of $\cH_h$ as a function of the mass scale of the supersymmetric sector $m_h$, and for different values of the  shift $\delta \mu^2|_{\text{max}}\equiv \sqrt{3} \scS+3 B_{\text{max}}$. From top to bottom: $\delta \mu^2|_{max} = -0.01,\, -0.005, \, -0.002, \, -0.0005$. In the plot the index has been rescaled to display the percentage of tachyonic fields $100 \times \cI_H/(2 \scN_h)$.}
 \label{fig:QsepM4}
\end{figure}
%%%%%%%%%%%%%%%%%%%%%%%%%%%%%%%%%%%

Let us first consider the typical vacua in the regime $m_h>1$.   As we have seen in the previous subsection, to zero order in perturbation theory there can be a significant fraction of scalar fields in the supersymmetric sector with a mass much lower than the gravitino, which can become   tachyonic after including the non-separable corrections to the K\"ahler potential. This is illustrated by the left plot of Fig. \ref{fig:QsepM4}, where we have displayed the typical spectral density of the scalar masses in the case $m_h = 2$ for two different choices   for the   bisectional curvatures. For simplicity we have set the fermion mass derivatives to zero $\pd m_\lambda/\pd X=0$, and the bisectional curvatures to  be independent of the  direction  in field space, i.e. constant for all choices of    $Z_{\pm \lambda}$, so that the shifts \eqref{massShiftsM4} are also constant.    In the right plot  we see that when these corrections  give a negative contribution, the spectral density shifts towards ne\-ga\-ti\-ve values and it becomes non-zero in a small range where $\mu^2<0$, indicating the presence of tachyons in the spectrum with a small tachyonic mass of order $|\mu^2| \sim \epsilon$.\\

 In order to clarify how large is the fraction of scalar fields which are tachyonic in these circumstances, we will now make a simple estimate. From the density function for the zero-order mass parameters $(\mu_{\pm \lambda}^2)^0$  \eqref{totaldensityM4}, 
 and using that the first order corrections can not be larger than
 \be
  \delta \mu^2|_{\text{max}}\equiv \sqrt{3} \scS+3 B_{\text{max}}
  \ee
 (see eq. \eqref{shiftsBounds}), we can  calculate a lower bound on the fraction of tachyons for negative values of this shift. The  right plot of Fig. \ref{fig:QsepM4}  shows the result  for different values of the shifts $\delta \mu^2|_{\text{max}}$, and as a function of $m_h$. Note that the fraction of tachyonic fields in the supersymmetric sector peaks around $m_h =\sqrt{2}$. At this value of $m_h$ the spectral density of fermion masses \eqref{SCMP} evaluated at $m=1$ acquires its largest value, indicating that, at zero order, the fraction of light scalar fields should also be large according to  \eqref{HessianDecomp3b}. In the limit  $m_h\to \infty$, where there is a large hierarchy between the mass scale of the supersymmetric sector and the supersymmetry breaking effects,  the fraction of tachyonic  fields approaches to zero, as would be  expected in a KKLT type of construction where this fine-tuning is present. Indeed, in that case the supersymmetric sector is stabilised  at a supersymmetric minimum of the scalar potential and thus, since  all fermion masses must  satisfy $m_\lambda \ge2$,   from  \eqref{HessianDecomp3b} it follows that the masses of the scalar fields must be bounded below by the gravitino mass $\mu_\text{min} \ge 1$.\\

  The right plot of Fig. \ref{fig:QsepM4} also shows that the KKLT mechanism to stabilise the supersymmetric sector is clearly different from the one in Large Volume Scenarios, since  the latter case corresponds to  the limit $\epsilon\sim1/\cV \to 0$,  and the mass parameter is of order one,  $m_h \sim 1$\footnote{In large Volume Scenarios all the fermion masses in the truncated sector are comparable to the gravitino mass \cite{Conlon:2005ki}, and thus $m_h \sim m_\lambda \sim 1$. The parameter $\epsilon$, which is determined by the volume of the internal space, is $\epsilon \sim 1/\cV\sim 10^{-10}$ in the $\mathbb{P}_{[1,1,1,6,9]}^4$ model \cite{Balasubramanian:2005zx,Gallego:2011jm}. }. The effect of taking this limit is illustrated in the plot  by calculating the lower bound on the number of tachyons for different  values of $\delta \mu^2|_{\text{max}} \sim \cO(\epsilon)$ decreasing in magnitude (from top to bottom in Fig. \ref{fig:QsepM4}). In particular, in the limit $\epsilon\to 0$ we expect the supersymmetric sector to be completely stable regardless of the value of $m_h$, since this corresponds to the separable case discussed above. In consequence,  if we were to consider the complex structure moduli and dilaton in isolation in a Large Volume Scenario, that is, without considering the interactions with the K\"ahler moduli or supersymmetry breaking ($\gamma \to -1$), we would find that the configuration of the truncated sector  corresponds to any type of supersymmetric critical point, not necessarily a minimum  as it is required in KKLT type of constructions.  This result has already been confirmed using numerical techniques in   \cite{BP}, where the authors  study the  statistical properties of vacua in  type-IIB string theory compactified on  $\mathbb{P}_{[1,1,1,6,9]}^4$.  \\
 
Interestingly, as the mass scale of the supersymmetric sector approaches the gravitino mass from above $m_h \to 1$, the lower bound on the number of tachyons always drops to zero. This is related to the fact  that,  in the regime $m_h<1$, the supersymmetric sector is typically stabilised at a minimum of the K\"ahler function $G$, and  to zero order in $\epsilon$ the mass density function of these configurations has a finite gap around $\mu=0$.  Thus, as long as the non-separable corrections are sufficiently small, $\epsilon \ll |1-m_h|$,  the shifts to the zero order masses \eqref{massShiftsM4} will not be able to produce any tachyon. There could still be atypical vacua where such tachyons are present, however the appearance of  such tachyons requires  some  scalar fields $H^{\alpha}$ to be much lighter than the gravitino to zero order in $\epsilon$, and as we have argued in the previous section, the probability of having light fields in the truncated sector is exponentially suppressed  when $m_h<1$.  Therefore, in models where the parameters satisfy the hierarchy $m_h<1$ and $1-m_h\gg \epsilon$,  Minkowski critical points are expected to be stable with order one probability $\mathbb{P}\sim \cO(1)$  after including the non-separable corrections to the K\"ahler potential. The robust stability properties of the minima of the K\"ahler function can be of interest to explore new stabilisation mechanisms. For example in cases where the supersymmetric sector is not protected by a large mass hierarchy,  as in KKLT constructions, or where the   non-separable corrections in the  K\"ahler function \eqref{quasiSeparableG} are small but not exponentially suppressed, as  in Large Volume Scenarios. \\
  
Finally, as we mentioned above, the contributions to the K\"ahler function from $\alpha'$, string loop, and non-perturbative corrections in KKLT constructions and LVS will not satisfy \eqref{Gmix} in general. The main consequence of this is that the expectation values of the truncated fields, the complex structure moduli in this case, will shift when the supersymmetry breaking effects are taken into account. Consequently, the Hessian $\cH_h$  will receive an extra $\cO(\epsilon)$ correction not included in \eqref{pertHessian} (see e.g. \cite{Rummel:2011cd}). However, since these additional corrections are of order $\epsilon$, using the same reasoning as in the previous paragraph, it can be argued that  in these models the truncated sector will also remain metastable with  $\mathbb{P}\sim \cO(1)$  in the regime where $m_h<1$  and $1-m_h\gg \epsilon$.

\subsection{Stability of inflationary trajectories}
%%%%%%%%%%%%%%%%%%%%%%%%%

In this section we  will consider the stability analysis of the truncated sector in the case when the vacuum energy of the field configuration is driving a phase of slow-roll inflation, and thus  $\gamma\simeq H^2/m_{3/2}^2 >0$.  The size of $\gamma$ depends on the particular inflationary model under consideration. For instance for inflationary models based on the standard KKLT construction, the stability of the volume modulus requires the Hubble parameter to be at most of order of the gravitino mass, $\gamma \approx 1$ \cite{Kallosh:2004yh}, however there are modifications of this framework which allow for values of the Hubble parameter much larger than the gravitino mass  $\gamma \gg1$ \cite{BlancoPillado:2005fn}.\\

As in the previous section we will assume that the K\"ahler function satisfies the condition (\ref{conditionTruncation}), and we  restrict ourselves to situations where the $H^{\alpha}$  fields act as spectators, they do not participate neither in  supersymmetry breaking or inflation. In particular, the configuration $H_0^\alpha$ is still an extremum of the scalar potential $V_\alpha|_{H_0}=0$, and the inflaton direction belongs to the supersymmetry breaking sector.  In this setting the full configuration $(H_0^\alpha,L^i)$ represents a point of the inflationary trajectory, and the  truncated sector must be  tachyon-free for this trajectory to be stable. The Hessian of the scalar potential along the supersymmetric sector reads
\be
\cH_h =\cG_h \big(\cG_h + 3 \gamma\,  \unity\big)
+\sqrt{3(\gamma+1)} \, D_X \cM_h  - 3(\gamma+1) \cR_h.
\vspace{.1cm}
\label{HessianDecomp5}
\ee
As we also argued in the case of Minkowski vacua, in general $\cH_h$ will not be diagonal in the basis of eigenvectors of $\cM_h$. Thus, the parameters $\mu_{\pm\lambda}^2$ in \eqref{constraints} (the diagonal elements of  $\cH_h$ in this basis) can no longer be identified with the squared-masses of the fields in the supersymmetric sector. But still, the necessary conditions \eqref{constraints} can be used to constrain the type of couplings between the truncated and the supersymmetry-breaking sectors.\\

We begin considering the bound on the bisectional curvatures \eqref{mildConstraint} which  must be sa\-tis\-fied for all the fermion masses in the truncated sector at the configuration $H_0^{\alpha}$. At  a typical configuration, the expectation value of the smallest fermion mass is zero, cf. eq. \eqref{limitingEig}, and thus,  for the configuration to be tachyon-free, the largest bisectional curvature $B[X,\lambda]$ at $H_0^{\alpha}$ must satisfy:
\be
B_\text{max}\geq-\frac{3\gamma+1}{3(\gamma+1)}.
\label{boundR3}
\ee
For example,  when the Hubble scale is of the order of the gravitino mass, $H\approx m_{3/2}$ ($\gamma \approx 1$), and for a large inflationary scale $H\gg m_{3/2}$ ($\gamma \gg 1$), the previous bound reduces to: 
\be
H \approx m_{3/2}: \quad B_\text{max}\gtrsim -\ft{2}{3}\ , \qquad  \qquad H\gg m_{3/2}: \quad B_\text{max}\gtrsim-1.
\ee
In addition, if we want to consistently neglect the effect of the supersymmetric sector during inflation, the masses of the truncated fields should remain larger than the Hubble parameter ($\mu_{\pm \lambda}^2 \gtrsim \gamma$), so that primordial fluctuations are sourced solely by the supersymmetry breaking sector.
In this case, these constraints  become slightly tighter, and in particular for  $\gamma\approx 1$ we would have the bound $B_\text{max} \geq-1/2$. As we have also discussed thoroughly in previous sections, there could be  atypical vacua where the lightest fermion mass has a large value comparable to the gravitino mass, but the pro\-ba\-bi\-li\-ty of this type of critical points is exponentially suppressed as in \eqref{fluctuations}, unless there is a large mass hierarchy between the truncated sector and the supersymmetry breaking scale, i.e. $m_h \gg  \scN_h\gg1$.  \\

From the study of the complete set of stability conditions \eqref{constraints} it is possible to derive more restrictive bounds on the bisectional curvatures, 
but this analysis requires a detailed knowledge of the spectrum of fermion mass derivatives $\pd m_\lambda/\pd X$, which depends on the couplings between the truncated and the supersymmetry breaking sectors, and therefore it is model dependent.    
However, we will see that it is still possible to extract some information from them without considering particular models. Recall that at a typical configuration of the truncated sector $H_0^\alpha$, the magnitude of the fermion mass  derivatives are bounded \eqref{boundDerivatives}, and provided the K\"ahler manifold is regular at this point, the bisectional curvatures will take values  in a  range $B[X,\lambda]\in \[ B_\text{min}, B_\text{max}\]$ with finite bounds (see section \ref{section:upliftedVacua}).   This implies that the value that can be attained by the parameters $\mu_{\pm\lambda}^2$  at $H_0^\alpha$ is bounded above by 
\be
 \mu_{\pm\lambda}^2|_\text{max} \equiv  (\mu_{\pm\lambda}^2)^0 + \sqrt{3 (\gamma +1)} \scS + 3 (\gamma +1)B_\text{max}  \; \ge \; 0  \quad \text{for all}  \quad \lambda = 1, \ldots ,\scN_h \ ,
\label{inflationConditions}
\ee
where $(\mu_{\pm \lambda}^2)^0$ are the eigenvalues of the first term in  \eqref{HessianDecomp5},  which only depend on the fermion masses, eq. \eqref{upliftedmasses}.  Clearly, for the configuration of the truncated sector  to be tachyon-free, all the previous expressions \eqref{inflationConditions} should  be  non-negative, since this is the most favourable situation. \\

\begin{figure}[t]
\vspace{-1cm}
 \centering \includegraphics[width=0.4\textwidth]{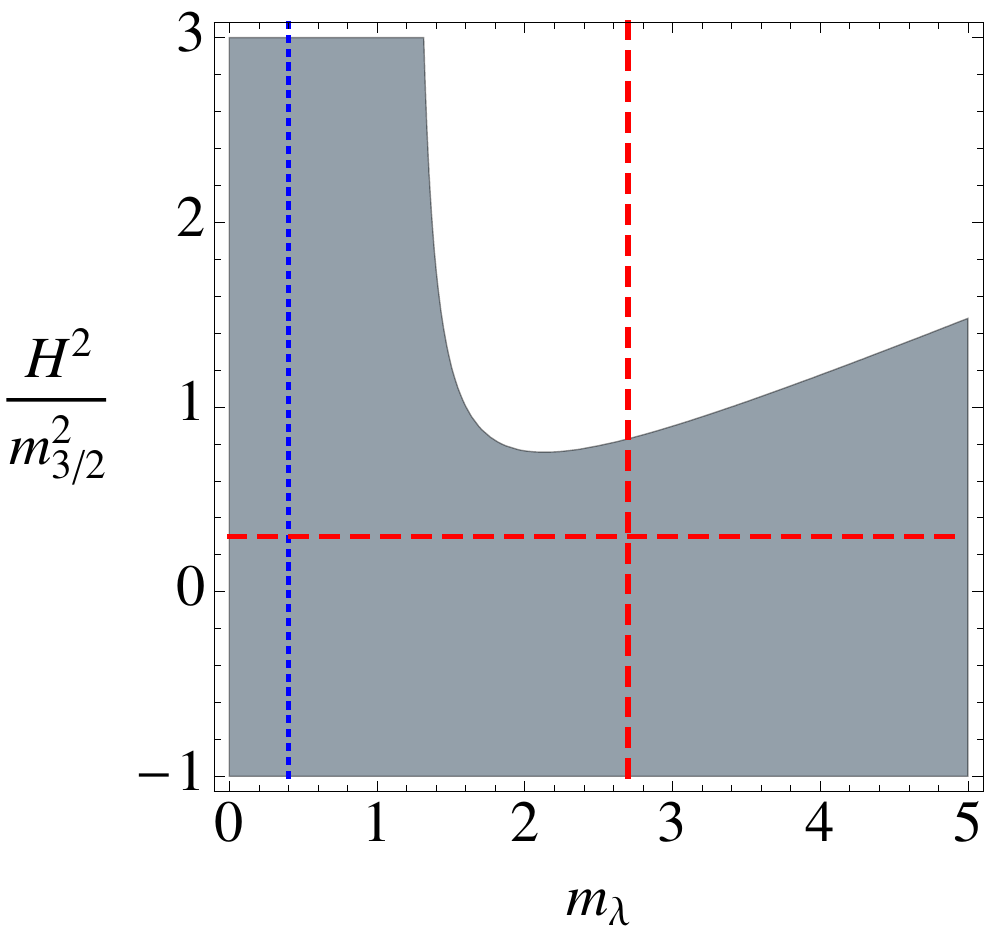}
\hspace{1cm}
\centering \includegraphics[width=0.4\textwidth]{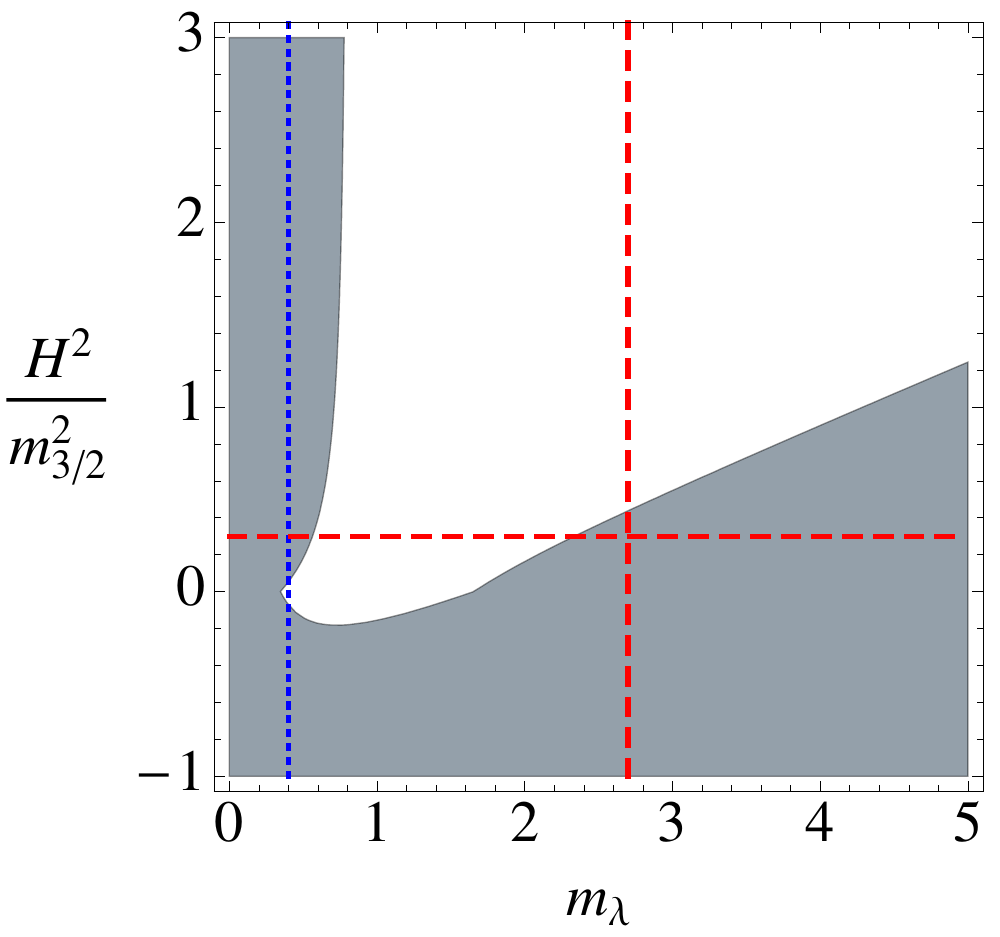}
    \caption{Stability diagram of a field configuration $H_0^\alpha$ of the supersymmetric sector  for a general K\"ahler function satisfying \eqref{conditionTruncation}.   The grey area is the region where the necessary conditions for stability \eqref{inflationConditions} are satisfied. The vertical lines correspond to two different values of the mass scale of the truncated sector with $m_h=0.4$ (blue dotted), $m_h=2.7$ (red dashed), and the horizontal line to  $\gamma=0.3$. In both diagrams    $\scS=0.1$, while the maximum value of the bisectional curvatures is set to $B_\text{max}=0.2$ (left plot) and  $B_\text{max}=-0.2$ (right plot). } 
 \label{fig9}
\end{figure}

In Fig. \ref{fig9} we have plotted the stability diagrams associated to the condition \eqref{inflationConditions} corresponding to some arbitrary pair of eigenvectors $Z_{\pm \lambda}$ of the fermion mass matrix $\cM_h$, and the parameters set to  $\scS =0.1$, $B_\text{max} = 0.2$ (left plot) and  $B_\text{max} = -0.2$ (right plot).
%%%%%%%%%%%%%%%%%%%%%%%%%%%%%%%%%%%%
As we discussed in detail in section \ref{nonsusyconf}, in this  type of diagrams the shaded regions   re\-pre\-sent the values of $m_\lambda$ and $\gamma$  where the stability conditions are met.  Typically the  fermionic  masses in the truncated sector will be distributed  along the interval $m_\lambda \in \[0,m_h\]$ according to the Mar\v{c}enko-Pastur law \eqref{SCMP}. Therefore, given a particular value of $\gamma$, whenever the line $(m_\lambda, \gamma)$ crosses a white region in the interval  $m_\lambda \in \[0,m_h\]$, the scalar spectrum will contain one or more tachyons with order one probability. This is true  even if the point ($m_h$,$\gamma$) falls into the grey region. In the diagrams we have indicated  two points of parameter space $(\gamma,m_h)$, both with $\gamma=0.3$ and mass parameters $m_h=0.4$ (blue dotted) and $m_h=2.7$ (red dashed), and in the right plot ($B_\text{max}=-0.2$)  it can be seen that for the largest choice of $m_h$ the configuration $H_0^\alpha$ will have tachyonic instabilities with order one probability. Note that  when  this analysis signals the presence of instabilities at typical configurations, it is still possible to have  atypical fluctuated fermionic spectra where all the fermionic masses satisfy the conditions \eqref{inflationConditions}, but in  general those configurations have an exponentially suppressed probability.     \\

The plots also show  that  inflationary models with a large Hubble scale $H\gg m_{3/2}$, i.e.  for $\gamma\gg 1$, favour  a small mass value for the mass scale  $m_h$. This is consistent with  the discussion in section \ref{subsec:nonsusy2} where we argued that, for a sufficiently large inflationary scale, the only configurations of the supersymmetric sector which can be stable   during inflation  are those where  the largest fermionic mass (which has expectation value equal to $m_h$ \eqref{limitingEig}) satisfies the upper bound  \eqref{largeGbound} set by the bisectional curvatures. In this regime, those configurations minimising  the K\"ahler function seem to be  promising candidates to stabilise the supersymmetric sector, since all the fermion masses are bounded above by the gravitino mass, $m_\lambda <1$.    \\

 In the next  subsections we will discuss the statistical properties of the mass spectrum of the supersymmetric sector for models where the K\"ahler function has the separable form \eqref{separableG1}, and for small deformations of this type of structure.

\subsubsection{Separable K\"ahler function}
\label{separableinflation}
%%%%%%%%%%%%%%%%%%%%%%%%%%%%

As discussed above, when the K\"ahler function has the separable structure \eqref{separableG1}, the supersymmetric sector is consistently decoupled from the non-supersymmetric sector. In order to have a positive vacuum energy to drive inflation, $\gamma>0$,  after the truncation of the $H^\alpha$ fields, the K\"ahler function of the sector surviving the truncation $G_l(L,\bar L)$ must satisfy: 
\be
G_l^{i \bar j} G_{l|i} G_{l|\bar j} \, \big|_L>3
\ee
at every point of the inflationary trajectory $(H^\alpha_0,L^i)$ \cite{Achucarro:2007qa,Achucarro:2008fk,Achucarro:2012hg}.  Note that this type models can also be seen as a deformation of the no-scale structure which  preserves  the su\-per\-sym\-me\-tric decoupling of the truncated fields. In \cite{Achucarro:2012hg} a class of inflationary models where inflation is driven by the supersymmetry breaking sector was explicitly analysed, and stability conditions for a supersymmetric sector with a single field were considered. Therefore, the stability ana\-ly\-sis we perform in the following completes the analysis in \cite{Achucarro:2012hg} by adding an arbitrarily large number of fields in the supersymmetric sector. Other realisations with a separable K\"ahler function are presented for instance in \cite{Davis:2008sa}, where they include, among others, a KKLT-style moduli sector and study the viability of the resulting inflationary models.\\

%%%%%%%%%%%%%%%%%%%%%%%%%%%%%%%%%%%%%%%
\begin{figure}[t]
\vspace{-1cm}
\centering \includegraphics[width=0.45\textwidth]{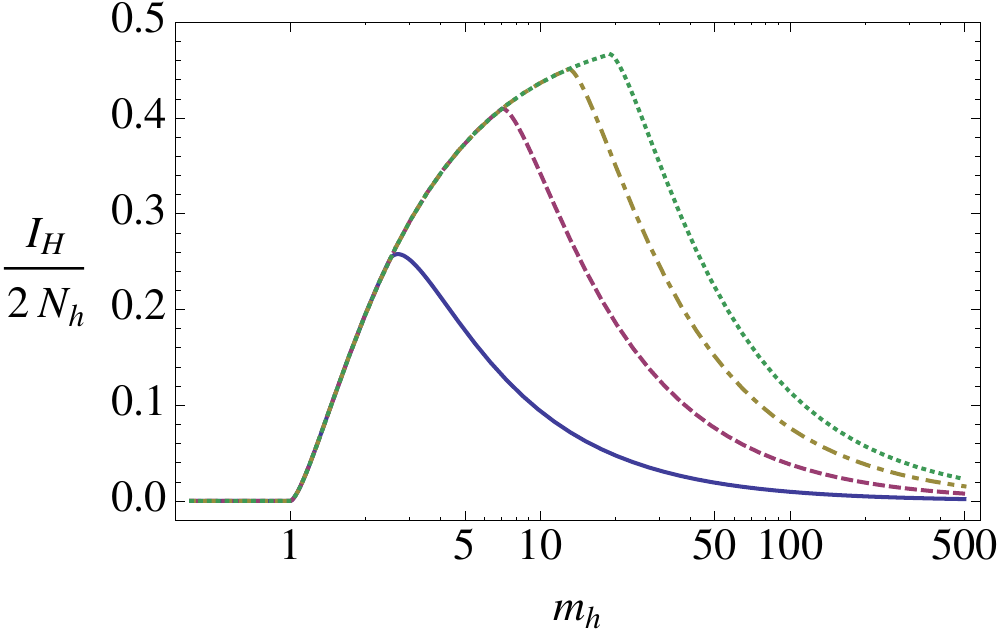}
 \caption{Expectation value of the Morse index of the Hessian of the supersymmetric sector $\cH_h$,  normalised by the total number of eigenvalues, $\cI_\cH/2 \scN_h$. The index is plotted as a function of the parameter $m_h$, and  for different values of the squared ratio between the Hubble parameter and the gravitino mass, $\gamma \equiv H^2/m_{3/2}^2$. From bottom to top $\gamma = 0.5, \, 2,\,  4, \, 6$.}
 \label{Fig:index2}
\end{figure} 
%%%%%%%%%%%%%%%%%%%%%%%%%%%%%%%%%%%%%%%
We have previously seen in section \ref{sec:M4Separable} that for separable K\"ahler functions, the Hessian  of the scalar potential along the supersymmetric sector $\cH_h$ can be expressed only in terms of the Hessian of the K\" ahler function $\cG_h = \unity + \cM_h$, since the contributions $D_X \cM_h$ and  $\cR_h$ are identically zero, \eqref{separableHessian}. For $\gamma>0$  it reads simply:
\be
\cH_{h} =\cG_h \big(\cG_h + 3 \gamma\,  \unity\big).
\vspace{.1cm}
\label{HessianDecomp4}
\ee
Since $\cH_h$ is diagonalised in the basis of eigenvectors of the fermion mass matrix $\cM_h$, the parameters $\mu_{\pm\lambda}^2$, given in \eqref{upliftedmasses}, coincide with the squared-masses of the truncated scalar fields.  Therefore, the conditions \eqref{VupStability}, represented in Fig. \ref{separableDiagram}, guarantee the stability of this sector.\\
  
As in the case of supersymmetric  critical points, in order to characterise the spectral density of  the Hessian at a typical point of the inflationary trajectory $(H_0^\alpha, L^i)$ , we can calculate the Morse index of $\cH_h$, that is, the number of negative eigenvalues of the spectrum at this point.  Using the Wigner's law \eqref{SCMP}, which gives the spectral density of the fermion masses,  in combination with the expression (\ref{upliftedmasses}), relating the fermion masses to the eigenvalues of the Hessian (see Fig. \ref{Fig:index2}), we find:
\begin{equation}
\cI_\cH(m_h)= \left\{ \begin{array}{ll}
\cI_G(m_h) &\qquad  \textrm{if $m_h\le 3 \gamma +1$},\\
\cI_G(m_h) -\cI_G(\ft{m_h}{3 \gamma+1}) &\qquad  \textrm{if $m_h > 3 \gamma +1\ge 1$\ ,}
\end{array} \right. 
\label{indexInflation}
\end{equation}
where $\cI_\cG$ is the  index of $\cG_h$, which is given by the formula \eqref{indexG} with the total number of scalar fields $\scN$ replaced by $\scN_h$. Note that the index of $\cG_h$ is a very good indicator of the stability of the supersymmetric configuration for values of $m_h$ smaller than  $3 \gamma +1$.    Proceeding as in the previous section, we can also calculate the eigenvalue density function for the scalar masses after the uplifting. First, by inverting \eref{upliftedmasses} we can find a multivalued expression for the fermion masses $m^2$ in terms of the eigenvalues of the Hessian $\mu^2$.  The two branches read:
\be
m_{\pm}^2(\mu) = \[ \ft{1}{2}(3 \gamma +2)\pm \sqrt{ \mu^2 +\ft94 \gamma^2}\,\]^2\ , \qquad \Bigg|\frac{d\,  m_{\pm}^2}{ d \, \mu^2}\Bigg| = \frac{| m_{\pm}|}{ \sqrt{ \mu^2 +\ft94\gamma^2}}\ .
  \label{spectrum3}
\ee 
As in the previous section, we use the Mar\v{c}enko-Pastur law \eref{SCMP} and substitute it in \eqref{totaldensity} to find the eigenvalue density function of the scalar masses.\\

%%%%%%%%%%%%%%%%%%%%%%%%%%%%%%%%%%%%
\begin{figure}[t]
\vspace{-1cm}
 \centering \includegraphics[width=0.45\textwidth]{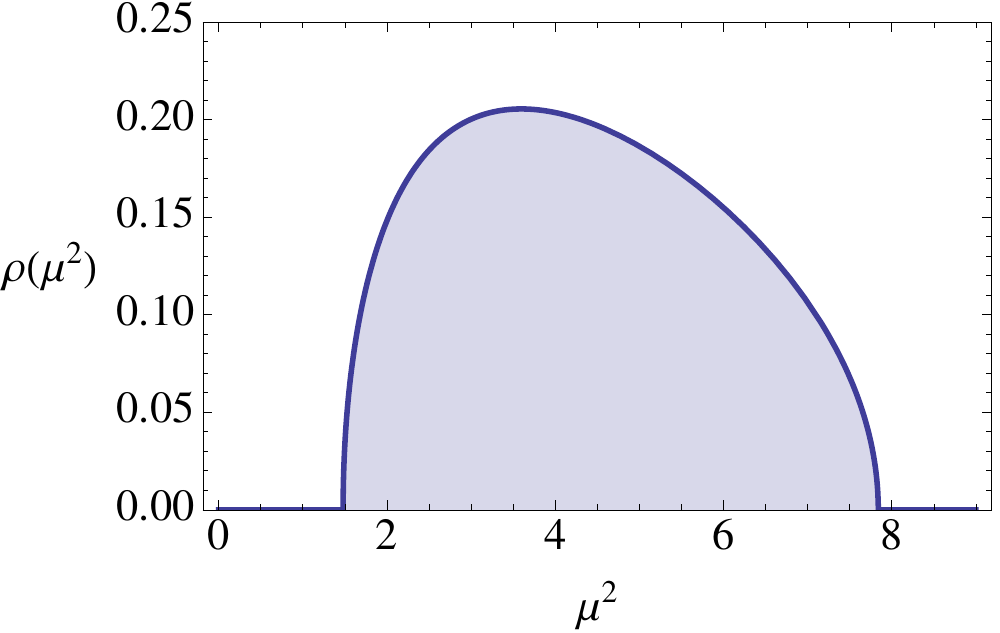}
\hspace{1cm}
\centering \includegraphics[width=0.45\textwidth]{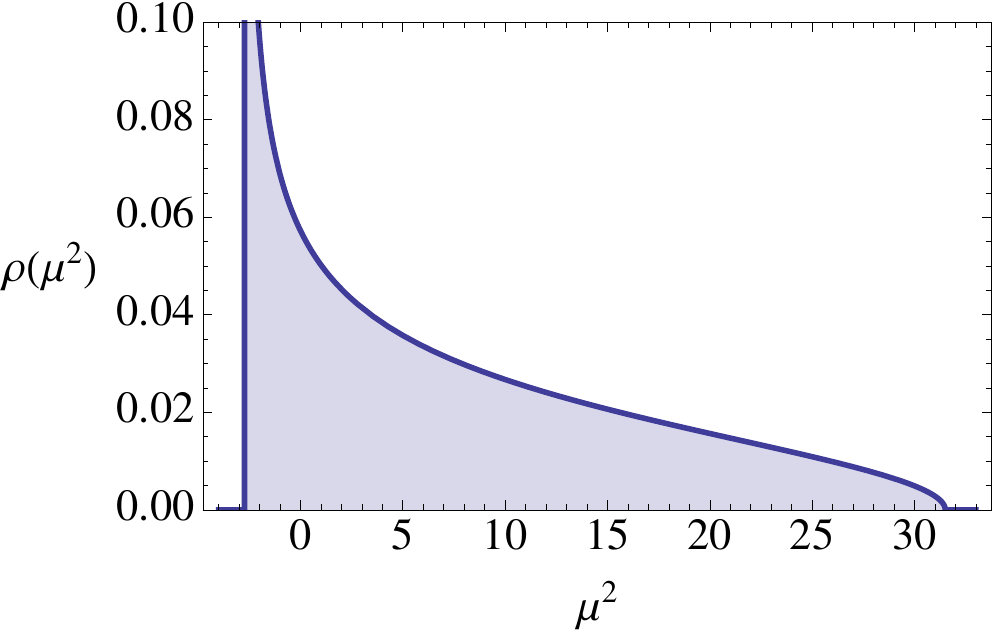}
    \caption{Spectral density of the Hessian of the supersymmetric sector $\cH_h$, when the K\"ahler function has the separable  form \eqref{separableG1}. In the plot the squared ratio of the Hubble scale to the gravitino mass is $\gamma\equiv H^2/m_{3/2}^2=1.1$. LEFT: When the mass scale of the truncated sector is small, i.e. $m_h\le 1$, the typical configuration of the truncated sector contains no tachyons (in the plot $m_h= 0.6$). RIGHT:  When the mass scale $m_h$ is larger than the gravitino mass, $m_h\ge 1$, the typical mass spectrum of the truncated sector  contains tachyons, and  thus $H_0^\alpha$ is unstable (in the plot $m_h = 3.2$).}
 \label{fig3}
\end{figure}
%%%%%%%%%%%%%%%%%%%%%%%%%%%%%%%%%%%

  From the plot of the index of the Hessian $\cI_\cH$ in   Fig. \ref{Fig:index2} it can be seen that, regardless of the value of $\gamma>0$,   the supersymmetric sector only remains tachyon-free with order one probability for $m_h<1$. Recall that in this regime  all the fermion masses are lighter than the gravitino, and thus the configuration $H_0^\alpha$ minimises the K\"ahler function $G$. The typical mass spectrum in this case is displayed in the left plot of fig. \ref{fig3}, which shows that the mass gap   present in  Minkowski vacua also survives for $\gamma>0$. Actually, from equation \eqref{upliftedmasses}, it follows that the mass of the lightest scalar field when $m_h<1$  is bounded below by:
\be
\mu_{\pm  \lambda}^2|_\text{min} \ge (m_h-1)\big(m_h- (3\gamma +1) \big).
\label{massGapInflation}
\ee
This is quite interesting since it implies that the corresponding mass gap becomes wider the higher the 
value of the Hubble parameter,  and moreover,  for $\gamma \gg1$ the scalar fields   in the supersymmetric sector have  large masses which scale as $\mu^2_{\pm \lambda} \sim \gamma$. This analysis provides further support to the claim that the stability of configurations $H^\alpha_0$ which minimise $G$ is particularly robust, and can survive deformations of this class of models. \\ 

In the regime $m_h>1$ the Morse index $\cI_H$ is non-zero for all $\gamma>0$  (Fig. \ref{Fig:index2}), and thus the configuration $H_0^\alpha$ always presents tachyonic instabilities with order one probability. An example of the typical   spectrum of  masses  $\mu_{\pm \lambda}^2$ is displayed  in the right plot in fig. \ref{fig3}.   In this regime there is also an exponentially suppressed fraction of configurations $H_0^\alpha$ where the truncated sector remains stable. Indeed, it follows from \eqref{VupStability} that  atypical configurations where either the heaviest fermion is lighter than the gravitino, $m_{\scN_h}<1$, or  the mass of the lightest fermion satisfies $m_1 > 3 \gamma +1$, are tachyon-free\footnote{There are other configurations which also remain tachyon-free, where the fermion mass spectrum satisfies a combination of the two conditions, but for the sake of simplicity we will not consider them here.}.
The probability of those fluctuated spectra can be estimated using the expression derived from the Tracy-Widom distribution \eqref{fluctuations}. For $m_h>1$ and to leading order in $1/\scN_h$ they read:
\be
 \mathbb{P}\big(m_1>3\gamma +1\big) \sim \rme^{- \frac{2(3 \gamma +1)^2}{m_h^2} \scN_h^2}\ , \qquad \quad \mathbb{P}\big(m_{\scN_h}<1\big) \sim \rme^{- \ft{1}{6}|x|^3 \scN_h^2}\ , 
\label{largest2}
\ee
 where $x\equiv  -1 +m_h^{-2}$.  In this situation, the regime similar to  KKLT constructions  occurs when the mass scale of the supersymmetric sector satisfies the hierarchy $m_h\gg ( 3 \gamma +1) \, \scN_h\gg1$,  which ensures that the fraction of configurations where all the fermion masses are larger than $3 \gamma+1$ is of order one.  In other cases where this hierarchy is not present, it is easy to check that for sufficiently large values of the  Hubble parameter ($\gamma\gg m_h$), the largest number of stable configurations  always corresponds to the minima of the K\"ahler function $G$, i.e. $m_{\scN_h}<1$.   The expressions \eqref{largest2} are just estimates valid to describe the probability distribution of the smallest and largest fermion masses close to their mean values, and a more accurate calculation can be found in appendix \ref{RMT}.  \\

Note that, in general, the Hubble  parameter, and thus $\gamma$,  will vary during inflation. Therefore, in simple models with a separable K\"ahler function \eqref{separableG1},  the point representing the  configuration $H_0^\alpha$ on the stability diagram of Fig. \ref{separableDiagram} will move vertically along the inflationary trajectory. The different possibilities in the simplest case  where the supersymmetric sector contains a single field  were studied in \cite{Achucarro:2012hg},  but  here we just want to emphasise that, in a broad region of the parameter space, the most stable configurations  during inflation are those where all the truncated fermions are lighter than the gravitino, that is, the minima of the K\"ahler function $G$ along the supersymmetric directions. Even in that case, demanding enough amount of inflation imposes additional constraints on the geometry and couplings \cite{Achucarro:2012hg,Borghese:2012yu}.

%%%%%%%%%%%%%%%%%%%%%%%%%%%%%% 
\subsubsection{Quasi-separable K\"ahler function}
%%%%%%%%%%%%%%%%%%%%%%%%

For completeness,  in this section we will briefly discuss the class of K\"ahler functions  which has the almost separable structure defined in section \ref{sec:Qseparable}.
Here we will also assume the hierarchical setting where the parameters satisfy $m_h\sim \cO(1-10)$ and $|1 -m_h| \gg \epsilon$. 
Using the same argument as for  Minkowski vacua, it is possible to prove that, when $\gamma>0$, the parameters $\mu_{\pm\lambda}^2$ in \eqref{constraints} can also be identified with the approximate eigenvalues of $\cH_h$ to first order in $\epsilon$. Therefore, to this level of accuracy, satisfying the conditions \eqref{constraints} is both necessary and sufficient to guarantee the stability of the truncated sector. We can distinguish two regimes: in the first one  $\gamma \sim \epsilon$,  so that the $\gamma$ parameter is comparable to the size of the corrections to the separable ansatz, and in the second one  $\gamma \gg \epsilon$.    \\

The first case,  $\gamma \sim \epsilon$, is closely related to the situation in models of modular inflation based in  Large Volume Scenarios,  where  the size of $\epsilon$ is suppressed by the volume of the internal space $\epsilon\sim 1 /\cV$ \cite{Gallego:2011jm}, and the maximum  scale of inflation that can be realised is also of order $\gamma \sim1 /\cV$ \cite{Conlon:2005jm,Covi:2008cn}. In this regime  the stability analysis of the truncated sector is very similar to the case of Minkowski vacua in section \ref{sec:Qseparable}. To see this, first note that the zero-order terms of $\mu_{\pm\lambda}^2$ coincide with the eigenvalues of $\cH_h$ in the fully separable case at a Minkowski configuration \eqref{HessianDecomp3b}.  On the one hand, for $m_h<1$ the squared-masses of the scalar fields are positive and they are not expected to become tachyonic after adding the corrections from perturbation theory, since they are protected by a mass gap of size much larger than $\epsilon$  \eqref{massGap}.  On the other hand, when $m_h>1$ there is a potentially large fraction of light scalar fields in the truncated sector which might become tachyonic after adding the first order corrections, but the condition  \eqref{sufficientCondition} derived in the case of Minkowski vacua is also sufficient to ensure the stability of the truncated  sector. Indeed,  to order  $\epsilon$ in perturbation theory,  the mass parameters satisfy:
\be
\mu_{\pm\lambda}^2 \gtrsim   \pm \sqrt{3}\,  \frac{\pd m_\lambda}{\pd X} + 3 \, B[X,\lambda]   + \cO(\epsilon^2), 
\ee
as the zero order term is bounded below by  $-\ft94 \gamma^2 \sim \cO(\epsilon^2)$ \eqref{upliftedmasses}, and both the derivatives of the fermion masses  and the bisectional curvatures are of order $\epsilon$. Consequently, the bound \eqref{sufficientCondition}, which ensures that the right hand side is non-negative, also implies that $\mu_{\pm\lambda}^2\gtrsim0$ to this level of accuracy,  as we claimed above. \\

The second case, $\gamma \gg\epsilon$, is  simpler to analyse provided we also maintain the hierarchy $|1- m_h| \gg \epsilon$.  In this case the stability is determined by  the  zero order  term in perturbation theory, which was analysed in section \ref{separableinflation}. In the regime $m_h<1$ the squared-masses are all  positive and protected by a mass gap \eqref{massGapInflation}, which is large compared to $\epsilon$, and thus  the truncated sector remains stable with order one probability  after taking into account the $\cO(\epsilon)$ corrections. When $m_h>1$, according to the expectation value for the index  of $\cH_h$  \eqref{indexInflation}, the  truncated sector has tachyonic instabilities with order one probability to zero order in $\epsilon$, and the squared-mass of the most unstable mode  is  given by the expression: 
\begin{equation}
\mu_{\pm \lambda}^2|_\text{min}= \left\{ \begin{array}{ll}
 (m_h-1)\big(m_h- (3\gamma +1) \big)  &\qquad  \textrm{if $m_h \le \ft12(3 \gamma +2)$},\\
 -\ft94 \gamma^2 &\qquad \textrm{otherwise,}
\end{array} \right. 
\end{equation}
with probability close to one. The magnitude of this mass parameter is  always much larger than the size of the corrections induced by the mixing terms in the almost separable ansatz  \eqref{quasiSeparableG}, and thus the truncated sector will also remain tachyonic to order $\cO(\epsilon)$ in perturbation theory.  \\

Summarising, for this class of K\"ahler functions we have also checked that those configurations $H_0^\alpha$ minimising the K\"ahler function have very robust stability properties. Indeed,  the supersymmetric sector always remains tachyon-free with order one probability in the regime  $m_h<1$,  which is precisely when  $H_0^\alpha$ typically corresponds to a minimum of   $G$.  
 
\section{Summary and discussion}
\label{sec:summary}

%%%%%%%%%%%%%%%%%%%%%%%%%%%%%%%%%%%%
\begin{table}[t]
\vspace{-1cm}
\begin{center}
\begin{tabular}{| c |  c | c  |}
\hline 
%\vspace{-.3cm}
%&& \\
 $H\ll m_{3/2}$ \ \ (Dark energy) &  $H\approx m_{3/2}$ \ \ (Inflation) & $H \gg m_{3/2}$ \ \ (Inflation)\\
%\vspace{-.3cm}
%&&\\
\hline
&&\\
$B_\text{max} \ge - \frac{1}{3}$ & $B_\text{max}\gtrsim -\frac{2}{3}$ & 
$B_\text{max} \gtrsim -1$
\\
&&\\
\hline
\end{tabular}
\caption{Necessary conditions for the stability of a typical configuration of the supersymmetric sector. $B_{\text{max}}$ is the maximum value that can be attained by the bisectional curvature at $(H_0^\alpha,L_0^i)$ varying the sGoldstino direction $z_X$, and the direction of  the  fermion mass eigenstate $z_\lambda$. }
\end{center}
\label{bisectionalTable}
\end{table}
%%%%%%%%%%%%%%%%%%%%%%%%%%%%%%%%
%%%%%%%%%%%%%%%%%%%%%%%%%%%%%%%%%%%%%
%%%%%%%%%%%%%%%%%%%%%%%%%%%%%%%%%%%%%%%

In this paper we have studied the type of couplings  which are necessary  for the stability of a supersymmetric sector which is embedded in a theory with broken supersymmetry.   In particular we have considered $\cN=1$ supergravity models  involving only chiral multiplets,  which are consistent with the supersymmetric truncation of the fields preserving supersymmetry. This class of theories are characterised by  a K\"ahler function $G$ satisfying:
\be
\partial_\alpha G(H, \bar H,L, \bar L)|_{H_0}=0 \quad \text{for all}\quad L^i\ , \nonumber 
\ee
where $L^i$ are the fields in the supersymmetry breaking sector, and $H^\alpha$ are the fields in the supersymmetric sector which are  frozen at a configuration $H^\alpha_0$.   In addition, following \cite{Marsh:2011aa,Bachlechner:2012at},  we have treated the couplings in the decoupled  sector as random variables, and we studied the Hessian of the scalar potential using tools from random matrix theory.  This analysis is motivated by  the supergravity description of the dilaton  and complex structure moduli   in KKLT constructions and Large Volume Scenarios of type-IIB flux compactifications, where  the complex structure and dilaton fields are truncated from the theory before considering the stability of the K\"ahler moduli, 
and the truncation is done in a way that leaves  supersymmetry approximately unbroken. The stability of the truncated sector is crucial to ensure the via\-bi\-li\-ty of cosmological  models based on  supergravity theories with this structure, both when they describe  the present vacuum with a small cosmological constant (dark energy), as well as for scenarios of slow-roll inflation. The main conclusion of our analysis is that,  in broad range of parameter space, the configuration of a decoupled  supersymmetric sector  $H_0^\alpha$ remains free of tachyons with order one probability, $\mathbb{P}_{\text{stable}}\sim \cO(1)$. \\
 
 In order to perform the analysis   we have derived the set of necessary conditions \eqref{constraints} for the stability of the supersymmetric sector.  These conditions, which can be seen as constraints on the K\"ahler potential $K$ and the superpotential $W$,  are expressed in terms of the ratio of the Hubble parameter to the gravitino mass ($\gamma=H^2/m_{3/2}^2$), the masses the chiral fermions and their derivatives along  the sGoldstino direction ($m_\lambda$ and $\partial m_\lambda/\partial X$, respectively), and  the bisectional  curvatures of the K\"ahler manifold
\be
B[X,\lambda] \equiv - R_{I\bar J K \bar L} \,  z_X^{\phantom{,,}I}\,  \bar z_{X}^{\phantom{,,}\bar J}\,  z_\lambda^{\phantom{,} K} \, \bar z_{ \lambda}^{\phantom{,} \bar L} . \nonumber
\ee
Here $R_{I\bar J K \bar L}$ are the components of the Riemann tensor, $z_X$ is a unit vector along the sGoldstino direction, and the $z_\lambda$ form an orthonormal basis on the supersymmetric sector.  In general,  the conditions \eqref{constraints} are necessary but can not guarantee  the stability of the truncated  supersymmetric sector.  Still, assuming that the  number of $H^\alpha$ fields is large, $\scN_h\gg1$, and  that there is no large hierarchy between the masses in the truncated sector and the supersymmetry breaking scale (as in LVS), we were able to derive generic constraints on the geometry of the moduli space. In particular, we have shown that positive values of the bisectional curvatures $B[X,\lambda]$ considerably improve the stability of the supersymmetric sector, as is  summarised in Table 1.\\

We have analysed in detail a class of models which includes  physically relevant sce\-na\-rios, and for which the conditions \eqref{constraints} are \emph{necessary and sufficient} for the stability of the supersymmetric sector. This also allows us to perform a detailed study of the scalar mass spectrum and establish specific criteria to achieve tachyon-free spectra. This class of models is characterised by an almost separable K\"ahler function:
\begin{enumerate}

\item[]\centering $G(H,\bar H,L^i,\bar L)=G_h(H,\bar H)+G_l(L,\bar L)+\epsilon\, G_{mix}(H,\bar H,L,\bar L)$,

 with \quad$\partial_\alpha G_h|_{H_0}=\partial_\alpha G_{mix}|_{H_0}=0$\quad\   and \quad$\epsilon\ll 1$.

\end{enumerate}
When the parameter $\epsilon$ is set to zero, this type of K\"ahler functions describes a large class of no-scale models, and in particular it includes  the effective supergravity description of type-IIB compactifications to zero-order in $\alpha'$ and non-perturbative corrections. In the latter case the fields $H^\alpha$ are identified with the dilaton and complex structure moduli, and $L^i$ with the K\"ahler sector. When $\epsilon$ is small but non-zero, this type of theories   include models with  a similar structure to the effective description of LVS, where the magnitude of $\epsilon$ is suppressed by the volume of the compact space, $\epsilon \sim 1/ \cV$. We have then studied case by case the typical scalar mass distribution of the supersymmetric sector and its dependence on the parameters appearing in \eqref{constraints}.   A fundamental quantity that determines the stability of field configurations is the mass scale of the supersymmetric sector in units of the gravitino mass, denoted by $m_h$. This parameter is defined in such a way that typically the fermion masses are distributed in the interval $m_\lambda\in\[0,m_h\]$.   When the mass scale of the supersymmetric sector is larger than the gravitino mass, $m_h>1$, the typical spectrum has the following general features:

\begin{itemize}

\item At Minkowski vacua the spectrum has no tachyons in the fully separable case $(\epsilon=0)$.  However, depending on the parameters, the scalar mass spectrum may contain  a significant fraction of fields with can be much lighter than the gravitino (Fig. \ref{fig:sepM4}).  

\item Minkowski configurations are very susceptible to become tachyonic when the K\"ahler function has a small non-separable term. However, in LVS  the parameter $\epsilon\sim1/ \cV$ is exponentially small, and in practice we recover the fully separable case, for which Minkowski vacua are metastable (Fig. \ref{fig:QsepM4}). 

\item In the regime $m_h\sim \cO(1- 10^2)$, typical field configurations always become tachyonic for sufficiently large values of the ratio $\gamma = H^2/m_{3/2}^2$ (Fig. \ref{Fig:index2}).

\item When the mass scale $m_h$ is much larger than the gravitino mass, $m_h\gg\scN_h \gg1$, the supersymmetric sector is always stable.  This is precisely the hierarchy needed  in KKLT type of stabilisation mechanisms.

\end{itemize}

Conversely, when the mass scale of the supersymmetric sector is smaller than the gra\-vi\-ti\-no mass, $m_h<1$,  the spectrum displays a mass gap which protects the stability of the truncated sector (Figs. \ref{fig:sepM4} and \ref{Fig:index2}). Therefore, provided the non-separable corrections to the K\"ahler function are small, $\epsilon \ll |1-m_h|$, \emph{the truncated sector is typically stable in the regime $m_h<1$, regardless of the ratio $H/m_{3/2}$}. Interestingly, in this regime, the configuration of the supersymmetric sector always corresponds to minima of the K\"ahler function $G$.  The presence of the  mass gap in the spectrum suggests that the robust stability properties of the minima of $G$ will survive in more realistic models where the truncation of the supersymmetric sector is only approximate. Therefore, the minima of the K\"ahler function can be of interest to explore new stabilisation mechanisms,   for example in cases where the supersymmetric sector is not protected by a large mass hierarchy,  as in KKLT constructions, or where the   non-separable corrections in the  K\"ahler function  are small but not exponentially suppressed, as  in Large Volume Scenarios.  \\

 From our analysis  it follows that, in general, \emph{supersymmetric AdS minima are more difficult to realise than the stabilisation of a   supersymmetric sector}, when this sector is embedded in a theory where supersymmetry is spontaneously broken.  In other words, a decoupled supersymmetric sector can be stabilised at configurations which do not correspond to supersymmetric minima in the absence of supersymmetry breaking. This discussion already allows us to understand the situation in LVS in relation to the random supergravity analysis that we have performed here and the one in \cite{Bachlechner:2012at}. Recall that in LVS all moduli can be stabilised in a very model-independent way, while  
  the fraction of configurations of the complex structure moduli which correspond to supersymmetric  minima of the potential induced by fluxes  is exponentially small  $\mathbb{P}_{\text{susy}}\sim \text{exp}(- 8 \scN_h^2 /m_h^2)$ \cite{Bachlechner:2012at}. Conversely, our results indicate that there can be a large fraction of stable  configurations for the supersymmetric moduli   when   the supersymmetry breaking effects are included, $\mathbb{P}_{\text{stable}} \sim \cO(1)$.  This can be easily understood  from the discussion  in the previous paragraph: by considering only  supersymmetric AdS minima of the scalar potential induced by the fluxes,  we are discarding most configurations where  the supersymmetric sector is stable.  Actually,  \emph{the  non-generic couplings between the supersymmetric sector and the remaining  moduli  can turn  all the Breitenlohner-Freedman allowed tachyons of   supersymmetric critical points into stable modes}. In Large Volume Scenarios the no-scale structure of the K\"ahler sector is fundamental to achieve the stabilisation of the supersymmetric moduli. 
This effect has already been confirmed in a numerical analysis of the  $\mathbb{P}_{[1,1,1,6,9]}^4$ model of   type-IIB compactifications  \cite{BP}.   It is also worth mentioning the stabilisation of supersymmetric saddle points and maxima, which was previously  discussed in  the context  of $F-$term uplifting mechanisms in   \cite{Achucarro:2007qa,Achucarro:2008fk}.  \\

A possible extension of this work would be to consider a more general class of supergravity theories along the lines of  \cite{GomezReino:2007qi,Achucarro:2008fk}, that is, to study models involving vector multiplets in addition to the chiral multiplets, which could gauge the local isometries of the K\"ahler manifold. The results of these works suggest that the inclusion of vector multiplets would improve the stability of the supersymmetric sector at dS and Minkowski vacua. As  was shown in \cite{Achucarro:2008fk}, this is certainly the case for separable K\"ahler functions of the form \eqref{separableG1},  but a more detailed analysis is still needed for more general K\"ahler functions. Many interesting questions regarding the implications for inflation in supergravity arise from our analysis. In particular, it is important to understand in which situations the scalar fields in  the  supersymmetric sector have masses  much larger than the Hubble scale, since in this case it is consistent to neglect completely the presence of this sector during inflation.  More precisely, in this regime we avoid the contribution from the truncated sector to primordial fluctuations, and inflation can be entirely described considering only  the dynamics of the supersymmetry breaking sector. 
It would also be useful to study the inflationary constraints in particular models where inflation is driven by a field orthogonal to the sGoldstino direction, or in other words, a field contained in the supersymmetric sector, as in \cite{Kallosh:2010xz}. We intend to address these questions  in a future work \cite{Geodesics}. 

%\newpage
\acknowledgments

We are grateful to A. Ach\'ucarro, J. J. Blanco-Pillado and J. Urrestilla for very useful suggestions and discussions, to J. Frazer for helpful comments and collaboration in the early stages of this work, and also to J.W. van Holten, D. Marsh, L. McAllister, T. van Riet, D. Roest, Y. Sumitomo and  I. Zavala  for conversations. We particularly thank D. Marsh for very useful comments on the manuscript. KS acknowledges financial support from the University of the Basque Country UPV/EHU via the programme ``Ayudas de Especializacion al Personal Investigador (2012)'',  from the Basque Government (IT-559-10), the Spanish Ministry (FPA2009-10612), and the Spanish Consolider-Ingenio 2010 Programme CPAN (CSD2007-00042). PO is supported by F.O.M. (Dutch organization for Fundamental Research in Matter). PO thanks the University of the Basque Country UPV/EHU for hospitality during the completion of this work.

\appendix

\section{Atypical minima and fluctuated spectra}
%%%%%%%%%%%%%%%%%%

\label{RMT}

In this appendix we review the expressions for the probability of occurrence of  atypical fluctuations of the fermionic mass spectra, and in particular we  will discuss  the probability distribution of the lightest and largest fermion masses. As we have discussed in the main text, the CI-ensemble describes the statistical properties of the fermion mass matrix $\cM_h$ for a generic supersymmetric sector. The CI-ensemble is closely related to the set of Wishart ensemble \cite{Bachlechner:2012at} for which there are many results in the literature regarding fluctuated spectra. For this reason we will first discuss known results for the Wishart ensemble, and then we will translate them  into properties of the fermion mass spectrum in a generic supersymmetric sector. 

\addtocontents{toc}{\protect\setcounter{tocdepth}{1}}
\subsection{Typical spectral density in the  Wishart and CI-ensembles}
%%%%%%%%%%%%%%%%%%%%%%%%%
\addtocontents{toc}{\protect\setcounter{tocdepth}{2}} 

The Wishart ensemble is composed of matrices of the form ${\cal W}= A  A^\dag$, where $A$ is an $\scN \times \scM$  real or complex matrix, (with $\scM \ge \scN$), whose entries are independent and identically distributed (i.i.d.)\ random variables drawn from a statistical distribution with zero mean and variance $\sigma^2$: $A_{IJ} \in \Omega(0,\sigma)$. When $\Omega= N(0,\sigma)$ is a normal distribution, the joint probability distribution for the ordered  eigenvalues  $\lambda_1\le \lambda_2, \ldots ,\le \lambda_{\scN}$ is  \cite{1928}:
\be
f(\lambda_1, \ldots, \lambda_{\scN}) = {\cal C}\ \exp\left(- \frac{\beta}{2} \left( \frac{1}{\sigma^2} \sum_{a=1}^{\scN} \lambda_a \ - 2\sum_{a < b}^{\scN}{\rm{ln}}|\lambda_b - \lambda_a| - \xi\,\sum_{a=1}^{\scN} {\rm{ln}}\,\lambda_a \right)\right) \, , \label{eq:Wishart}
\ee
where $\xi= \scM-\scN+1-2/\beta$, and $\beta=1,2$ for real and complex matrices, respectively. The eigenvalue density function  for the eigenvalues of ${\cal W}$ is given by the
Mar\v{c}enko-Pastur law \cite{MPLaw},
\begin{equation} \label{eq:wishartapp}
\rho_\text{MP}(\lambda) \, d\lambda  =  \frac{1}{2\pi \sigma^2 \lambda}\sqrt{(\lambda_+-\lambda)(\lambda-\lambda_-)} \, d\lambda\,,
\end{equation}
with support $\lambda \in \[\lambda_-,\lambda_+\]$, where
\begin{equation} \label{etadefinition}
\lambda_\pm = \scN \sigma^2 (1\pm\sqrt{\eta})^2\,, \qquad  \textrm{and} \qquad \eta=\scM/\scN \geq 1.
\end{equation}

The joint probability distribution for the eigenvalues of a matrix from the CI-ensemble was given in  eq. \eqref{eq:CI}. As was pointed out in \cite{Bachlechner:2012at}, the p.d.f. of the eigenvalues of a Wishart matrix \eqref{eq:Wishart} reduces to \eqref{eq:CI} for $\beta=1$ and $\scM = \scN+1$ after doing the identification $\lambda_a \leftrightarrow m_\lambda^2$. Moreover, as we are interested in results to leading  leading order in $1/\scN$, it will be sufficient to discuss square  Wishart matrices $\scN \approx \scM$.  Thus, since the fermion mass matrix of a generic supergravity theory can be identified with an element of the CI ensemble,  the typical spectral density of the fermion masses $m_\lambda$ is also given by the Mar\v{c}enko-Pastur law \eref{eq:wishartapp} with $\lambda = m^2$.  Defining $m_h \equiv 2 \sqrt{\scN} \sigma$, we have that  to leading order in $1/\scN$  the fermion mass density function reads:  
\be
\rho_\text{MP}(m^2) dm^2=\frac{2\scN}{\pi m_h^2m}\sqrt{m_h^2 - m^2}\ dm^2, 
\ee
which has support in   $m^2 \in\[0,m_h^2\]$. In the limit $\scN \to \infty$  the bounds of the support  coincide with the expectation value of the smallest and largest fermionic masses squared, $m_1^2$ and $m_{\scN}^2$ respectively  \cite{Johnstone01onthe}:
\be
\mathbb{E}[ m^2_1] =0, \qquad\qquad  \mathbb{E}[m_{\scN}^2]  \approx m_h^2.
\ee

\addtocontents{toc}{\protect\setcounter{tocdepth}{1}}
\subsection{Probability distributions of  the limiting eigenvalues of a Wishart matrix}
%%%%%%%%%%%%%%%%%%%%%%%%%%%%%%%%%%%%%%%
\addtocontents{toc}{\protect\setcounter{tocdepth}{2}} 

Let us first discuss the probability distribution of the largest eigenvalue $\lambda_{\scN}$ of a real, almost square  Wishart matrix, $\beta=1$, $\scM\approx \scN$.  The probability distribution of large $\cO(\sigma^2 \scN)$ fluctuations of $\lambda_{\scN}$ far to the right  and left of its mean value $\lambda_+$ was calculated in \cite{Johansson} and \cite{Vivo2007}, respectively, and are given by:
\bea
t >\lambda_{+} :\quad \lim_{\scN\to \infty}\mathbb{P}(\lambda_{\scN}\le t) &\approx&1-\left( \sqrt{x+1} + \sqrt{x}\right)^{2 \scN} \rme^{-2 \scN \sqrt{x (x+1)}}, 
\nonumber \\
t <\lambda_{+} : \quad \lim_{\scN\to \infty}\mathbb{P}(\lambda_{\scN}\le t) &\approx& \left(\frac{ x+1}{\sqrt{\rme} } \right)^{\frac{\scN^2}{2}} \rme^{-\frac{\scN^2}{4}\left( 1-x\right)^2}\ ,
\label{largeFluct}
\eea
where $x \equiv (t-\lambda_+)/\lambda_+$. For large but finite values of $\scN$, the maximum value of a Wishart matrix, $\lambda_{\scN}$, \emph{typically} fluctuates over a region of size $\cO(\sigma^2 \, \scN^{-1/3})$, and the corresponding probability distribution for these small fluctuations can be approximated by the Tracy-Widom distribution $F_1(x)$ \cite{Johnstone01onthe,Johansson,Tracy:1992rf}:
\be
\mathbb{P}(\lambda_{\scN}\le t) \approx F_1\left(\frac{\eta^{\frac{1}{12}} \scN^{\frac{1}{3}}\, (t - \lambda_+ )}{\sigma^{\frac{2}{3}} \lambda_+^{\frac{2}{3}}} \right) \approx F_1\left(2^{\frac{2}{3}} \scN^{\frac{2}{3}}\, \frac{  (t - \lambda_+ )}{\lambda_+} \right)\ ,
\label{TWdist}
\ee
where we have used the leading order approximation  $\eta=1$ and $\lambda_+ \approx 4 \scN \sigma^2$  for large $\scN$ in the last step.  For the asymptotic values of the probability \eref{TWdist}, see \cite{TracyWidom} and references therein.  In particular, to leading order in $1/\scN$, the cumulative probability distribution for the largest eigenvalue $\lambda_{\scN}$ is:
\bea
t >\lambda_{+} : \quad \lim_{\scN\to \infty}\mathbb{P}_{\scN}(\lambda_{\scN}\le t) &\approx& 1- \frac{\rme^{-\ft43 \scN\, x^{\frac{3}{2}}}}{8 \sqrt{\pi} \,\scN \, x^{\frac{3}{2}}} - \frac{\rme^{-\ft83 \scN\, x^{\frac{3}{2}}}}{64 \pi \scN\,  x^{\frac{3}{2}}}\ , \nonumber \\ \nonumber \\
t <\lambda_{+} : \quad \lim_{\scN\to \infty}\mathbb{P}_{\scN}(\lambda_{\scN}\le t)& \approx& \tau_1 \, \frac{\rme^{- \ft{1}{6}|x|^3 \scN^2}}{2^\frac{1}{24} \scN^\frac{1}{24} |x|^{\frac{1}{16}}\,}\ ,\hspace{1cm}
\label{largest2A}
\eea
where $\tau_1\equiv 2^{-11/48}\,\rme^{\frac12\zeta '(-1)}$,  and $\zeta '(-1)=-0.16542\dots$ is the derivative of the Riemann zeta function evaluated at $-1$. It is easy to check that, to leading order in $\cO(1/\scN)$,  the probability distributions  in \eqref{largeFluct} match the tail behaviour of the Tracy-Widom distributions in the limit $t \to \lambda_+$, which describes small fluctuations.\\

The probability distribution of the smallest eigenvalue $\lambda_1$ of a  real square Wishart matrix  was derived in \cite{Edelman}. To leading order in $1/\scN$ it is given by:
 \be
  \lim_{\scN\to \infty}\mathbb{P}(\lambda_1\ge t) \approx \frac{\lambda_+}{4 \scN^2} \, \rme^{-\frac{2 \scN^2}{\lambda_+} t}.
 \label{Edelman}
\ee

\vspace{-.5cm}
\addtocontents{toc}{\protect\setcounter{tocdepth}{1}}
\subsection{Probability of field configurations with atypical  fermionic mass spectra}
\addtocontents{toc}{\protect\setcounter{tocdepth}{2}} 

In section \ref{sec:M4Separable}, where we study the stability of a consistently truncated supersymmetric sector in models with a separable K\"ahler function, we estimated the probability of  occurrence of  critical points with light scalar fields, i.e. with a mass $\mu^2|_{min}\le \alpha^2$, in the regime $m_h<1-\alpha$. We argued that, due to the relation between scalar and fermion masses, this would require the largest fermion mass fermion to be above its expectation value $m_{\scN} \ge 1- \alpha > m_h$. Using the first equation in \eqref{largeFluct}, and taking into account the relation between the Wishart and CI-ensembles, we find
\be
1- \alpha >m_h :\quad \lim_{\scN\to \infty}\mathbb{P}(m_{\scN}\ge 1- \alpha) \approx \left( \sqrt{x+1} + \sqrt{x}\right)^{2 \scN} \rme^{-2 \scN \sqrt{x (x+1)}},
\ee
with $x = (1 - \alpha)^2/m_h^2 - 1$. The Tracy-Widom distribution gives an accurate description in the limit $m_h \to (1-\alpha)^-$, where the deviations of $m_{\scN}$ from its expectation value  are small, and thus the previous probability distribution reduces to \eqref{PlightFields} to leading order in $1/\scN$, which can also be derived from the first  equation in \eqref{largest2A}.\\

In section \ref{separableinflation} we discuss the stability of the truncated sector, when the  sector surviving the truncation is driving a period of inflation, and the K\"ahler function is also separable in the two sectors. In general, in the regime where the mass scale $m_h$ is larger than the gravitino mass ($m_h>1$), the typical spectrum contains tachyons  (see Fig. \ref{Fig:index2} and right plot in Fig. \ref{fig3}). However, as illustrated by \eref{largest2}, there is an exponentially suppressed probability that the fermionic spectrum fluctuates in such a way that the scalar spectrum is free of tachyons. There are two possible types of configurations which are non-tachyonic: when the fermion masses are confined to $m_\lambda<1$, or to $m_\lambda>3\gamma+1$, for all $\lambda$. It is interesting to check, for a configuration with a Hubble parameter given in terms of $\gamma$, in what regime of parameters one of these types of critical points becomes more abundant than  the other. The probability that the fermion masses are bounded below as $m_\lambda \ge 3 \gamma +1$,  can be calculated from \eqref{Edelman}:     
\be
\lim_{\scN\to \infty}\mathbb{P}\Big(m_1\ge 3 \gamma +1\Big) \approx \frac{m_h^2}{4 \scN^2} \, \rme^{-\frac{2 \scN^2}{m_h^2} (3 \gamma +1)^2}\ .
\label{fluctuationsmall}
\ee
On the other hand, the probability that the fermion masses are bounded above by $m_{\scN}= 1$, also to leading order in $1/\scN$, can be derived from the second equation in \eqref{largeFluct}:
\be
\lim_{\scN\to \infty}\mathbb{P}\Big(m_{\scN}\le 1\Big) \approx \frac{\rme^{-\scN^2/4}}{m_h^{\scN^2}}\,\rme^{-\frac{\scN^2}{4} \(2-\frac{1}{m_h^2}\)^2}\ .
\label{fluctuationlarge}
\ee
From the above expressions it can be seen that a fluctuation to a minimum of the K\"ahler function, where all $m_{\lambda} <1$  \eqref{fluctuationsmall}, becomes more likely than a fluctuation to large fermionic masses \eqref{fluctuationlarge} when the following condition is satisfied:
\be
(2+\scN^2)\log m_h-2\log (2\scN)-\frac{2\scN^2}{m_h^2}(1+3\gamma)^2+\frac{\scN^2}{4}+\frac{\scN^2}{4}\(2-\frac{1}{m_h^2}\)^2<0\ .
\ee
In the regime $\scN\gg m_h\gg1$ and with $\gamma$ comparable to $m_h$ the expression above simplifies to
\be
\gamma^2 \ge \frac{m_h^2}{18} \, (\log m_h+\tfrac{5}{4})\ ,
\label{conditionfluct1}
\ee
which can be rewritten in terms of the Hubble parameter as follows:
\be
H^2\gtrsim \frac{m_hm_{3/2}}{3 \sqrt{2}}\sqrt{\log m_h+\frac{5}{4}}\ . 
\label{conditionfluct1H}
\ee
We can see that regardless of the value of the mass scale of the supersymmetric sector $m_h$, 
there is always a value of the Hubble parameter \eqref{conditionfluct1H} above which the largest fraction of stable  critical points corresponds to a minimum of the   K\"ahler function. This is particularly important for cosmological models which involve a large inflationary scale $H$ and a low supersymmetry breaking scale. This result depends strongly on the value of the mass scale of the supersymmetric sector   $m_h$ which, as we discussed in section \ref{randomSugra}, should also be regarded as a random variable  depending on the value of the gravitino mass $m_{3/2}$, and its statistical properties should be derived from a  realistic characterisation of the String Theory Landscape. 

%%%%%%%%%%%%%%%%%%%%%%%%%%%%%%%%%%%%%%%%
\section[Quasi-separable  K\"ahler functions]{Mass spectrum for quasi-separable  K\"ahler functions}
\label{appendix1}
%%%%%%%%%%%%%%%%%%%%%%%%%%%%%%%%%%%%%%%%

In this appendix we derive in full detail the mass spectrum of the supersymmetric sector in models characterised by an almost separable K\" ahler function of the form \eqref{quasiSeparableG}. We will briefly review eigenvalue perturbation theory, we also re-derive the result \eref{upliftedmasses} for separable K\"ahler functions, and finally we will use this to calculate the perturbed eigenvalues for a K\"ahler function with a small mixing between sectors.

%%%%%%%%%%%%%%%%%%%%%%%%%%%%%%%%%%
%%%%%%%%%%%%%%%%%%%%%%%%%%%%%
\addtocontents{toc}{\protect\setcounter{tocdepth}{1}}
\subsection{Perturbation theory}
\addtocontents{toc}{\protect\setcounter{tocdepth}{2}} 
\label{s:perturbation}
%%%%%%%%%%%%%%%%%%%%%%%%%%%%

Consider a $2 \scN\times 2 \scN$ square matrix $\cH=\cH_0+ \epsilon \,  \delta\cH$, where $\epsilon\ll1$ is a small parameter. Let us denote by $\mu^2_{p,0}$ the eigenvalues of $\cH_0$, where $p=1,\dots,2 \scN_h$. The eigenvectors corresponding to those eigenvalues form a orthonormal basis with which one can build the matrix $A$ that diagonalises $\cH_0$, that is, $A^\dagger \cH_0 A=\text{diag}\big(\mu^2_{p,0}\big)$. Then, the eigenvalues of the full matrix $\cH$  to first order in perturbation theory will be given by
\be
\mu^2_{p}=\(A^\dagger\cH A\)_{pp}=(\mu^2_{p})^0+\epsilon \, \(A^\dagger\delta\cH A\)_{pp} + \cO(\epsilon^2 ) \  ,\quad p=1,\dots,2\scN_h\ .
\ee
In other words, the perturbation over the zero-order eigenvalues is given by the diagonal elements of the matrix perturbation in the basis that diagonalises the zero-order matrix.

%%%%%%%%%%%%%%%%%%%%%%%%%%%%%%%%%%%
%%%%%%%%%%%%%%%%%%%%%%%%%%%%%%%%%%
 \addtocontents{toc}{\protect\setcounter{tocdepth}{1}}
\subsection{Perturbed eigenvalues}
\addtocontents{toc}{\protect\setcounter{tocdepth}{2}}
\label{s:eigen}
%%%%%%%%%%%%%%%%%%%%%%%%%%%%
Let us consider an almost separable K\"ahler function $G$ of the form \eqref{quasiSeparableG}:
\be
G(H,\bar H,L,\bar L) =G_h(H,\bar H) +G_l(L,\bar L)  + \epsilon \, G_\text{mix}(H,\bar H,L,\bar L)\ ,
\ee
where the  fields $H^\alpha$ are consistently truncated at the supersymmetric critical point such that\footnote{Since we want to impose this condition for any value of $\epsilon$, both functions $G_h$ and $G_{mix}$ must satisfy this requirement.}
\be
\pd_\alpha G_{h}|_{H_0}= \pd_\alpha G_{\text{mix}}|_{H_0}=0\ .
\ee
Consider the block of the Hessian of the supersymmetric sector $\cH_h$ \eqref{HessianDecompSusy} whose elements are:
\bea
\nabla_\alpha \nabla_\beta V|_{H_0,L_0} &=&    \rme^G \left[(3\gamma+2) \nabla_\alpha G_{ \beta} +G^i\, \nabla_{i} (\nabla_{\alpha} G_\beta) \right] \ ,\\
\nabla_\alpha \nabla_{\bar \beta} V|_{H_0,L_0} &=& \rme^G \left[G_{\alpha \bar \beta}\,  (3\gamma+1)  +G^{\gamma \bar \sigma} \(\nabla_\gamma G_{ \alpha}\)\( \nabla_{\bar \sigma}G_{ \bar \beta }\) - R_{i \bar j \alpha \bar \beta}\, G^i G^{\bar j} \right]\ .
\eea
After the rescalings \eqref{reescalings}, and switching to an orthonormal basis $G_{\alpha \bar \beta}|_{H_0,L_0} = \delta_{\alpha \bar \beta} $ where one of the axes points in the direction of the sGoldstino (see section \ref{sec:structureH}),  the mass matrix $\cH_h$ can be rewritten in terms  of the quantities above as follows:
\be
\cH_h=\(\begin{array}{ccc}  (3\gamma+1)\unity_{\scN_h}+MM^*&\ \  &(3\gamma+2)M^*\\(3\gamma+2)M& &(3\gamma+1)\unity_{\scN_h}+M^*M \end{array}\)+\(\begin{array}{ccc}\Omega&\ \  &Y^*\\Y& &\Omega^\dag\end{array}\)=\cH_{h} ^0+\epsilon \delta\cH_h\ ,
\label{ap1}
\ee
where  $M_{\alpha \beta}$ is the rescaled fermion mass matrix   \eqref{fermionMasses} in  the supersymmetric sector, and we have also used the definitions
\be
Y_{\alpha\beta}\equiv \sqrt{3 (\gamma +1)}\,\nabla_X \(M_{\alpha \beta}\)=Y_{\beta\alpha}, \qquad \qquad
\Omega_{\alpha\beta} \equiv- 3 (\gamma +1) R_{X \bar X \alpha \bar \beta} =\Omega_{\beta\alpha}^*\ .\label{YKOmegaK}
\ee
Note that the second contribution to $\cH_h$  in \eqref{ap1} is at least of order  $\cO(\epsilon)$.  Using the freedom of field redefinition left after choosing an orthonormal basis, it is still possible to transform the  fields so that the fermion mass matrix transforms as: 
\be
M\to\tilde M=UM U^t\ ,
\label{transf1}
\ee
where $U$ is a unitary matrix, and $U^t$ its transpose.  Using this freedom, it is possible to bring $M$ to a diagonal form, this is the so-called  Takagi's factorisation, i.e. $\tilde M=U m \, U^t$, where  $m$ is a diagonal matrix with non-negative entries $m=\text{diag}(m_\lambda)$, with $m_\lambda\ge0$ for $\lambda = 1, \ldots, \scN_h$. After rewriting   the unperturbed mass matrix ${\cal H}^0_h$ in a new basis, it has four blocks of size $\scN_h \times\scN_h$ each, which are diagonal:
\be
\cH^0_h=\(\begin{array}{ccc|ccc}{\scriptstyle (3\gamma+1)+m_1^2}& &{\bf 0} &{\scriptstyle(3\gamma+2)m_1}& &{\bf 0}\\ &\ddots&  & &\ddots& \\{\bf 0} & &{\scriptstyle (3\gamma+1)+m_{\scN_h}^2 }\vspace{0pt} &{\bf 0}& &{\scriptstyle(3\gamma+2) m_{\scN_h} }\\ \hline {\scriptstyle(3\gamma+2)m_1}& &{\bf 0}&{\scriptstyle (3\gamma+1)+m_{1}^2 }& &{\bf 0}\\ &\ddots&  & &\ddots& \\{\bf 0} & &{\scriptstyle(3\gamma+2)m_{\scN_h}} &{\bf 0}& &{\scriptstyle (3\gamma+1)+m_{\scN_h}^2}\end{array}\)\ .\hspace{1cm}
\label{h0}
\ee
We can always solve the eigenvalue problem by rearranging rows and columns to make the mass matrix block diagonal, with blocks of dimension $2 \times2$ given by the matrices
\be
\cH^{0 (\lambda)}_h=\(\begin{array}{cc}\(m_{\lambda}^2+3\gamma+1\)&(3\gamma+2)m_{\lambda}\\(3\gamma+2)m_{\lambda}&\(m_{\lambda}^2+3\gamma+1\)\end{array}\)\ ;\ \cH^{0}_h=\(\begin{array}{ccc}\cH^{0 (1)}_h& &{\bf 0}\\ &\ddots& \\{\bf 0} & &\cH^{0 (\scN_h)}_h\end{array}\)\ .
\label{rearrange}
\ee
Although the previous step is not strictly necessary, it makes the eigenvalue problem more visual to solve it for each subspace. We easily find the eigenvalues, which  are given by:
\be
(\mu_{\pm \lambda}^2)^0=(m_\lambda \pm 1 )(m_\lambda \pm (3\gamma+1) )= \( m_{\lambda}\pm\ft12 (3\gamma+2)\)^2-\ft94\gamma^2\ .
\label{eigen0}
\ee
Note that this expression is precisely the one in  \eref{upliftedmasses}. The eigenvectors $u_{\pm\lambda }$ are easily found through the equation:
\be
\(\begin{array}{cc}\mp(3\gamma+2)m_{\lambda}&(3\gamma+2)m_{\lambda}\\(3\gamma+2)m_{\lambda}&\mp (3\gamma+2)m_{\lambda}\end{array}\)\(\begin{array}{c} a_\lambda\\ b_\lambda \end{array}\)=\(\begin{array}{c}0\\0\end{array}\)\Rightarrow  u_{\pm\lambda }=\frac{1}{\sqrt{2}}\(\begin{array}{c} 1 \\\pm 1\end{array}\)\ .
\ee
The matrix of change of basis for the eigenspace associated to the fermion mass $m_\lambda$    is then:
\be
A_\lambda=\frac{1}{\sqrt{2}}\(\begin{array}{cc}1&1\\1&-1\end{array}\)\Rightarrow A_\lambda^\dagger {\cal H}^{0(\lambda)}_hA_\lambda=\text{diag}\(\mu_{+\lambda}^2,\mu_{-\lambda}^2\)\ .
\ee
We can repeat the process for every value of $\lambda$, which leads to our final result for the unperturbed mass matrix:
\be
A=\(\begin{array}{ccc}A_1& &{\bf 0}\\ &\ddots& \\{\bf 0} & &A_{\scN_h}\end{array}\)\Rightarrow A^\dagger{\cal H}^0_hA=\(\begin{array}{ccccc}\mu_{+1}^2& & & &{\bf 0}\\ &\mu_{-1}^2& & & \\ & &\ddots& & \\ & & &\mu_{+\scN_h}^2& \\{\bf 0}& & & &\mu_{-\scN_h}^2\end{array}\)\ .
\label{basis}
\ee
We retrieve the results of \cite{Achucarro:2012hg,Achucarro:2007qa,Achucarro:2008fk}. Now we just have to express the perturbation matrix $\delta\cH$ in the basis that diagonalises $\cH^0_h$. In order to do that, we first have to undo the rearranging of rows and columns we did to get to \eref{rearrange} (which was only done to facilitate the discussion). Instead of rearranging, it is easy to realise that the matrix that diagonalises $\cH^0_h$ \eref{h0} is
\bea
A&=&\frac{1}{\sqrt{2}}\(\begin{array}{cc}\unity_{\scN_h}&\unity_{\scN_h}\\\unity_{\scN_h}&-\unity_{\scN_h}\end{array}\)\Rightarrow\\\nonumber \\\Rightarrow A^\dagger {\cal H}^0_hA&=&\text{diag}\(\mu_{+1}^2,\dots ,\mu_{+\scN_h}^2,\mu_{-1}^2,\dots ,\mu_{-\scN_h}^2\)\ .
\eea
Therefore, the leading order correction to the eigenvalues \eref{eigen0} due to $\delta\cH$ in \eref{ap1} is given by the diagonal elements of the matrix 
\be
A^\dagger \delta\cH_h\,A=\frac12\(\begin{array}{ccc}\[\(\Omega+\Omega^\dag\)+\(Y+Y^*\)\]&\ \  &\[\(\Omega-\Omega^\dag\)+\(Y-Y^*\)\]\vspace{5pt}\\\[\(\Omega-\Omega^\dag\)-\(Y-Y^*\)\]& &\[\(\Omega+\Omega^\dag\)-\(Y+Y^*\)\]\end{array}\)\ ,
\ee
where the matrices  $\Omega$ and $Y$ are now written in the basis where $M$ is diagonal \eref{transf1}. To first order in perturbation theory the squared-mass parameters  read
\be
\mu_{\pm\lambda}^2\approx (m_\lambda \pm 1 )(m_\lambda \pm (3\gamma+1) )\pm \sqrt{3 (\gamma +1)} \frac{\pd m_\lambda}{\pd X } +  3 (\gamma +1) B[X,\lambda]+ \cO(\epsilon^2),
\ee
which coincide with the parameters  we defined in \eqref{constraints}. To derive this expression  we   used  
\bea
 \Omega_{\lambda \bar \lambda } & =&-  3 (\gamma +1) R_{X\bar X \lambda \bar \lambda} =  3 (\gamma +1) \, B[X,\lambda], \nonumber \\ 
\Re (Y_{\lambda \lambda}) &=&  \sqrt{3 (\gamma +1)} \, \Re ( (\nabla_X\, m)_{\lambda \lambda} )\equiv\sqrt{3 (\gamma +1)}\,  \frac{\pd m_\lambda}{\pd X }.  
\label{OY}
\eea
 and here  $(\nabla_X\, m)_{\lambda \lambda} $ are the diagonal elements of $\nabla_X M_{IJ}$ in the basis that diagonalises $M$:
\be
(\nabla_X\, m)_{\lambda \lambda} = (U (\nabla_X\, M)  U^t)_{\lambda \lambda} . 
\label{dmdx}
 \ee
 This result was derived in \cite{Achucarro:2012hg} for the simplest case of one light field and one heavy field\footnote{The notation in \cite{Achucarro:2012hg} is slightly different, it corresponds to $\gamma\to b/3$.}. We emphasise that for almost separable K\"ahler functions \eqref{quasiSeparableG}, the parameters $B[X,\lambda]$ and $ \pd m_\lambda / \pd X$ are of order $\cO(\epsilon)$ or higher, but this does not need to be the case in general.
Note that the parameters $\pd m_\lambda/ \pd X$ involve the derivatives of the fermion masses along the sGoldstino direction together with additional terms which appear when  the Levi-Civita connection of the K\"ahler manifold   is expressed in the local frame associated to  Takagi's factorisation.\\

{\bibliography{randomSugraV4}}

\end{document}